\documentclass[letterpaper,11pt]{article}
 
\usepackage[margin=1in]{geometry}

\usepackage[T1]{fontenc}

\usepackage{amsmath, amsthm, amssymb, mathtools, bm, dsfont}
 
\usepackage[comma,longnamesfirst]{natbib}
\usepackage[dvips]{epsfig}         % For EPS figures
\usepackage{graphicx}              % For general graphics inclusion
\usepackage[table,usenames,dvipsnames]{xcolor}  % Enable table coloring (e.g. \rowcolor)

\usepackage{dcolumn, hhline, multirow, booktabs}
\usepackage[flushleft]{threeparttable}

\usepackage[percent]{overpic}
\usepackage{rotating}
\usepackage{wrapfig}
\usepackage{adjustbox}

\usepackage{enumerate}
\usepackage{enumitem}
\usepackage{paralist}  % Provides compactitem/compactenum

\usepackage{afterpage}
\usepackage{placeins}
\usepackage{lscape}  % (Load only once)

\usepackage{xr}
\usepackage[toc,page]{appendix}
\usepackage{chngcntr}

\usepackage[breaklinks]{hyperref}
\usepackage{breakurl}

\usepackage{caption}
\usepackage{subcaption}

\usepackage{algorithm}
\usepackage{algorithmicx}
\usepackage{algpseudocode}

\usepackage[thinc]{esdiff}
 
\usepackage{diagbox}
\usepackage{slashbox}

\usepackage{titlesec}    % (Loaded once only)
\usepackage{sectsty}
\usepackage{setspace}

\usepackage{eurosym}
\usepackage{quiver}

\usepackage{dsfont}

\DeclareUnicodeCharacter{0301}{\'{e}}

\def \dsP {\text{$\mathds{P}$}}

\def \dsR {\text{$\mathds{R}$}}

\def \bbP {\text{$\mathbb{P}$}}

\DeclareMathOperator{\DIC}{DIC}

\DeclareMathOperator{\diag}{diag}

\DeclareMathOperator{\iid}{i.i.d.}

\DeclareMathOperator{\vecl}{vecl}

\newcommand{\lrc}[1]{{\left\{#1\right\}}}
\newcommand{\lrb}[1]{{\left[#1\right]}}
\newcommand{\lrp}[1]{{\left(#1\right)}}

\newcommand{\tinytxt}[1]{{\scalebox{0.6}{#1}}}

\newcommand{\dd}{\mathrm{d}}

\def \bvec {\text{\boldmath$b$}}

\def \lvec {\text{\boldmath$l$}}    
\def \mvec {\text{\boldmath$m$}}

\def \svec {\text{\boldmath$s$}}    
    
\def \uvec {\text{\boldmath$u$}}    
    
\def \wvec {\text{\boldmath$w$}}    
\def \xvec {\text{\boldmath$x$}}    
\def \yvec {\text{\boldmath$y$}}    
\def \zvec {\text{\boldmath$z$}}

\def \mL {\text{\boldmath$L$}}

\def \mU {\text{\boldmath$U$}}

\def \mX {\text{\boldmath$X$}}
\def \mY {\text{\boldmath$Y$}}
\def \mZ {\text{\boldmath$Z$}}

\def \mtildeL {\text{$\widetilde{\mL}$}}

\def \mtildeX {\text{$\widetilde{\mX}$}}

\def \alphavec        {\text{\boldmath$\alpha$}}

\def \deltavec        {\text{\boldmath$\delta$}}

\def \thetavec        {\text{\boldmath$\theta$}}

\def \muvec           {\text{\boldmath$\mu$}}

\def \tauvec          {\text{\boldmath$\tau$}}

\def \thetahatvec        {\text{\boldmath$\hat \theta$}}

\def \tautilde          {\text{$\tilde \tau$}}

\def \zerovec {\mathbf{0}}

\def \Omegabar {\bar\Omega}

\newcommand{\independent}{\perp\!\!\!\!\perp}

\def\N{\mathrm{N}}
\def\T{\mathrm{T}}
\def\Unifdst{\mathrm{Unif}}
\def\EL{\mathrm{EL}}

\def\USE{\mathrm{USE}}

\def\UST{\mathrm{UST}}

\def\TRUSE{\mathrm{TrUSE}}
\def\TRUST{\mathrm{TrUST}}
\def\TRUSN{\mathrm{TrUSN}}
\def\ETRUST{\mathrm{ETrUST}}

\def\Gammadst{\mathrm{Gamma}}
\def\Unifdst{\mathrm{Unif}}

\def\theequation{\thesection.\arabic{equation}}  
\def\abstract{\if@twocolumn
\section*{Abstract}
\else \normalsize 
\begin{center}
{\bf Summary\vspace{-.5em}\vspace{0pt}} 
\end{center}
\quotation 
\fi}
\def\endabstract{\if@twocolumn\else\endquotation\fi}

\makeatletter
\newcommand{\myappendix}[1]{
	\setcounter{section}{1}
        \renewcommand{\thesection}{A\arabic{section}}}

\usepackage{color}
\usepackage{colordvi}
\newlength{\breite}
\breite\textwidth
\newcounter{aufg}[section]
  {\refstepcounter{aufg}\noindent\textbf{Exercise \arabic{aufg}:}
   \\*[1ex]\noindent}{\vspace{.5cm}}
   
 \newcounter{notes}[section]
  {\refstepcounter{aufg}\noindent\textbf{}
   \\*[1ex]\noindent}{\vspace{.5cm}}
   
\usepackage{amsthm}  % Theoreme
\newtheorem{assumption}{Assumption}
\newtheorem{corollary}{Corollary}
\newtheorem{theorem}{Theorem}

\newtheorem{definition}{Definition}

\newtheorem{lemma}[]{Lemma}
\newtheorem*{beisp*}{Example}
\newtheorem{Proof}{Proof}


\newtheoremstyle{break}% name
  {}%         Space above, empty = `usual value'
  {}%         Space below
  {}% Body font
  {}%         Indent amount (empty = no indent, \parindent = para indent)
  {\bfseries}% Thm head font
  {.}%        Punctuation after thm head
  {\newline}% Space after thm head: \newline = linebreak
  {}%         Thm head spec
  
\theoremstyle{break}

\newcommand{\head}[2]%
 {\hrule \vspace{.15cm} {\sfbold Advanced Statistical Inference, Summer Term 2012, Georg-August-University G\"ottingen}\hfill
{\sfbold Sheet #1}\\
{\sfbold Prof. Dr. Thomas Kneib, Nadja Klein}\hfill {\sfbold #2}

\vspace{.2cm}
\hrule

\vspace{1cm}

}

\newcounter{auf}
{\refstepcounter{auf}
\begin{center}
\fcolorbox[gray]{0}{.95}{
\makebox[\breite]{
\textbf{Exercise \arabic{auf}}
}}\\*[1ex]\noindent
\end{center}
}{\vspace{.5cm}}

\newcounter{loes}[section]
{\stepcounter{loes}
\begin{center}
\fcolorbox[gray]{0}{.95}{
\makebox[\breite]{
\textbf{L"osung \arabic{loes}}
}}\\*[1ex]\noindent
\end{center}
}{}

{\begin{center}
\fcolorbox[gray]{0}{.95}{
\makebox[\breite]{
\textbf{Zu Aufgabe #1}
}}\\*[1ex]\noindent
\end{center}\vspace{1cm}
}{\vspace{1cm}}

\newcounter{ka}
{\refstepcounter{ka}
\begin{center}
\framebox[\textwidth]{
\textbf{Aufgabe \arabic{ka}} \hfill #1 Punkte
}\\*[1ex]\noindent
\end{center}
}{\vspace{1cm}}

\newcounter{lka}
{\refstepcounter{lka}
\begin{center}
\framebox[\textwidth]{
\textbf{L\"osung \arabic{lka}} \hfill #1 Punkte
}\\*[1ex]\noindent
\end{center}
}{\vspace{1cm}}

\titlespacing*\section{0pt}{0pt plus 4pt minus 2pt}{0pt plus 2pt minus 2pt}
\titlespacing*\subsection{0pt}{0pt plus 4pt minus 2pt}{0pt plus 2pt minus 2pt}
\titlespacing*\subsubsection{0pt}{0pt plus 4pt minus 2pt}{0pt plus 2pt minus 2pt}

\definecolor{myblue}{RGB}{0,73,114}

\renewenvironment{itemize}[1]{\begin{compactitem}#1}{\end{compactitem}}

\usepackage[skins,breakable]{tcolorbox}
\newtcolorbox{cvbox}[2][]{%
  blanker,
  after skip=8mm,  % enlarge distance to the next box
  title=#2,
  coltitle=cyan,
  #1 
}

\makeatletter
\def\@seccntformat#1{%
  \@ifundefined{#1@cntformat}%
    {\csname the#1\endcsname\quad}% default
    {\csname #1@cntformat\endcsname}%
}
\let\oldappendix\appendix
\renewcommand\appendix{%
  \oldappendix
  \newcommand{\section@cntformat}{\appendixname~\thesection\quad}%
} 
\makeatother

\begin{document}
  % Adjust display spacing as needed:
  \setlength{\abovedisplayskip}{0.15cm} 
  \setlength{\belowdisplayskip}{0.15cm}
  
  \pagestyle{empty}
  \begin{titlepage}

\newcommand{\titlename}{\LARGE\bfseries\color{myblue}  
%Tractable Unified Skew-t Distribution and Copula: Specification, Bayesian Inference and Application}
A Tractable Unified Skew-t Distribution and Its Copula for Heterogeneous Asymmetries}
 
\title{\titlename}
\author{Lin Deng, Michael Stanley Smith and Worapree Maneesoonthorn}
\date{\today}
\maketitle
\noindent
{\small Lin Deng is a Post-doctoral Fellow and Michael Smith is Chair of Management (Econometrics), both at the Melbourne Business School, University of Melbourne, Australia.
Worapree Maneesoonthorn is Associate Professor at the Department of Econometrics and Statistics at Monash University, Australia.  Correspondence should be directed to Michael Smith at {\tt mikes70au@gmail.com}. 
\\

\noindent \textbf{Acknowledgments:} This research was supported by The University of Melbourne’s Research Computing Services and the Petascale Campus Initiative. Michael Stanley Smith's research 
has been partially supported by the Australian Research Council (ARC) Discovery Project grant DP250101069, and that
of Worapree Maneesoonthorn by ARC Discovery Project Grant DP200101414. The authors are grateful to two referees and an Associate Editor whose thoughtful comments improved the paper.}\\

\noindent \textbf{Data Availability Statement:} Two datasets support the findings of this study. The first is available in the public domain at the AEMO website at {\tt www.aemo.com.au} and also included in the supplementary materials. The second is from the LSEG DataScope Select Database, which is a subscriber-only service. MATLAB code and data to implement the methodology in this study can be found at {\tt https://github.com/lindenglab/trust-distribution-copula}. 

%\normalsize
\newpage
\begin{center}
\mbox{}\vspace{2cm}\\
{\title{\titlename}}\\
\vspace{1cm}
{\Large Abstract \\} 
\end{center}
\vspace{-1pt}
\onehalfspacing
\noindent
{
Multivariate distributions that allow for asymmetry and heavy tails are important building blocks 
in many statistical models. The Unified Skew-t (UST) is a promising
choice because it is scalable and allows for a high level of flexibility in the asymmetry of the distribution.
However, it suffers from parameter identification and computational hurdles that have to date inhibited its use for modeling data. 
In this paper we propose a new tractable variant of the unified skew-t (TrUST) distribution
that addresses both challenges. 
Moreover, the copula of this distribution is shown to also
be tractable, while allowing for greater heterogeneity in asymmetric dependence over variable pairs than the popular skew-t copula. 
We show how Bayesian posterior inference for both the distribution and its copula can be computed using an extended likelihood derived from a generative representation of the distribution. The efficacy of this Bayesian method, and the enhanced flexibility of both the TrUST distribution and its implicit copula, is first demonstrated using simulated data. Applications of the TrUST distribution to highly skewed Australian electricity prices, and the TrUST copula to intraday U.S. equity returns, demonstrate how our proposed distribution and its copula can provide substantial increases in accuracy in practice. 
}
\vspace{20pt}
 
\noindent
{\bf Keywords}: Asymmetric Dependence, Bayesian Data Augmentation, Electricity Prices, Implicit Copulas, Intraday Equity Returns, Markov chain Monte Carlo, Skew-Elliptical Distributions.

\end{titlepage}
%\doublespacing
  
  \newpage
   
  \pagestyle{plain}
  \setcounter{equation}{0}
  \renewcommand{\theequation}{\arabic{equation}}
  
  % \tableofcontents
  % \newpage
  % \setcounter{page}{1}
  
  \section{Introduction}\label{sec:intro}
Skew-elliptical distributions are models of multivariate asymmetry that scale well. 
While there are different variants~(for some of these see~\citealp{genton2004skew}) 
those formed by conditioning on a latent truncated variable as suggested by~\cite{branco2001general} are prominent. This approach is often called ``hidden truncation'' and is used to form the skew-t distribution of~\cite{Azzalini_Capitanio_2003} (AC hereafter).
The AC skew-t distribution is a particularly popular choice in applied data analysis, with diverse applications in
renewable energy~\citep{Hering01032010}, actuarial studies~\citep{eling2012}, macroeconomic modeling~\citep{adrian2019vulnerable}, atmospheric science~\citep{morris2017space},
and in flow cytometric analysis~\citep{PyneetalPNAS2009,fruhwirth2010bayesian} among others. However, in $d$ dimensions skew-elliptical distributions typically have only $d$ parameters to control asymmetry across
$d(d-1)/2$ variable pairs, which can limit their effectiveness for modeling asymmetry. Unified skew-elliptical (USE) 
\footnote{\cite{arellano2006unification} use the abbreviation SUN for the Unified Skew-Normal distribution. In this paper, we break with this convention and adopt the word-ordered acronyms USN for this distribution, USE for the Unified Skew-Elliptical distribution,
	and UST for the Unified Skew-t distribution.}   
distributions~\citep{arellano2010multivariate} are an extension that introduces more latent variables and parameters to allow for a much more flexible form of asymmetry.
In particular, the unified skew-t (UST) discussed by~\cite{wang2024multivariate} is a promising
generalization of the impactful AC skew-t distribution. 
However, despite the strong potential of USE distributions, they have not been adopted for data analysis because (i)~the parameters are unidentified, and (ii)~likelihood evaluation and optimization is computationally difficult even in low dimensions. 
 
The current paper addresses these problems by 
specifying a novel subclass of the USE that is tractable.
The parameters of a general USE are unidentified with respect to permutation of the latent variables~\citep{wang2023non}. Our subclass resolves this problem by
ordering the eigenvalues of the conditional scale matrix of the latent variables.
%which is a major problem for its application in data analysis.   
%For this proposed subclass, it is shown that the USE parameters are identified
%under an ordering of the eigenvalues of the conditional scale matrix of the latent variables. 
%The parameters of a general USE are unidentified with respect to permutation of the latent variables~\citep{wang2023non}, which is a major problem for its application in data analysis.
The special case of a tractable UST distribution, which we label a TrUST distribution, is considered in detail. 
A generative representation is given for the proposed TrUST distribution that 
is used to define an extended likelihood that is easier to evaluate than the regular likelihood. From this, a Bayesian augmented posterior can be defined and computed using a
proposed Markov chain Monte Carlo (MCMC) sampler that is both fast and efficient, and 
for which
the identifying eigenvalue constraint is easy to impose. 
The proposed TrUST distribution nests the popular AC skew-t as a special case, but generalizes it 
to allow for richer multivariate asymmetry. An application of the TrUST distribution to both simulated data and highly skewed regional daily Australian electricity prices shows it captures asymmetry more 
accurately than the popular AC skew-t and the alternative skew-t of~\cite{sahu_dey_branco_2003}, and that this improves the quality of fit considerably. 

However, the main advantage of the proposed TrUST distribution is that its implicit copula is also tractable. Every multivariate 
continuous distribution has a unique implicit copula; see~\cite{smith2021implicit} for 
an introduction to this class of copulas. Currently, implicit copulas of differing skew-t distributions are used to capture the strong 
asymmetric dependence often found in financial data; see~\cite{smith_gan_kohn_2010}, \cite{Christoffersen_Errunza_Jacobs_Langlois_2012}, \cite{creal2015high}, \cite{lucas2017}, \cite{opschoor2021}, \cite{oh_patton_2023} and \cite{deng2024large} for examples. Of these,~\cite{deng2024large} show that the implicit 
copula of the AC skew-t distribution allows for the greatest level of asymmetric dependence. But the implicit copula of the TrUST distribution (hereafter the TrUST copula) allows for a much greater level of
heterogeneity in asymmetric dependencies over variable pairs than the AC skew-t copula. We show here that by doing so, the TrUST copula can better
capture the dependence between financial returns data.
To estimate the TrUST copula parameters a Bayesian augmented posterior is specified using 
a modification of the extended likelihood of the TrUST distribution, which is then evaluated using an MCMC sampler. Expressions for the Kendall and Spearman
rank correlations for this TrUST copula are also derived. 
As far as we are aware, ours is the first paper to construct the implicit copula of any USE 
distribution and show how to compute statistical inference for its parameters.

\cite{deng2024large} use the implicit copula of the AC skew-t distribution to capture dependence between intraday equity returns. 
In our main application, we extend their analysis to show that the TrUST copula is more effective at capturing such 
dependence than the AC skew-t copula. This is done for five large equities and an 
equity market volatility index (VIX) using data collected both before and during the COVID-19 pandemic. The performance of the TrUST copula is compared with that of both the symmetric Student-t copula and the AC skew-t copula using the Deviance Information Criterion (DIC) and cumulative log-score to assess accuracy. The results indicate that the TrUST copula better captures heterogeneity in asymmetric dependence across variables. High levels of heterogeneity are also found
by~\cite{le2021covid} and \cite{ando2022quantile} using network models of tail dependencies. Finally, we note that~\cite{durnte19} shows that the posterior of the coefficients of a probit model is USN. This is very different than what we undertake in this paper, which is to specify a tractable UST distribution and its copula for modeling data.

The rest of this paper is organized as follows. Section~\ref{sec:02} briefly introduces the USE distribution and the special case of the UST, after which our proposed tractable USE variant is outlined.
Section~\ref{sec:03} considers the case of the TrUST distribution in detail. Section~\ref{sec:04} outlines estimation methodology for the TrUST distribution,
and illustrates its efficacy using simulated and real data. Section~\ref{sec:05} presents the TrUST copula, derivation of its rank correlations, Bayesian inference and a simulation example. Section~\ref{sec:06} shows how our TrUST copula captures heterogeneity in asymmetric dependence between high-frequency financial variables; Section~\ref{sec:07} concludes. Extra results are given in the Appendix, while an Online Appendix contains additional material and all proofs. MATLAB code and data to implement the methodology in this study can be found at {\tt https://github.com/lindenglab/trust-distribution-copula}.

  \section{Unified Skew-Elliptical Distributions}\label{sec:02}
In this section, the unified skew-elliptical (USE) distribution is introduced, including the special case of the unified skew-t (UST). Our tractable variant of 
the USE
distribution is then outlined in detail. We consider the standardized case (i.e. with zero mean and 
unit scale) because this is used in forming the implicit copula in Section~\ref{sec:05}. Generalization of the distribution by
multiplying the random vector by a diagonal scaling matrix and adding a location parameter is straightforward.

\subsection{Unified Skew-Elliptical Distribution}
\cite{arellano2010multivariate} extend the popular skew-elliptical distribution
to allow for additional skew parameters.
It can be specified for the standardized case as follows.
Let $\mX\in \dsR^d$ and $\mL \in \dsR^q$, for $q\geq 1$, be jointly (radially symmetric)  elliptically 
distributed as
\begin{equation}\label{eq:sue_rep}
	\lrp{ \begin{array}{c} \mX \\ \mL \end{array} } \sim \EL \lrp{\zerovec, R, g},
	\quad R  = \lrp{ \begin{array}{ll} \Omega & \Delta^\top \\ \Delta & \Sigma \end{array} }, 
\end{equation}
where $R$ is a $(d+q) \times (d+q)$ correlation matrix partitioned to be consistent with $(\mX^\top,\mL^\top)^\top$, $\zerovec$ is a zero mean vector\footnote{Throughout this paper we use $\zerovec$ to denote a conformable vector or 
	matrix of zeros without explicitly denoting its dimension.},
and $g: [0, \infty) \to [0, \infty)$ is the density generator function; see~\citet[Chp. 2]{Fang_Kotz_Ng_1990} for specification of an elliptical distribution.
Then the USE distribution can be defined as
\begin{equation}\label{eq:cond_trunc}
	\mZ \overset{\dd}{=} \lrp{\mX | \mL > \zerovec}\,,  
\end{equation}
where ``$\overset{\dd}{=}$'' denotes equality in distribution, and the inequality $\mL>\zerovec$ holds element-wise. The elements of $\mL$ are latent variables and the
specification at~\eqref{eq:cond_trunc} 
is often called ``hidden truncation'' \citep{arnold2000hidden,arnold2004elliptical}. 
The $(q\times d)$ matrix $\Delta$ contains parameters that control the level and direction
of skew. When $\Delta$ is a full matrix (which is what we assume in this paper) then $q$ is typically low, such as $q\in\{1,2,3,4\}$.

Let 
$f_{\EL}(\yvec;V,g)$ and $F_{\EL}(\yvec;V,g)$ denote the density and distribution 
functions, respectively, of a zero-mean elliptically distributed random variable $\mY\sim \EL \lrp{\zerovec, V, g}$ with scale matrix $V$ and density generator $g$. Also, for the partition $\mY=(\mY_1^\top,\mY_2^\top)^\top$, the conditional $\mY_1 \mid \mY_2$ is elliptically distributed~\citep{Fang_Kotz_Ng_1990} with density generator
denoted here as \( g_{\mY_2} \).
Then, the density of $\mZ$ is obtained by using Bayes' theorem to evaluate the 
conditional density on the righthand side of~\eqref{eq:cond_trunc}, giving 
\begin{equation}\label{eq:sue_pdf}
	f_{\USE}\lrp{\zvec; \Omega, \Delta, \Sigma, g } = f_{\EL}\lrp{\zvec; \Omega, g} \frac{ F_{\EL}\lrp{\Delta \Omega^{-1} \zvec;  \Sigma - \Delta \Omega^{-1} \Delta^\top, g_{\mX} } }{ F_{\EL} \lrp{\zerovec;  \Sigma, g } }\,.
\end{equation}
If a random variable $\mZ$ has this density, then we write
$\mZ \sim \USE_{q}\lrp{ \Omega, \Delta, \Sigma, g} $.

The number of elements in $\Delta$ increases linearly with $q$. In this way, $q$ plays a role for multivariate asymmetry that is analogous to that of the number of factors in a traditional factor model for the covariance. 
When \(q = 1\), the USE distribution reduces to the skew-elliptical distribution introduced by \cite{branco2001general} and \cite{Azzalini_Capitanio_2003}. In this case there is only a single skew parameter for each dimension, so that
\(\Delta \equiv \deltavec^\top = \lrp{\delta_1,\ldots,\delta_d}^\top \). 
Because $\deltavec$ is constrained, for computing inference it is common to 
transform it to an unconstrained parameterization via the one-to-one transformation 
$\alphavec = (1 - \deltavec^{\top} \Omega^{-1} \deltavec)^{-1 / 2} \Omega^{-1} \deltavec$ with inverse $\deltavec = (1+\alphavec^{\top} \Omega \alphavec)^{-1 / 2} \Omega \alphavec$. 

\subsection{Unified Skew-t Distribution}\label{sec:ust}
%\cite{arellano2006unification} consider the special case where~\eqref{eq:sue_rep}
%is a normal distribution, 
%leading to the unified skew-normal distribution. Beyond this, 
The elliptical distribution
with greatest applied potential is the t distribution, which
has density generator function $g(x)=(1+x/\nu)^{-(\nu+d)/2}$ that introduces the parameter $\nu>0$.
Assuming   
$(\mX^\top,\mL^\top)^\top$ is multivariate t with location zero, scale matrix $R$
and $\nu$ degrees of freedom, the density at~\eqref{eq:sue_pdf} is given by
\begin{equation}\label{eq:pdf_SUT}
	f_{\UST,q}(\zvec; \Omega,\Delta,\Sigma,\nu)= t_d(\zvec ; \Omega, \nu) \frac{T_q\lrp{ \sqrt{\frac{\nu+d}{\nu+Q(\zvec)}} \Delta \Omega^{-1} \zvec; \Sigma - \Delta \Omega^{-1} \Delta^\top,\nu+d }}{T_q\lrp{\zerovec; \Sigma, \nu}}.
\end{equation}
Here, $Q(\zvec) = \zvec^\top\Omega^{-1}\zvec$, while \( t_d\lrp{\yvec; V, \upsilon} \) and \( T_d(\yvec; V, \upsilon) \) denote the density and distribution functions, respectively, of a zero mean $d$-dimensional multivariate t random variable with scale matrix $V$ and degrees of freedom \(\upsilon\) evaluated at $\yvec$. If a random variable $\mZ$ has density at~\eqref{eq:pdf_SUT}, then we write
$\mZ \sim \UST_{q}\lrp{\Omega, \Delta, \Sigma, \nu} $.
 
%Under the hidden truncation framework, the Unified Skew-t (UST) distribution is derived by extending the approach of , originally developed for the Unified Skew-Normal (USN) distribution. In this formulation, the latent variable is generalized from a univariate to a multivariate setting by introducing an unobserved vector \(\mL \in \dsR^q\). The components of \(\mL\) are assumed to be jointly distributed with the observable vector \(\mX\), thereby generalizing the density generator \(g\) to a Student-t density with finite degrees of freedom (\(\nu > 2\)).
%Consequently, the UST density function is given by:
%\begin{equation}\label{eq:pdf_SUT}
%	f_{\UST,q}(\zvec; \Omega,\deltavec,\nu)= t(\zvec ; \Omega, \nu) \frac{T\lrp{ \sqrt{\frac{\nu+d}{\nu+Q(\zvec)}} \Delta \Omega^{-1} \zvec; \Sigma - \Delta \Omega^{-1} \Delta^\top,\nu+d }}{T\lrp{\zerovec; \Sigma, \nu}}.
%\end{equation}
%Here, \( t_d\lrp{\xvec; \Sigma, \nu} \) and \( T_d(\xvec; \Sigma, \nu) \) represent the \(d\)-dimensional density and distribution functions, respectively, evaluated at \(\xvec\), with covariance matrix \(\Sigma\) and degrees of freedom \(\nu\). And $Q(\xvec) = \xvec^\top\Sigma^{-1}\xvec$ is the squared Mahalanobis distance. 

Three special cases of the UST distribution are:
(i)~when $\nu\rightarrow \infty$, the UST converges to the unified skew-normal of~\cite{arellano2006unification} (which is a re-parameterization of the closed skew normal of~\citealp{gonzalez2004closed});
%gupta2013usn} 
%and~\cite{arellano2022USN};
(ii)~when $\Delta$
is a matrix with all elements equal to zero, then $\mZ$ is distributed symmetric multivariate t with location zero, scale matrix $\Omega$
and $\nu$ degrees of freedom; and (iii)~when \(q=1\), the UST distribution is the AC skew-t distribution. 
\cite{wang2024multivariate} provide a comprehensive overview of the UST distribution and its properties.
%We denote by
%\(
%\mZ \sim \ST\lrp{\zerovec, \Omega, \alphavec, \nu}
%\)
%a random vector following the skew-$t$ distribution, whose density function is given by:
%\begin{equation}\label{eq:st_pdf}
% 	f_{\ST}(\zvec; \Omega,\deltavec,\nu)=2 t(\zvec ;\Omega, \nu) T \lrp{ \sqrt{\frac{\nu+d}{\nu+Q(\zvec)}} \alphavec^{\top} \zvec  ; 1, \nu+d }.
%\end{equation}
%
%When $\nu \rightarrow \infty$, the distribution degenerate to skew-normal distribution, and when $\alphavec = \zerovec$, it degenerates to Student-$t$ distribution. When both $\nu \rightarrow \infty$ and $\alphavec = \zerovec$ applied, the distribution tends to the Gaussian distribution.

%Compared to the SE distribution, the USE distribution increases the number of skewness parameters from \(1\) to \(q\) for each margin in the model, providing greater flexibility in capturing asymmetry. However, this enhancement comes at the cost of losing the direct transformation between unconstrained and constrained skewness parameters. Additionally, the skewing term in Equation \eqref{eq:sue_parts} significantly increases computational complexity, as it now involves solving a 
%$q$-dimensional integral with observation $\zvec$ in a sample size $n$. 

\subsection{Tractable Unified Skew-Elliptical Distribution}\label{sec:truse}
The novel subclass of the USE distribution is now outlined,
that we call here a Tractable Unified Skew-Elliptical (TrUSE) distribution. It employs
two assumptions for the distribution of $\bm{L}|\bm{X}$. The first is an assumption that the scale matrix is diagonal, which reduces the parameter space. 
The second is an ordering of the elements of this diagonal scale matrix, which identifies
the latent variable permutation. Under these two assumptions
$\Omega$ is an unconstrained correlation matrix, $\Delta$ is a full matrix with a unique row order, and $\Sigma$ is a full matrix that is a function of $\Omega$ and $\Delta$. 
%which 
%reduces the computational burden of evaluating the density at~\eqref{eq:sue_pdf}
%while maintaining the ability of the distribution to capture asymmetry.

%In this research, we introduce a novel subclass of the unified skew elliptical distribution, termed the Tractable Unified Skew-Elliptical (TrUSE) distribution. This distribution still builds upon the existing unified skew elliptical framework by leveraging the hidden truncation mechanism. However, the TrUST distribution distinguishes itself by incorporating the assumptions in the latent conditional independence (CI) property. This generalization allows for the decomposition of $q$-dimensional skewing term into more manageable, $q$ separate $1$-dimensional components. And it significantly reducing the computational burden while maintaining the ability to capture asymmetries within the data.

\begin{assumption}[Diagonal Conditional Scale Matrix]\label{asmp:CI} Let $(\mX^\top,\mL^\top)^\top$ follow
the joint elliptical distribution at~\eqref{eq:sue_rep}, so that  
%Then assume each component of $\mL$ is independent, conditional on $\mX$, so that $L_i \independent L_j | \mX $, for all $\lrp{i,j} \subset \lrc{1,\ldots,q}$ and $q > 1$. 
$\mL|\mX$ is also elliptically distributed. Then we assume this conditional distribution has a diagonal scale matrix $H$. 
\end{assumption}
Under this assumption, the scale matrix $\Sigma$ is a deterministic function of 
$\{\Omega,\Delta\}$, and the density at~\eqref{eq:pdf_SUT}
is simplified, as summarized by the following two lemmas.
\begin{lemma}\label{lem1}
If Assumption~\ref{asmp:CI} holds, then the scale matrix of the marginal elliptical distribution of $\mL$ in~\eqref{eq:sue_rep} is given by
$\Sigma = I_q + \left(M-\diag(M)\right)$, with $M=\Delta \Omega^{-1}\Delta^\top$.
\end{lemma}
\begin{lemma}\label{lem2}
If Assumption~\ref{asmp:CI} holds, then
\[H=\Sigma-\Delta \Omega^{-1} \Delta^\top = \mbox{diag}\left((1 - \deltavec_1^\top \Omega^{-1} \deltavec_1),\ldots,(1 - \deltavec_q^\top \Omega^{-1} \deltavec_q)\right)\,,\]
is a diagonal matrix, and the distribution function of $\mL|\mX$ is given by
\[
F_{\EL}\lrp{\Delta \Omega^{-1} \zvec;  H, g_{\mX} }=
%\prod_{k=1}^{q} F_{\EL} \lrp{\alphavec_k^\top \zvec;1, g_\mX}
F_{\EL}\lrp{A^\top \zvec;  I_q, g_{\mX} }
\]
where $\Delta^\top = [\deltavec_1|\deltavec_2| \cdots| \deltavec_q]$, $A=[\alphavec_1|\alphavec_2|\cdots|\alphavec_q]$ with
$\alphavec_k = \lrp{1 - \deltavec_k^\top \Omega^{-1} \deltavec_k}^{-1/2} \Omega^{-1} \deltavec_k$ for $k=1,\ldots,q$.
\end{lemma}
Proofs of both lemmas are found in Online Appendix~\ref{app:proof}, and together they can be used to define our proposed tractable USE distribution as follows.

%Note that CI is also the dual problem of maximum entropy completion \citep{vandenberghe1998determinant}, defined as the maximum determinant for completing the positive definite correlation matrix parameter \(\Sigma\) given known \(\Omega\) and \(\Delta\). In other words, the explicit solution is given by  
%\[
%\tilde\Sigma = \underset{R \succ 0}{\argmax}\,\log\det\left(\Sigma - \Delta\,\Omega^{-1}\Delta^\top\right).
%\]

\begin{definition}[TrUSE Distribution]
	\label{def:truse_dist}
 	If $\mZ \sim \USE_{q}\lrp{ \Omega, \Delta, \Sigma, g }$ and Assumption~\ref{asmp:CI} holds, then $\mZ$ is said to be distributed Tractable Unified Skew-Elliptical (TrUSE) with density function 
 	%and is
 	%written as $\mZ \sim \TRUSE_{q}\lrp{\zerovec, \Omega, \mA, g }$ has density function
\begin{equation}\label{eq:fse_pdf}
	f_{\TRUSE, q}\lrp{\zvec; \Omega, A, g }  
	% = \dsP \lrp{\mX \leq \zvec} \frac{ \prod_{k=1}^{q} \dsP \lrp{L_k > 0 | \mX = \zvec} }{ \dsP\lrp{} }
	= f_{\EL}\lrp{\zvec; \Omega, g} 
	\frac{ F_{\EL}\lrp{A^\top \zvec;  I_q, g_{\mX} } }{ F_{\EL} \lrp{\zerovec;  \Sigma, g } }.
\end{equation}
where $\Sigma$ is given in Lemma~\ref{lem1}, $\alphavec_k$ is given in Lemma~\ref{lem2}, $g$ is the density generator function at~\eqref{eq:sue_rep},
and $g_\mX$ is the density generator function of the elliptical distribution of $\mL|\mX$.
We write 
$\mZ \sim \TRUSE_{q}\lrp{\Omega, A, g }$.
%We denote each vector in the transpose of constrained skewness matrix as \(\Delta^\top = \lrp{\deltavec_1, \ldots, \deltavec_q}  \). For \(k = 1, \ldots, q\), there exists a one-to-one transformation from \(\deltavec_k\) to an unconstrained skewness vector \(\alphavec_k\), given by:
%\begin{equation}\label{eq:alphavec2deltavec}
%	\alphavec_k = \lrp{1 - \deltavec_k^\top \Omega^{-1} \deltavec_k}^{-1/2} \Omega^{-1} \deltavec_k, \quad \text{and} \quad \deltavec_k = \lrp{1 + \alphavec_k^\top \Omega \alphavec_k}^{-1/2} \Omega \alphavec_k,
%\end{equation}
%and denoting the unconstrained skewness matrix as $\mA = \lrp{\alphavec_1, \ldots, \alphavec_q}$. 
\end{definition}

Parameterization of the TrUSE distribution in terms of $A$ is attractive because each $\alphavec_k\in \mathbb{R}^d$ is unbounded given $\{\Omega,\alphavec_{j\neq k}\}$, whereas $\deltavec_k$ has 
complex nonlinear bounds given $\{\Omega,\deltavec_{j\neq k}\}$. We show later that this property is useful for constructing MCMC schemes to generate $\alphavec_k$, rather than $\deltavec_k$, from its posterior.\footnote{Note that there is a one-to-one transformation between 
$\Delta$ and $A$, and that in this paper we denote the USE/USN/UST distributions in terms of $\Delta$, while we
denote the TrUSE/TrUSN/TrUST distributions in terms of $A$.}

The TrUSE distribution is more tractable than the general USE for three reasons. First, it simplifies
the parameter space because $\Sigma$ is not a parameter, but instead a deterministic function of $\{\Omega,A\}$. In contrast, 
for a general USE $\Sigma$ is a parameter with off-diagonal elements that have complex nonlinear bounds given $\{\Omega,\Delta\}$.   
Second, it is more computationally tractable to compute inference because (i) the density at~\eqref{eq:sue_pdf} is simplified, and (ii) efficient MCMC sampling is possible from the augmented posterior constructed using an extended likelihood based on the joint of $(\mX,\mL)$. Third, a parameter identification constraint that is straightforward to impose exists as outlined later. 
In contrast, it is hard to compute statistical inference
for the unconstrained parameters of the general USE distribution. In particular, the tractability of the TrUST distribution is illustrated in Section~\ref{sec:04}.

Adopting
Assumption~\ref{asmp:CI} alone is insufficient to identify the latent variable $\mL$ permutation for the USE distribution. \cite{wang2023non} discuss this problem in detail, which is characterized by the lemma below.

\begin{lemma}[Permutation Un-identification]
	\label{theo:PI}
	Let \(G(q) = \{\pi : \pi \text{ is a bijection from } \{1, \ldots, q\} \text{ to itself}\}\) be the set of all possible orderings of the indices \(\{1, \ldots, q\}\). Then, the distribution of $ \mZ \stackrel{\dd}{=} \lrp{\mX | \mL > \zerovec} \sim \TRUSE\lrp{ \Omega, A, g}$ is unidentified under permutation of latent variable $\mL$. 
	That is, for every permutation $\pi \in G(q) $, $\mZ \overset{\dd}{=} \mZ_{\pi}$, where $ \mZ_{\pi} \overset{\dd}{=} \mX | \mL_\pi > \zerovec$ and  $\mL_\pi = \lrp{L_{\pi(1)},\ldots,L_{\pi(q)}}^\top$. 
\end{lemma}

To identify the latent variable permutation, a constraint based on the following ordering of
the eigenvalues of the scale matrix $H$ of the conditional 
distribution of $\mL|\mX$ is used.
 
\begin{assumption}[Latent Permutation Constraint]\label{asmp:LP}
Let \(\mZ \overset{\dd}{=} \lrp{\mX | \mL > \zerovec}\) follow a USE distribution, and
$H =\Sigma-\Delta \Omega^{-1}\Delta^\top$ be the scale matrix of the conditional 
distribution of $\mL|\mX$ with eigenvalues $\lambda_1,\ldots,\lambda_q$. Then we assume the permutation \(\pi^*\in G(q)\) of the latent vector $\mL$ and associated columns of $\Delta^\top$, satisfies:
\begin{equation*}
	\lambda_{\pi^*(1)}\leq \ldots \leq \lambda_{\pi^*(q)}\,.
\end{equation*}
% solves the problem
%	\begin{equation*}
%		\pi^* = \argmax_{\pi \in G(q)} \sum_{k=1}^q \pi(k) \lambda_{\pi(k)}.
%	\end{equation*}
\end{assumption}

This assumption, when applied to the TrUSE distribution under Assumption~\ref{asmp:CI}, results in a permutation of
the latent vector $\mL$ that identifies the columns of \(A\) and \(\Delta^\top\), as below.

\begin{theorem}[Latent Permutation Identification]\label{thm1}
	Under Assumption~\ref{asmp:CI} , $H=\Sigma-\Delta \Omega^{-1} \Delta^\top=\mbox{diag}(h_1,\ldots,h_q)$ is a diagonal matrix, so that $\lambda_k=h_k$ for $k=1,\ldots,q$. 
	Under the additional Assumption~\ref{asmp:LP}, the permutation of $\mL$ (and therefore also the columns of $A$ and $\Delta^\top$) in the TrUSE distribution is based on the optimal ordering:
	\begin{equation*}
		% \pi^*: 0 \leq \lambda_{\pi^*(1)} \leq \lambda_{\pi^*(2)} \leq \cdots \leq \lambda_{\pi^*(q)} \leq 1,
		\pi^*: 0 \leq h_{\pi^*(1)} \leq \cdots \leq h_{\pi^*(q)} \leq 1,
	\end{equation*}
	% where \(\lambda_k(\bfH) = 1 - \deltavec_k^\top \Omega^{-1} \deltavec_k\) then \( \lambda_{\pi^*(k)}(\bfH) = 1 - \deltavec_{\pi^*(k)}^\top \Omega^{-1} \deltavec_{\pi^*(k)}\), for \(k = 1, \ldots, q\).
	where \( h_{\pi^*(k)} = 1 - \deltavec_{\pi^*(k)}^\top \Omega^{-1} \deltavec_{\pi^*(k)}\), for \(k = 1, \ldots, q\). Moreover, when these inequalities are strict, the ordering is unique.
\end{theorem}

In practice, because the posterior distribution of $(h_1,\ldots,h_q)^\top$ is continuous, the inequalities in Theorem~\ref{thm1} are strict, so that the ordering is unique. 
Moreover, the order of $(h_1,\ldots,h_q)^\top$ corresponds to a unique order of the rows of $\Delta$ because from Lemma~\ref{lem1}, $H=\Sigma-\Delta\Omega^{-1}\Delta^\top=I_q-\mbox{diag}(\Delta \Omega^{-1} \Delta^\top)$.
In the remainder of the paper Assumption~\ref{asmp:LP} is adopted to identify the latent 
coordinate ordering of $\Delta, A$ and $\Sigma$ in
the TrUSE (and TrUST) distributions with respect to $\pi^*$. 
Imposing this constraint is straightforward within
an MCMC scheme as discussed in Section~\ref{sec:bayesdist}.

\subsection{Discussion of TrUSE Distribution}
We finish this section with three additional comments on the TrUSE distribution. First,
 Assumption~\ref{asmp:CI} is on the distribution of $\bm{L}|\bm{X}$, not on that of $\bm{X}$, so that the TrUSE distribution can capture the same degree of rank correlation as the
USE distribution. Second, Assumptions~\ref{asmp:CI} and~\ref{asmp:LP} are identifying assumptions that enable the USE distribution to be useful in data analysis. In comparison, \cite{wang2024multivariate} suggest (but do not implement) some other ways to identify the parameters
in a UST distribution that involve much stronger restrictions on $\Delta$ and/or $\Omega$. 
However, these 
reduce substantially the ability to capture variation in pairwise correlations and/or asymmetry across variable pairs, which is the main advantage of the UST over skew-t distributions.  
Third, the  unified skew normal subclass of~\cite{arellano2006unification} (equivalent to  the closed skew normal of~\citealp{gonzalez2004additive}) has been used previously to model data. However, these applications employ direct constraints on $\Omega,\Sigma,\Delta$ based on the structure of spatial or time series data~\citep{zareifard2013non,zareifard2025skew} that are stronger
than those suggested here.
  \section{Tractable Unified Skew-t Distribution}\label{sec:03}
The special case of the tractable unified skew-t (TrUST) distribution that is the
focus of our paper is now outlined. 
%Its marginal 
%distribution is given, along with a generative representation that is used
%to construct the extended likelihood in Section~\ref{sec:04}.
 
\subsection{Joint Distribution}
The TrUST distribution results from selecting a t distribution with $\nu$ degrees of freedom as the choice of elliptical distribution in Section~\ref{sec:truse}. In this case, the density at Definition~\ref{def:truse_dist} is
\begin{equation}\label{eq:trust_joint_pdf}
    f_{\TRUST,q} \lrp{\zvec; \Omega, A, \nu} = t_d(\zvec; \Omega, \nu) \frac{ T_q\lrp{\sqrt{\frac{\nu+d}{\nu+Q(\zvec)}} A^\top \zvec; I_q, \nu+d }}{T_q\lrp{\zerovec; \Sigma, \nu}},
\end{equation}
where $Q(\zvec) = \zvec^\top\Omega^{-1}\zvec$, $A=[\alphavec_1|\alphavec_2|\cdots|\alphavec_q]$, $\Delta^\top = [\deltavec_1|\deltavec_2| \cdots| \deltavec_q]$, 
$\alphavec_k = \lrp{1 - \deltavec_k^\top \Omega^{-1} \deltavec_k}^{-1/2} \Omega^{-1} \deltavec_k$ for $k=1,\ldots,q$. Under Assumption~\ref{asmp:CI}, $\Sigma = I_q + \left(M-\diag(M)\right)$ with $M=\Delta \Omega^{-1}\Delta^\top$. 
Under Assumption~\ref{asmp:LP} the rows of $\Delta$ are ordered so that 
the elements of $H=\Sigma-\Delta \Omega^{-1} \Delta^\top = \mbox{diag}\left((1 - \deltavec_1^\top \Omega^{-1} \deltavec_1),\ldots,(1 - \deltavec_q^\top \Omega^{-1} \deltavec_q)\right)$ are in ascending order.
If a random vector $\mZ$ has the above density, then we write $\mZ \sim \TRUST_{q}\lrp{\Omega, A, \nu}$. If $q=1$ then the TrUST, UST and 
AC skew-t all coincide.

If $\nu \rightarrow \infty$, then this is the tractable unified skew-normal (TrUSN) distribution, with density 
\begin{equation*}
    f_{\TRUSN,q} \lrp{\zvec; \Omega, A} = \phi_d(\zvec; \Omega) \frac{\prod_{k=1}^{q} \Phi_1 \lrp{\alphavec_k^\top \zvec; 1}}{\Phi_q \lrp{\zerovec; \Sigma}},
\end{equation*}
where $\phi_d\lrp{\xvec; \Omega}$ and $\Phi_d\lrp{\xvec; \Omega}$ are density and distribution functions, respectively, for a $N\lrp{\zerovec, \Omega}$ distribution. If a random vector $\mZ$ has the above density, then
we write $\mZ \sim \TRUSN_{q}\lrp{\Omega,A}$.  

\begin{figure}[tbh]
	\centering
	\includegraphics[width=0.9\textwidth]{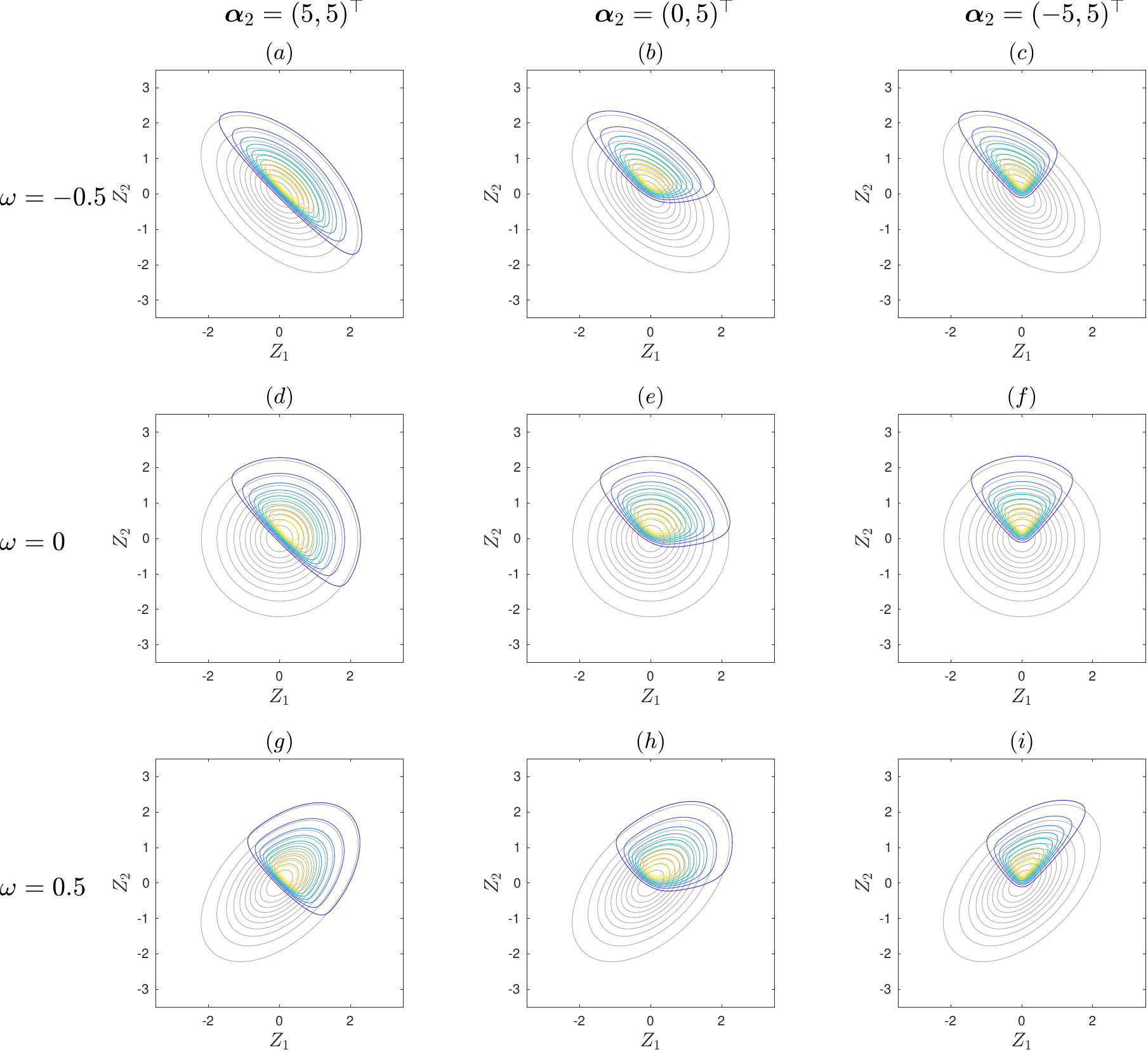}
	\caption{  
    Contour plots of the bivariate TrUST \((q = 2)\) density with  \(\nu = 5\) and \(\alphavec_1 = (5, 5)^\top\) fixed. Rows vary by correlation \(\omega \in \{-0.5, 0, 0.5\}\) (top to bottom), and columns by second skewness vector \(\alphavec_2 \in \{(5, 5)^\top, (0, 5)^\top, (-5, 5)^\top\}\) (left to right) which vary by only the first element. The shadowed contours represent the corresponding Student-t distributions for the same values of \(\omega\) and \(\nu\).  
	}
	\label{fig:fst_pdfs}
\end{figure}

Figure~\ref{fig:fst_pdfs} presents the bivariate density contours of the TrUST distribution, when $d=2$, $q=2$, $\nu = 5$ and the first skew parameter vector is \(\alphavec_1 = (5, 5)^\top\). Each row corresponds to a different correlation parameter \(\omega \in \lrc{-0.5, 0, 0.5}\), ordered from top to bottom. The columns correspond to \(\alphavec_2 \in \lrc{ (5, 5)^\top, (0, 5)^\top, (-5, 5)^\top }\) and show the impact of varying the second skew parameter vector. If $\alphavec_2=(0,0)^\top$ then the distribution is equivalent to the AC skew-t, and the contours are visually indistinguishable from that when $\alphavec_2=(5,5)^\top$, so is not presented. 
%The shadowed contour represents the Student-\(t\) distribution with the given correlation parameter \(\omega\), and degrees of freedom \(\nu\).  

\subsection{Marginal Distribution}\label{sec:fst_margin}
Because the UST distribution is closed under marginalization,
the marginals of a TrUST distribution are UST with parameters derived from those of the joint as below.

\begin{lemma}\label{lem:marginal}
Let $\mZ \sim \TRUST_{q}\lrp{\Omega, A, \nu}$ with $\Delta$ computed from $A$, and $\Sigma$ computed from $\{\Delta,\Omega\}$ as in Lemma~\ref{lem1}. If \( J \subset \lrc{1,\ldots,d} \) is a subset of $d_J<d$ indices, then the marginal distribution of the random vector $\mZ_J = (Z_{J(1)},Z_{J(2)},\ldots,Z_{J(d_J)})$ is \( \mZ_J \sim \UST_{ q} \lrp{\Omega_J, \Delta_J, \Sigma, \nu} \), 
%\begin{equation}\label{eq:fst_margin_pdf}
%    f_{\UST,q} \lrp{\zvec_J; \zerovec, \Omega_J,\Delta_J, \tildeSigma, \nu} = t (\zvec_J; \Omega_J, \nu) \frac{T \lrp{ \sqrt{\frac{\nu+d_J}{\nu+Q(\zvec_J)}} \Delta_J \zvec_J ;  \tildeSigma - \Delta_J \Omega_J^{-1} \Delta_J^\top, \nu+d_J }}{T \lrp{\nullvec;  \tildeSigma, \nu }},
%\end{equation}
%where $\Delta_{J} = \lrp{\deltavec_{J(1)}, \ldots, \deltavec_{J(d_J)}}$ contains the 
%columns of $\Delta$ indexed by $J$,
where $\Delta_{J}$ contains the corresponding columns of $\Delta$  
 and $\Omega_J=\{\omega_{ij}\}_{i,j\in J}$ is the sub-matrix of $\Omega=\{\omega_{ij}\}$ comprising the elements indexed by set $J$.
\end{lemma}

The case where $d_J=1$ is of particular importance when constructing the implicit
copula of a TrUST distribution in Section~\ref{sec:05}. Let $\Delta_j$ be the $j$th column of $\Delta$ (whereas $\deltavec_k^\top$ is the $k$th row of $\Delta$). Then, the marginal for the $j$-th element is $Z_j \sim \UST_{q}( 1, \Delta_{j}, \Sigma, \nu)$ for $j \in \lrc{1,\ldots,d}$ with density function
\begin{equation}\label{eq:trust_unimargin_pdf}
    f_{\UST,q} \lrp{z_j; 1,\Delta_j, \Sigma, \nu} = t_1(z_j;1, \nu) \frac{T_q \lrp{ \sqrt{\frac{\nu+1}{\nu+Q(z_j)}} \Delta_{j} z_j ;  \Sigma - \Delta_{j} \Delta_{j}^\top, \nu+1 }}{T_q\lrp{\zerovec;  \Sigma, \nu }}\,,
\end{equation}
where $Q(z_j)=z_j^2$. 
The distribution
and quantile functions are computed numerically from \eqref{eq:trust_unimargin_pdf} using the 
method given in~\cite{Yoshiba_2018} and~\cite{Smith_Maneesoonthorn_2018}.

\subsection{Generative Representation}\label{sec:gr}
For generating from the TrUST distribution, we use a scale mixture of normals representation of a t distribution for~\eqref{eq:sue_rep}. Let $W \sim \Gammadst \lrp{\nu/2, \nu/2}$ and
$(\mX^\top,\mL^\top)^\top|W\sim N(\zerovec,W^{-1}\odot R)$, where $R$ is defined
at~\eqref{eq:sue_rep} and
`$\odot$'
 is the product of a scalar with each element of a matrix. Then 
 the mixture of $(\mX^\top,\mL^\top)^\top|W$ over $W$ gives a multivariate $t$ 
 with location zero, scale matrix $R$ and $\nu$ degrees of freedom for $(\mX^\top,\mL^\top)$. Hidden truncation in~\eqref{eq:cond_trunc} defines $\mZ = \lrp{\mX | \mL > \zerovec}\sim \TRUST_{q}\lrp{\Omega, A, \nu}$.
 
From this representation, a draw of $\mZ$ can be obtained by completing the following three sequential
steps: (i) generate
$W \sim \Gammadst \lrp{\nu/2, \nu/2}$, (ii) draw $\mL|W \sim N(\zerovec,W^{-1}\odot\Sigma)$ constrained
so that $\mL>\zerovec$, and (iii) 
draw from the conditional distribution
\begin{equation*}
	\lrp{\mZ | \mL, W} \sim N \lrp{ \Delta^\top \Sigma^{-1} \mL , W^{-1} \odot \lrp{\Omega - \Delta^\top \Sigma^{-1} \Delta}}\,.
\end{equation*} 
Because the $(q\times 1)$ vector $\mL$ is low dimensional (e.g. $1\leq q \leq 3$ in our empirical work) generating from the constrained distribution at step~(ii) is both easy and fast using the approach of \cite{botev2017normal} or another method.

From this representation, the joint density of $(\mZ,\mL,W)$ can be written as the product
\begin{equation}\label{eq:jnt}
	f_{Z,L,W}(\zvec,\lvec,w)=\phi_d\left(\zvec;\Delta^\top \Sigma^{-1} \lvec, w^{-1} \odot (\Omega - \Delta^\top \Sigma^{-1} \Delta)\right) \phi_{\lvec>\zerovec}(\lvec ;\zerovec,w^{-1}\odot\Sigma) f_{Gam}(w;\nu/2,\nu/2)\,,
\end{equation} 
where  $f_{Gam}(w;\alpha,\beta)$ denotes the density of a $\mbox{Gamma}(\alpha,\beta)$ distribution, and $\phi_{\xvec>\zerovec}(\xvec ;\zerovec,V)$ is the density of a $N(\zerovec,V)$ distribution constrained so that $\xvec>\zerovec$. Equation~\eqref{eq:jnt} is used to define an
extended likelihood when computing Bayesian inference in Section~\ref{sec:04} below.
Two other alternative generative representations of the TrUST distribution are given in Part~A of the Online Appendix. However, that above is preferred because it provides
for a numerically stable extended likelihood. 

\subsection{Discussion of TrUST Distribution}
We make four additional comments on the proposed TrUST distribution. First, it is defined with zero mean and correlation matrix $R$ because it is convenient for formation of the implicit copula in 
Section~\ref{sec:05}, where location and scale are unidentified in the copula. Second, generalization of the distribution is straightforward by adding a location parameter 
and multiplying by scale parameters, as we do when modeling electricity prices in Section~\ref{sec:04}. 
Third, Assumptions~\ref{asmp:CI} and~\ref{asmp:LP} can also be adopted in
the ``extended UST'', where an additional parameter $\tauvec \in \dsR^q$
is introduced and $\mZ\overset{d}{=}\mX|\mL+\tauvec>\zerovec$ \citep{Azzalini_Capitanio_2003,arellano2006unification,wang2024multivariate}. Doing so further
generalizes the TrUST distribution; see Appendix~\ref{sec:etrust} for details. Fourth, the conditional distribution of a TrUST distribution is of an extended UST form, as discussed in Appendix~\ref{sec:condtrust}.

%\LD{This paragraph needs to rewrite:}
%The term \textit{extended} follows the naming in ST distribution from \cite{Azzalini_Capitanio_2003} that has hidden truncation on $ L + \tau > 0$. In the TrUST distribution with a hidden random vector, we denote the Extended-TrUST (E-TrUST) distribution for random vector \(\mZ \overset{\dd}{=} \mX | \mL + \tauvec > \zerovec\), where $\tauvec = \lrp{\tau_1,\ldots,\tau_q}^\top $. Derivations is in Appendix~\ref{app:extended}.

%
%Therefore, a draw from the TrUST distribution can be obtained by first generating
%$W \sim \Gammadst \lrp{\nu/2, \nu/2}$, then $\mL_0\sim N(\zerovec,\Sigma)$ 
%with posterior distribution for latent variables
%\begin{equation*}
%	L_k | \mZ, W, \thetavec \sim \N^+ \lrp{ m_k, W^{-1} r_k}, 
%	\quad \text{and} \quad
%	W|\mZ, \mL, \thetavec \sim \Gammadst\lrp{ a, b},
%\end{equation*}
%where $a = (d+q+\nu)/2$, $b = 1/2 \lrb{ \lrp{\mZ - \Delta^\top \tildeSigma^{-1}\mL}^\top\lrp{\Omega - \Delta^\top \tildeSigma^{-1} \Delta}^{-1}\lrp{\mZ - \Delta^\top \tildeSigma^{-1}\mL} + \mL^\top \tildeSigma^{-1} \mL + \nu }$.

  \section{Bayesian Inference and Application}\label{sec:04}
This section outlines how to compute Bayesian inference for the parameters of the TrUST distribution, and applies
the approach to both simulated data and highly skewed Australian regional electricity prices.  

\subsection{Extended Likelihood}\label{sec:bayesdist}
The inferential problem is for the location-scaled version of the TrUST distribution, where \(\mY = \lrp{\muvec + S \mZ}\), \( \mZ \sim \TRUST_{q}\lrp{\Omega, A, \nu} \), $\muvec = \lrp{\mu_1, \ldots, \mu_d}^\top$ denotes the location vector, and $S = \diag\lrp{\svec}$ is a diagonal scale matrix with leading diagonal $\svec = \lrp{s_{1}, \ldots, s_{d}}^\top$. The vector \(\mY\) has density
 \begin{equation}
 	\begin{aligned}
 		f\lrp{\yvec; \muvec, S, \Omega, A, \nu} 
 		&= \det\lrp{S}^{-1} f_{\TRUST,q}\lrp{S^{-1}\lrp{\yvec-\muvec}; \Omega, A, \nu}\,.
 	\end{aligned} \label{eq:fy_trust}
 \end{equation} 
% with \(f_{\TRUST,q}\) given at~\eqref{eq:trust_joint_pdf}.
 
The likelihood based on~\eqref{eq:fy_trust} can exhibit a complex geometry, so that both its direct optimization and evaluation of the resulting Bayesian posterior can be difficult. 
To simplify the problem the following extended likelihood is employed.
 Let $\yvec_{\mbox{\tiny obs}}=\{\yvec_i\}_{i=1}^n$ be the observed data, with $\yvec_i=(y_{i1},\ldots,y_{id})^\top $ the $i$th observation of $\mY$. Also let $\wvec = \{w_i\}_{i=1}^n$ and $\lvec = \{\lvec_i\}_{i=1}^n$, with  $\lvec_i = \lrp{l_{i1}, \ldots, l_{iq}}^\top$, be the corresponding $n$ values 
 of latent $W$ and $\mL$. Then if $\thetavec$ denotes the parameters (including $\muvec$ and $\svec$),
 an extended likelihood based on the joint of $(\mY,\mL,W)$ is 
 \[
 {\cal L}(\thetavec,\lvec,\wvec;\yvec_{\mbox{\tiny obs}})=\mbox{det}(S)^{-n}\prod_{i=1}^n f_{Z,L,W}(\zvec_i,\lvec_i,w_i;\thetavec)\,. 
 \]
 Here, $\zvec_i=S^{-1}(\yvec_i-\muvec)$, while the density $f_{Z,L,W}$ is given at~\eqref{eq:jnt} and written here as a function of $\thetavec$. Another advantage of the extended likelihood is that its evaluation does not involve computation of the 
 distribution functions in~\eqref{eq:trust_joint_pdf}, unlike with the conventional likelihood based on~\eqref{eq:fy_trust}.
 
\subsection{Parameterization, Prior and Augmented Posterior}\label{sec:augpost}
An effective parameterization of a correlation matrix is in terms of hyper-spherical 
angles as in \cite{Rebonato_Jäckel_1999} and~\cite{creal2011dynamic}.
We adopt this for \(\Omega = B B^\top\) by setting $B = \{b_{ij}\}$ to a lower triangular Cholesky factor with elements
\begin{equation*}
	b_{ii} = 
	\begin{dcases*} 
		1 & for $i = 1$, \\
		\prod_{k=1}^{i-1} \sin(\psi_{ik})& for $i > 1$,  
	\end{dcases*}
	\quad \text{and} \quad 
	b_{ij} = 
	\begin{dcases*} 
		\cos(\psi_{i1}) & for $j = 1$, \\
		\cos(\psi_{ij}) \prod_{k=1}^{j-1} \sin(\psi_{ik}) & for  $j= 2,...,i-1$.  
	\end{dcases*} ,
\end{equation*}
that are functions of the angles $\psi_{i(i-1)}\in(0,2\pi]$ and $\psi_{ij}\in (0,\pi]$ for $j<i-1$.
The bounds on each angle $\psi_{ij}$ are unchanged when conditioning on the other angles, making
them an attractive parameterization for MCMC sampling as in~\cite{creal2011dynamic}. We denote
the $d(d-1)/2$ unique angles as $\Psi=\{\psi_{ij}\}$, so the parameters \(\thetavec= \{\Psi, A, \nu,\muvec,\svec\}\) and $\Omega$ is computed from $\Psi$. 

The angles have uniform independent priors
\begin{equation}
	\psi_{ij}
	\sim \left\{\begin{array}{ll} 
		\Unifdst(\epsilon, 2 \pi - \epsilon) & \text { for } j = i-1  
		\\
		\Unifdst(\epsilon,   \pi - \epsilon) & \text { for } j < i-1\,, 
	\end{array}\right.
	\label{eq:psi_prior}
\end{equation}
with \(\epsilon = 0.03\) chosen to bound $\Omega$ away from singular values. \cite{branco2001general} show that the Jeffreys prior for the single shape parameter
of a univariate skew-t distribution is symmetric around zero and proper. While derivation of
the Jeffreys prior for the $(d\times q)$ matrix $A$ is difficult, it motivates the adoption
of independent proper $N_1(0,5^2)$ priors for the elements of $A$. In our empirical work we find this prior uninformative in practice.
%Independent proper Gaussian priors are assigned to each component of \(A\), such that \(\alpha_{jk} \sim N_1(0, 25)\).
The prior for \((\nu - 2) \sim \Gammadst(3, 0.2)\), where the location shift ensures \(\nu > 2\), and we adopt the reference prior $p(\muvec,\svec)\propto \prod_{j=1}^d s_j^{-1}$ as in \cite{liseo2006note} and others.
The prior density is $p(\thetavec)\propto p_0(\thetavec)\mathds{1}(\Delta\in {\cal P}(\Omega))$, where $p_0(\thetavec)$
is the product of  
the prior densities outlined above, and 
the indicator function $\mathds{1}(X)=1$ if $X$ is true, and zero otherwise. This indicator function
constrains $\Delta$ to a set ${\cal P}(\Omega)$, which contains all values of $\Delta$ that satisfy 
Assumption~\ref{asmp:LP}. This is where the elements of the diagonal matrix $H=\Sigma-\Delta \Omega^{-1}\Delta^\top$ (which is a deterministic function of $\{\Delta,\Omega\}$) are in ascending order. Note that ${\cal P}$ depends on the value of $\Omega$ and we write it as such.

The Bayesian augmented posterior
\[
p(\thetavec,\lvec,\wvec|\yvec_{\mbox{\tiny obs}})\propto {\cal L}(\thetavec,\lvec,\wvec;\yvec_{\mbox{\tiny obs}}) p_0(\thetavec)\mathds{1}(\Delta\in {\cal P}(\Omega))\,,
\]
is computed using MCMC. Let $\theta_i$ be a scalar element of $\thetavec$, and $\thetavec_{-i}$ be the remaining elements, then the sampler is:

\noindent {\underline {\bf Algorithm~1}} {\em (MCMC Sampler for TrUST Distribution Parameters)}\\
\indent Step 1: Generate from $p(\lvec|\wvec,\thetavec,\yvec_{\mbox{\tiny obs}})$ as a block.\\
\indent Step 2: Generate from $p(\wvec|\lvec,\thetavec,\yvec_{\mbox{\tiny obs}})$ as a block.\\
\indent Step 3: For each $i$, generate from $p(\theta_i|\lvec,\wvec,\thetavec_{-i},\yvec_{\mbox{\tiny obs}})$ in a random order.

\vspace{10pt}

At Step~1, $\lvec$ is sampled first as a block by exploiting the fact that each element is independently constrained Gaussian, where its conditional posterior density is
\[
p(\lvec|\wvec,\thetavec,\yvec_{\mbox{\tiny obs}})=\prod_{i=1}^n \prod_{k=1}^q p(l_{ik}|\zvec,w,\thetavec) =\prod_{i=1}^n \prod_{k=1}^q\phi_{l_{ik}>0}(m_{ik},w_i^{-1}r_k)\,,
\]
$m_{ik}= \deltavec_k^\top \Omega^{-1}\zvec_i$ and $r_k = 1 - \deltavec_k^\top \Omega^{-1} \deltavec_k$. The conditional independence of the elements of $\lvec$ given $\zvec,\wvec,\thetavec$
is a key feature of
the TrUST distribution, and does not apply to the general UST distribution.

At Step~2, $\wvec$ is sampled next, also as a block from its posterior
$p(\wvec|\lvec,\thetavec,\yvec_{\mbox{\tiny obs}})=\prod_{i=1}^n f_{Gam}(w_i;a,b_i)$,
with $a = (d+q+\nu)/2$ and
\[
b_i = \frac{1}{2}\lrb{ \lrp{\zvec_i - \Delta^\top \Sigma^{-1}\lvec_i}^\top\lrp{\Omega - \Delta^\top \Sigma^{-1} \Delta}^{-1}\lrp{\zvec_i - \Delta^\top \Sigma^{-1}\lvec_i} + \lvec_i^\top \Sigma^{-1} \lvec_i + \nu }.
\]

At Step~3, each $\theta_i$ is sampled conditional on $\thetavec_{-i}$, $\lvec$ and $\wvec$. A range of 
popular methods can be used here, and we simply sample each element\footnote{Each constrained or bounded element of $\thetavec$ is first transformed to the real line using 
a one-to-one transformation (e.g. logarithm for positive valued parameters) to 
simplify sampling, with care taken to correct the posterior for these transformations.} (in random order to improve mixing) using adaptive random walk Metropolis-Hastings. When generating each element of $\Psi$ and $A$ the parameter constraint in Assumption~\ref{asmp:LP} can be imposed on the draws as follows. 
Compute the diagonal matrix $H=\Sigma-\Delta \Omega^{-1} \Delta^\top$, where recall that $\Delta, \Omega$ and $\Sigma$ are all known functions of $\{\Psi,A\}$. Then simply reorder the rows of $\Delta$ to ensure the 
leading diagonal elements of $H$ are ordered as permutation $\pi^*$; i.e. in ascending order as discussed in Section~\ref{sec:truse} (and thus also the ordering of $A$ and $\Sigma$).
%To see that the order of the leading diagonal elements of $H$ corresponds to the order of the rows of $\Delta$, recall that $\Sigma$ is a function of $\Delta$ as in Lemma~\ref{lem1}.
This approach ensures each draw satisfies the bound $\Delta \in {\cal P}(\Omega)$ without rejection. 

\begin{table}[htbp]
	\begin{center}
		\caption{Posterior Mean Parameter Estimates for Simulation Example~1}	\label{tab:sim_dist_case_1}
		\resizebox{0.75\textwidth}{!}{
			% Table generated by Excel2LaTeX from sheet 'Sheet12'
\begin{tabular}{ccccccccccrc}
    \toprule
    \toprule
    Variable &     & \multicolumn{3}{c}{$\Omega$} &     & $\deltavec_1$ & $\deltavec_2$ &     & $\nu$ &     & DIC \\
    \cmidrule{1-1}\cmidrule{3-5}\cmidrule{7-8}\cmidrule{10-10}\cmidrule{12-12}    &     & \multicolumn{10}{l}{Panel A: DGP1, fitted with $q=1$ (Correctly Specified)} \\
    \cmidrule{3-12}$Z_1$ &     & 1.000 &     &     &     & -0.285 & --- &     & \multirow{3}[1]{*}{10.658} &     & \multirow{3}[1]{*}{\textbf{6883}} \\
    $Z_2$ &     & 0.504 & 1.000 &     &     & 0.606 & --- &     &     &     &  \\
    $Z_3$ &     & 0.294 & 0.815 & 1.000 &     & 0.802 & --- &     &     &     &  \\
        &     &     &     &     &     &     &     &     &     &     &  \\
        &     & \multicolumn{10}{l}{Panel B: DGP1, fitted with $q=2$ (Misspecified)} \\
    \cmidrule{3-12}$Z_1$ &     & 1.000 &     &     &     & -0.250 & -0.239 &     & \multirow{3}[2]{*}{10.734} &     & \multirow{3}[2]{*}{6887} \\
    $Z_2$ &     & 0.539 & 1.000 &     &     & 0.585 & 0.599 &     &     &     &  \\
    $Z_3$ &     & 0.341 & 0.814 & 1.000 &     & 0.793 & 0.757 &     &     &     &  \\
    \midrule
    \midrule
        &     & \multicolumn{10}{l}{Panel C: DGP2, fitted with $q=1$ (Misspecified)} \\
    \cmidrule{3-12}$Z_1$ &     & 1.000 &     &     &     & -0.270 & --- &     & \multirow{3}[1]{*}{12.880} &     & \multirow{3}[1]{*}{6275} \\
    $Z_2$ &     & 0.229 & 1.000 &     &     & 0.758 & --- &     &     &     &  \\
    $Z_3$ &     & -0.016 & 0.787 & 1.000 &     & 0.951 & --- &     &     &     &  \\
        &     &     &     &     &     &     &     &     &     &     &  \\
        &     & \multicolumn{10}{l}{Panel D: DGP2, fitted with $q=2$ (Correctly Specified)} \\
    \cmidrule{3-12}$Z_1$ &     & 1.000 &     &     &     & 0.761 & -0.864 &     & \multirow{3}[2]{*}{10.548} &     & \multirow{3}[2]{*}{\textbf{4778}} \\
    $Z_2$ &     & 0.530 & 1.000 &     &     & 0.901 & -0.133 &     &     &     &  \\
    $Z_3$ &     & 0.314 & 0.803 & 1.000 &     & 0.829 & 0.190 &     &     &     &  \\
    \bottomrule
    \bottomrule
    \end{tabular}%
    
		}
	\end{center}
	Note: Results are given for the four combinations of DGP and TrUST distribution fit to the simulated data. Numbers reported are the posterior means, except for final column which 
	reports the DIC values. Lower DIC indicates a better fit.
\end{table}

\subsection{Simulation Example}\label{sec:sim_dist}
The Bayesian method and increased flexibility of the TrUST distribution over the AC skew-t are first illustrated using a simulation, where we assume \(\muvec=\zerovec\) and \(\svec=(1,\ldots,1)^\top\)
for simplicity.
We generate \(n=1040\) observations from two data generating processes: DGP1 with $q=1$ (i.e. a skew-t), and DGP2 with \(q=2\). For both cases \(d=3\), \(\nu=10\), and elements of \(\Omega\) set to  $\omega_{21}=0.5$, $\omega_{31}=0.3$, and \( \omega_{32}=0.811 \). The values of $A$ are: 
\begin{itemize}
    \item DGP1:  \(A = (-5, 3, 5)^\top\), implying \(\Delta = \lrp{-0.271, 0.618, 0.805}^\top\). 
    \item DGP2:  \(A = [\alphavec_1|\alphavec_2]\), where \(\alphavec_1 = (5, 3, 5)^\top\), \(\alphavec_2 = (-10, 0, 5)^\top\), implying  $\Delta^\top = [\deltavec_1| \deltavec_2]$, where \(\deltavec_1 = \lrp{0.748, 0.894, 0.835}^\top\), and \(\deltavec_2 = \lrp{-0.868, -0.097, 0.204}^\top\).
\end{itemize} 

\begin{figure}[bth]
	\centering
	\includegraphics[width=0.8\textwidth]{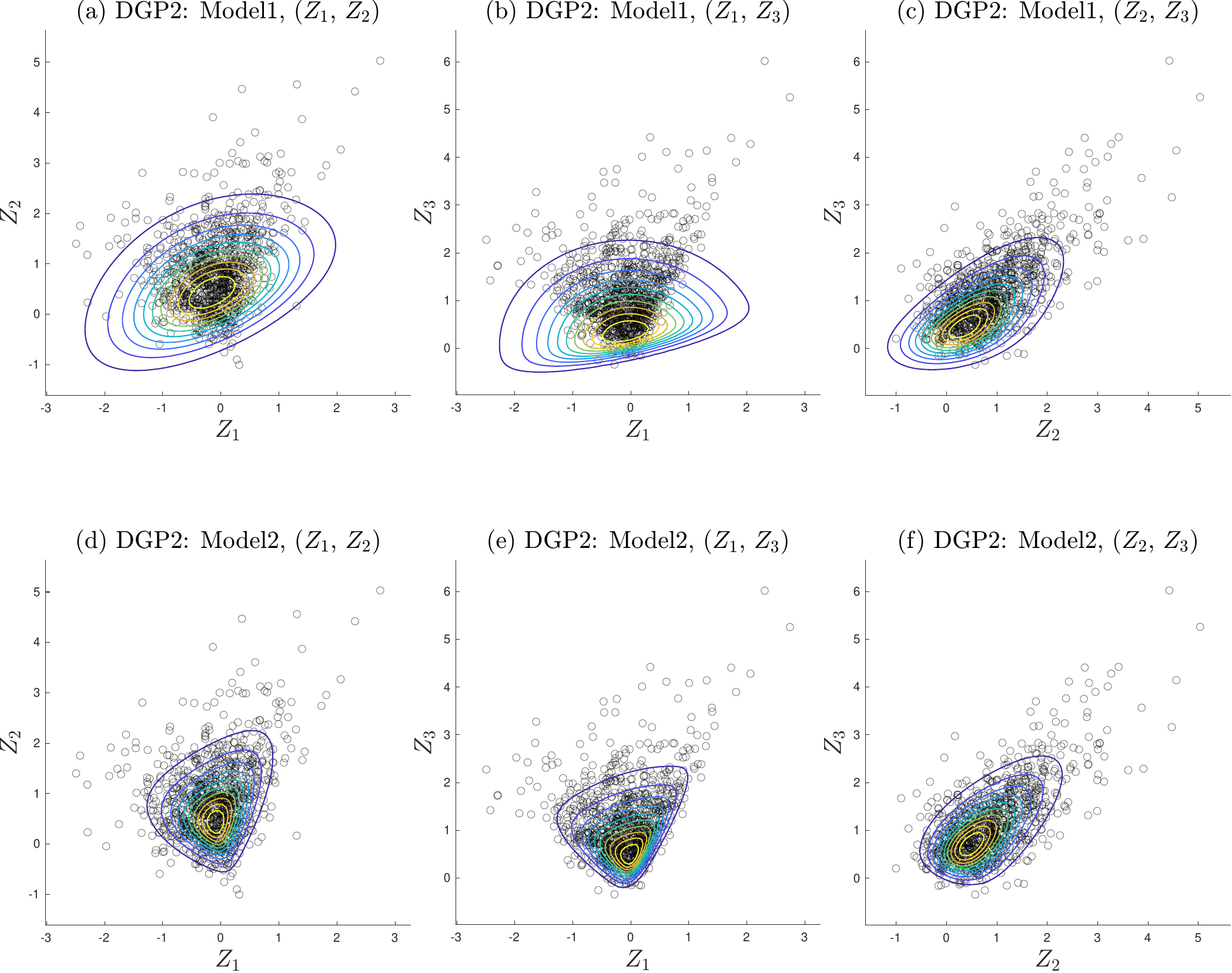}
	\caption{Contour plots of bivariate slices of the TrUST densities fitted to data from DGP2. Panels (a)-(c) give the contours when $q=1$, and panels (d)-(f) when $q=2$. The bivariate scatterplots are of the generated data and are the same in the first and second rows.}
	\label{fig:sim_dist_case_2}
\end{figure}

For each DGP, TrUST distributions with \(q =1 \) and \(q = 2\) are estimated and the quality of fit measured using 
the Deviance Information Criterion (DIC)  
\[ \DIC = -4\,\mathbb{E}_{\thetavec}[\log p(\yvec_{\mbox{\tiny obs}} \mid \thetavec)] + 2\log p(\yvec_{\mbox{\tiny obs}} \mid \thetahatvec), \] 
where \(p(\yvec_{\mbox{\tiny obs}}|\thetavec)=\prod_{i=1}^n f\lrp{\yvec_i; \muvec, S, \Omega, A, \nu} \) is evaluated as in~\eqref{eq:fy_trust}. The expectation \(\mathbb{E}_{\thetavec}\) is computed from the MCMC samples after burn-in and \(\thetahatvec\) denotes the posterior mean. 

Table~\ref{tab:sim_dist_case_1} presents the results. For DGP1 the misspecified TrUST distribution
with $q=2$ (Panel B) is almost as accurate as the correctly specified skew-t distribution with $q=1$ (Panel A). 
In contrast, when the model is misspecified with too few latent variables, the impact is substantial with a much higher DIC value in Panel C compared to Panel D.  Figure~\ref{fig:sim_dist_case_2} displays the major discrepancy in the two fitted densities for the case of DGP2, with the contours in the first row (Panels (a)–(c) for $q=1$) failing to align well with those in the second row (Panels (d)–(f) for $q=2$).

\subsection{Daily Australian Electricity Prices}\label{sec:dist_app}
Over \$17.5bn of electricity is traded annually in the Australian National Electricity Market (NEM). This wholesale market comprises the five interconnected regions of New South Wales (NSW), Victoria (VIC), Queensland (QLD), South Australia (SA), and Tasmania (TAS). There are separate 
prices in each region, which exhibit complex dependencies and extreme positive skew; see~\cite{panagiotelis2008bayesian} for a discussion of this market and the distribution of prices. 

The data comprise the $n=2070$ daily prices from 1 Dec. 2018 to 31 July 2024 for the \(d = 5\) regions. We consider eight TrUST distributions with values of $q\in\{0,1,2,3\}$ and for both Gaussian (i.e. $\nu \rightarrow \infty$) and unconstrained degrees of freedom $\nu$, fit to the logarithm of prices using the Bayesian method. The alternative skew-t of~\cite{sahu_dey_branco_2003} (SDB) fit as in~\cite{smith_gan_kohn_2010} is also included
as a further benchmark. 
Table~\ref{tab:nemdist} lists the distributions, the number of parameters ($\mbox{dim}(\thetavec)$),
DIC values and the (negative) 10-fold cross-validation predictive log score (LS).
The posterior means of $\nu$ are also given, indicating the extreme kurtosis in this data, and the means of the other parameters can be found in
the Online Appendix. Overall,
the TrUST distribution with \(q=2\) has the lowest DIC and LS values, and clearly dominates the other benchmarks, including the simpler AC and SDB skew-t distributions.

\begin{table}[htbp]
	\caption{Different TrUST Distributions fit to the Electricity Price Dataset}
	\label{tab:nemdist}
	\resizebox{1\textwidth}{!}{
		% \begin{tabular}{lccccccccccccccrc}
% \toprule
% \toprule
% $q$ &     & 0   &     & 1   &     & 2   & 3   &     & 0   &     & 1   &     & 2   & 3   &     & -- \\
% \cmidrule{1-1}\cmidrule{3-3}\cmidrule{5-5}\cmidrule{7-8}\cmidrule{10-10}\cmidrule{12-12}\cmidrule{14-15}\cmidrule{17-17}Dist. Name &     & Gaussian &     & \multicolumn{1}{l}{Skew-Normal} &     & \multicolumn{2}{c}{TrUSN} &     & Student-t &     & Skew-t &     & \multicolumn{2}{c}{TrUST} &     & SDB \\
% $\mbox{dim}(\thetavec)$ &     & 20  &     & 25  &     & 30  & 35  &     & 21  &     & 26  &     & 31  & 36  &     & 26 \\
% $\nu$ &    &  --   &    &  --   &     & --    &  --   &    & 2.189 &     & 2.024 &     & 2.029 & 2.031 &     & 2.083 \\
% DIC &     & -4943 &     & -5863 &     & -6280 & -6949 &     & -12704 &     & -13535 &     & \textbf{-13756} & -13683 &     & -12967 \\
% Log-Score &     & -1286 &     & -1647 &     & -1555 & -2194 &     & -5212 &     & -5532 &     & \textbf{-5567} & -5555 &     & -5331 \\
% \bottomrule
% \bottomrule
% \end{tabular}%

\begin{tabular}{lccccccccccccccrc}
\toprule
\toprule
$q$ &     & 0   &     & 1   &     & 2   & 3   &     & 0   &     & 1   &     & 2   & 3   &     & -- \\
\cmidrule{1-1}\cmidrule{3-3}\cmidrule{5-5}\cmidrule{7-8}\cmidrule{10-10}\cmidrule{12-12}\cmidrule{14-15}\cmidrule{17-17}
Dist. Name &     & Gaussian &     & \multicolumn{1}{l}{Skew-Normal} &     & \multicolumn{2}{c}{TrUSN} &     & Student-t &     & Skew-t &     & \multicolumn{2}{c}{TrUST} &     & SDB \\
$\mbox{dim}(\thetavec)$ &     & 20  &     & 25  &     & 30  & 35  &     & 21  &     & 26  &     & 31  & 36  &     & 26 \\
$\nu$ &     & -- &     & -- &     & -- & -- &     & 2.087 &     & 2.054 &     & 2.064 & 2.061 &     & 2.025 \\
DIC &     & -9324 &     & -10657 &     & -11163 & -11484 &     & -18510 &     & -19540 &     & \textbf{-19967} & -19697 &     & -- \\
Log-Score &     & -3440 &     & -4003 &     & -4021 & -4225 &     & -8586 &     & -9100 &     & \textbf{-9211} & -9210 &     & -8691 \\
\bottomrule
\bottomrule
\end{tabular}%

	}
	\caption*{Note: The eight TrUST distributions were fit using the Bayesian method outlined. The left-hand columns correspond to the Gaussian cases where $\nu\rightarrow \infty$. Lower values of DIC and LS indicate better fit, with the lowest value indicated in bold.
	The full set of parameter estimates are given in the Online Appendix. The final column provides results for the benchmark SDB skew-t distribution.}
\end{table}

\begin{figure}[htbp]
	\centering
	\includegraphics[width=1\textwidth]{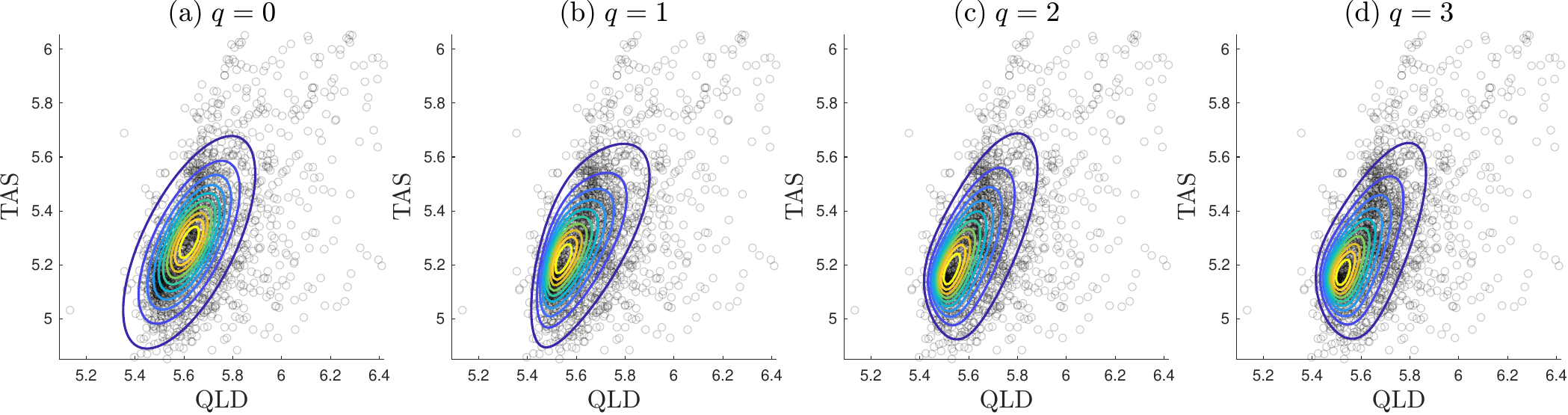} 
	\includegraphics[width=1\textwidth]{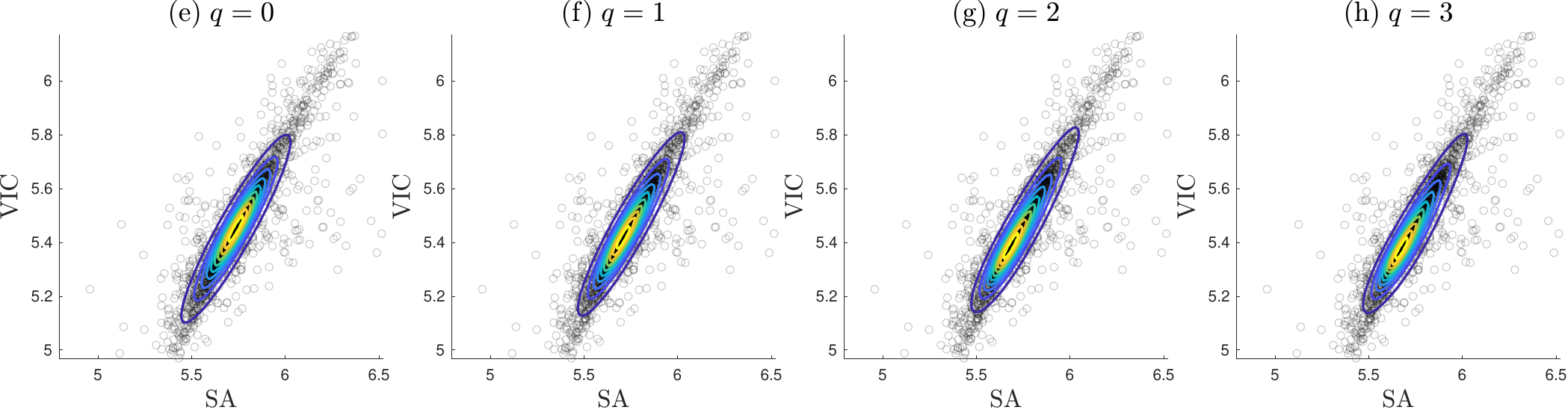} 
	\caption{
		Contour plots of bivariate marginal densities from the fitted TrUST distributions.
		These are given for the pairs (QLD, TAS) (top row) and (SA, VIC) (bottom row), while the columns correspond to TrUST distributions with \(q=0\) (Student-t), \(q=1\) (Skew-t), \(q=2\), and \(q=3\).}
	\label{fig:app_NEM}
\end{figure}

Figure~\ref{fig:app_NEM} gives bivariate marginal density contour plots for the fitted TrUST distributions with \(q \in \{0,1,2,3\}\) and variable pairs (a--d) (QLD, TAS), and (e--h) (SA, VIC). 
These are overlaid with scatterplots of the data, trimmed for extreme outliers for improved visualization only. 
Increasing $q$ impacts the asymmetry in the distribution substantially. Further contour plots for all pairwise slices of the TrUST distribution with $q=2$, along with five univariate marginals, are given in the Online Appendix, along with results for the SDB skew-t. Overall, this example highlights that the TrUST distribution fits this five-dimensional data more accurately than both the AC and SDB skew-t distributions.

  \section{TrUST Copula}\label{sec:05}
This section outlines the copula of the TrUST distribution, which  
extends the AC skew-t copula of~\cite{Yoshiba_2018} and~\cite{deng2024large} to allow for greater variability in asymmetric dependence across variable pairs.
% while remaining tractable. 
%We derive some of its 
%properties, show how to compute Bayesian inference 
%for the copula parameters, and
%demonstrate both the efficacy of the estimation method and proposed copula
%in comparison to the skew-t copula using simulated data.

\subsection{Copula Specification}
The copula of a continuous random vector $\mZ$ is known as an implicit copula and is obtained by inversion of Sklar's theorem; see~\citet[p.51]{nelsen2006introduction} and~\cite{smith2021implicit}. 
Let $\mZ = \lrp{ Z_1, \ldots, Z_d}^\top \sim \TRUST_q \lrp{\Omega, A, \nu}$, 
then from~\eqref{eq:trust_joint_pdf} and~\eqref{eq:trust_unimargin_pdf}, the TrUST copula density is
\begin{equation}\label{eq:trust_copula_pdf}
	c\lrp{\uvec; \thetavec} = \frac{
		f_{\TRUST, q}\lrp{\zvec; \Omega, A, \nu}
	}{
		\prod_{j=1}^{d} f_{\UST, q}\lrp{z_j; 1, \Delta_j, \Sigma, \nu}
	}\,.
\end{equation}
Here, $\uvec=(u_1,\ldots,u_d)^\top \in [0,1]^d$, $z_j=F_{\UST,q}^{-1}(u_j;1, \Delta_j, \Sigma, \nu)$ is the 
marginal quantile function of $Z_j$ evaluated at $u_j$ for $j=1,\ldots,d$, $\Delta_j$ is the $j$th
column of $\Delta$,
\footnote{Note that this is different from $\deltavec_k^\top$ which denotes the $k$th row of $\Delta$.} 
and $\zvec=(z_1,\ldots,z_d)^\top$. Parameterizing $\Omega$ in terms of the hyper-spherical angles $\Psi$ as in Section~\ref{sec:augpost},
	the copula parameters
are
 $\thetavec=\{\Psi,A,\nu \}$, and $\Sigma, \Delta$ are both known functions of $\thetavec$ as in Section~\ref{sec:03}.

To visualize the copula, Figure~\ref{fig:fstc_pdfs} in the Online Appendix plots contours of $c$ for the bivariate TrUST copulas of the bivariate distributions with $q=2$ in Figure~\ref{fig:fst_pdfs}.
%with $q=2$ and the parameters as those for the distributions in Figure~\ref{fig:fst_pdfs}.
This illustrates how the parameter $\alphavec_2$ (which is an additional copula parameter over that of the skew-t copula where $q=1$) affects the dependence asymmetries greatly, resulting in a more flexible copula.

\subsection{Rank Correlations}
\label{sec:copula_prop}
%To evaluate the dependence structure implied by the data, it is crucial to assess concordance, even in the presence of nonlinear transformations. Rank-based metrics, such as Kendall's and Spearman's correlations, are among the most commonly used measures for this purpose. These metrics are particularly well-suited for implicit copula models, which involve nonlinear transformations applied for twice. In this section, we present both Kendall's correlation \(\rho_K\) and Spearman's correlation \(\rho_S\) for the UST distribution, including the TrUST subclass, along with their derivations for different sizes of the latent vector $q$. 
We now present expressions for both Kendall's correlation \(\rho_K\) and Spearman's correlation \(\rho_S\) for the UST distribution, which includes the TrUST subclass. 
As far as we are aware, these general expressions are given for the first time in the literature.

% \subsubsection{Kendall Correlations}
% \begin{lemma}[Kendall's Correlation of UST]\label{theo:kendall}
% 	Let \( \lrp{Z_1, Z_2}^\top \sim \UST_{q}\lrp{\Omega, \Delta, \Sigma, \nu} \), then the Kendall's correlation for this pair is 
% 	\begin{equation}\label{eq:kendall}
% 		\rho_{K, \UST}\lrp{Z_1, Z_2} = 4 \frac{ T_{2+2q} \lrp{\zerovec; R_K, \nu }}{ \lrb{T_{q} \lrp{\zerovec; \Sigma, \nu}}^2 } - 1,
% 	\end{equation}
% 	where 
% 	% \begin{equation*}
% 	% 	R_K = \left(\begin{array}{rcc} 
% 	% 		2\Omega & &   \\
% 	% 		\Delta & \Sigma &   \\
% 	% 		-\Delta & \zerovec_{q\times q}  &\Sigma \\
% 	% 	\end{array}\right). 
% 	% \end{equation*}
% 	\begin{equation*}
% 		% \Omega = \left(\begin{array}{cc} 
% 		% 	1 & \omega  \\
% 		% 	\omega & 1  \\
% 		% \end{array}\right), \quad
% 		R_K = \left(\begin{array}{lc} 
% 			2\Omega & \Delta_K^\top  \\
% 			\Delta_K & I_2 \otimes \Sigma \\
% 		\end{array}\right)\,, 
% 		\quad \quad
% 		\Delta_K = \left(\begin{array}{c} 
% 			\Delta   \\
% 			-\Delta \\
% 		\end{array}\right)\,,
% 	\end{equation*}
% and \(  \otimes \) denotes the Kronecker product.
% \end{lemma}

\begin{lemma}[Kendall's Correlation of UST]\label{theo:kendall}
	Let \( \lrp{Z_1, Z_2}^\top \sim \UST_{q}\lrp{\Omega, \Delta, \Sigma, \nu} \), then the Kendall's correlation for this pair is 
	\begin{equation}\label{eq:kendall}
		\rho_{K, \UST}\lrp{Z_1, Z_2} =
		4 \frac{
			\mathbb{E}_{W,W'}\lrb{
			\Phi_{2+2q}\lrp{\zerovec; R_K(W,W')}
			}
		}{\lrb{\Phi_q\lrp{\zerovec;\Sigma}}^2} - 1,
	\end{equation}
	where \(W,W'\overset{\iid}{\sim}\Gammadst(\nu/2,\nu/2)\),
	\begin{equation*}
		R_K(w,w') = \left(\begin{array}{lc} 
			\Omega & B_K(w,w')^\top  \\
			B_K(w,w') & I_2 \otimes \Sigma \\
		\end{array}\right)\,, 
		\quad \quad
		B_K(w,w') = \left(\begin{array}{c} 
			\sqrt{\frac{w'}{w+w'}}\,\Delta   \\
			-\sqrt{\frac{w}{w+w'}}\,\Delta \\
		\end{array}\right)\,,
	\end{equation*}
and \(  \otimes \) denotes the Kronecker product. The expectation in~\eqref{eq:kendall} can be evaluated by Monte Carlo simulation of the Gamma variables.
\end{lemma}

\begin{lemma}[Spearman's Correlation of UST]
	\label{theo:spearman}
	Let $\lrp{Z_1, Z_2}^\top \sim \UST_{q}\lrp{\Omega, \Delta, \Sigma, \nu}$, then Spearman's correlation for this pair is
	\begin{equation*}
		\rho_{S}\lrp{Z_1, Z_2} =
		12 \frac{
			\mathbb{E}_{W,W_1,W_2}\lrb{
			\Phi_{2+3q}\lrp{\zerovec; R_S(W,W_1,W_2)}
			}
		}{\lrb{\Phi_q \lrp{ \zerovec; \Sigma }}^3} - 3,
	\end{equation*}  
	where \(W,W_1,W_2\overset{\iid}{\sim}\Gammadst(\nu/2,\nu/2)\) and
	\begin{equation*}
		R_S(w,w_1,w_2) = \left(\begin{array}{cc} 
			\Omega^\star_S(w,w_1,w_2) &  B_S(w,w_1,w_2)^\top  \\
			B_S(w,w_1,w_2) & I_3 \otimes \Sigma    \\
		\end{array}\right) \,,
	\end{equation*}
	\(\Omega^\star_S(w,w_1,w_2)\) has unit diagonal and off-diagonal element
	\(\omega\sqrt{w_1w_2/\{(w+w_1)(w+w_2)\}}\). If \(\Delta_j\) denotes column \(j\) of \(\Delta\), then
	\begin{equation*}
		B_S(w,w_1,w_2) =
		\left(\begin{array}{cc}
			\sqrt{\frac{w_1}{w+w_1}}\Delta_1 &
			\sqrt{\frac{w_2}{w+w_2}}\Delta_2 \\
			-\sqrt{\frac{w}{w+w_1}}\Delta_1 & \zerovec \\
			\zerovec & -\sqrt{\frac{w}{w+w_2}}\Delta_2
		\end{array}\right).
	\end{equation*}
	Here \(\omega\) is the off-diagonal element of the bivariate matrix \(\Omega\). The expectation can be evaluated by Monte Carlo simulation of the Gamma variables.
\end{lemma}

% \begin{corollary}[Spearman's Corrlelation of TrUST]\label{theo:spearman_trust}
% 	Let the random vector follows the TrUST distribution \( \lrp{Z_1, Z_2}^\top \sim \TRUST_q\lrp{\zerovec, \Omega, A, \nu}\). The Spearman's correlation for this pair has the form as:  
% 	\begin{equation*}
% 		\rho_{S, \TRUST}\lrp{Z_1, Z_2; \Omega, A, \nu} \equiv \rho_{S, \UST}\lrp{Z_1, Z_2; \Omega, \Delta, \tildeSigma, \nu}.
% 	\end{equation*}
% \end{corollary}

% For the specific case of $q=2$, $\rho_{S,\UST}\lrp{Z_1, Z_2} = 12\,\frac{T_{8}\lrp{\mathbf{0}; R_S, \nu}}{\left[1/4 + (2\pi)^{-1}\arcsin(\sigma)\right]^3} - 3$, where \(\sigma\) is the off-diagonal element of \(\Sigma\). For the specific case of $q=1$ (i.e. the skew-t distribution)
% $\rho_{S,\UST}\lrp{Z_1, Z_2} = 96\,T_{5}\lrp{\mathbf{0}; R_S, \nu} - 3$,
% where the skew parameter is \(\Delta = \deltavec^\top\). When $q=0$,  Lemmas~\ref{theo:kendall} and~\ref{theo:spearman} correspond to the usual expressions
% for an elliptical distribution.

When \(q=1\), the model reduces to the usual bivariate skew-t construction, with \(\Sigma=1\) and \(\Delta=\deltavec^\top\). Hence, the denominators in Lemmas~\ref{theo:kendall} and~\ref{theo:spearman} reduce to \(1/4\) and \(1/8\), respectively. When \(q=0\), Lemma~\ref{theo:kendall} reduces to the usual Kendall correlation formula for an elliptical distribution,
\( \rho_K=2/\pi\arcsin(\omega), \)
where \(\omega\) is the off-diagonal element of \(\Omega\), while Lemma~\ref{theo:spearman} gives the corresponding scale-mixture representation for the Spearman correlation of the bivariate Student-t copula.

%    \item When \(q=0\) the familiar expression for an elliptical distribution:
%    \[
%    \rho_{S,\EL}\lrp{Z_1, Z_2} = \frac{6}{\pi}\arcsin\left(\frac{\omega}{2}\right),
%    \]
%    where \(\omega\) denotes the off-diagonal element of \(\Omega\).

For the $d$-dimensional TrUST distribution, these expressions can be computed for a pair of elements in $\mZ$ using their bivariate marginal. When \(q=1\) the expressions have been presented previously in \cite{heinen2022kendall} and \cite{lu2024kendall}. Table~\ref{tab:rankcorr} reports the rank correlations for the nine bivariate distributions with $q=2$ shown in Figure~\ref{fig:fst_pdfs} and their implied copulas. 

\begin{table}[htbp]
    \centering
    \caption{Rank Correlations for the TrUST Distributions in Figure~\ref{fig:fst_pdfs} and their Implicit Copulas }
	\resizebox{0.9\textwidth}{!}{

\begin{tabular}{
  >{\centering\arraybackslash}p{1.2cm}
  >{\centering\arraybackslash}p{1.5cm}
  >{\centering\arraybackslash}p{1.5cm}
  >{\centering\arraybackslash}p{1.5cm}
  >{\centering\arraybackslash}p{1.5cm}
  >{\centering\arraybackslash}p{1.5cm}
  >{\centering\arraybackslash}p{1.5cm}}
\toprule
\toprule
& \multicolumn{3}{c}{Kendall Correlation}
& \multicolumn{3}{c}{Spearman Correlation} \\
\cmidrule(lr){2-4} \cmidrule(lr){5-7}
$\omega$
& \multicolumn{3}{c}{Second Skew Parameter $\alphavec_2^\top$}
& \multicolumn{3}{c}{Second Skew Parameter $\alphavec_2^\top$} \\
\cmidrule(lr){1-1} \cmidrule(lr){2-4} \cmidrule(lr){5-7}
& $[5,5]$ & $[0,5]$ & $[-5,5]$
& $[5,5]$ & $[0,5]$ & $[-5,5]$ \\
-0.5 & -0.563 & -0.385 & -0.289 & -0.732 & -0.534 & -0.415 \\
 0.0 & -0.307 & -0.144 & -0.001 & -0.416 & -0.207 &  0.006 \\
 0.5 &  0.013 &  0.117 &  0.289 &  0.031 &  0.190 &  0.415 \\
\bottomrule
\bottomrule
\end{tabular}
	}\label{tab:rankcorr}
		\caption*{Note: The parameter values are given in the caption to Figure~\ref{fig:fst_pdfs}, where the first skew vector is $\alphavec_1 = [5,5]^\top$, degrees of freedom \(\nu = 5\) in all nine panels.}

\end{table}

\subsection{Asymmetric Dependence Measurements}
To measure asymmetric dependence between any pair of variables $Y_1,Y_2$, 
differences 
between quantile dependence metrics  are computed as follows. Let $U_1=F_1(Y_1)$ and $U_2=F_2(Y_2)$, then for quantile $\kappa\in(0,0.5]$ the quantile
dependence metrics in the four quadrants are:
	\begin{align*}
	&  \text{Lower Left:} % &\quad (0, \kappa] \times (0, \kappa], 
	\;\lambda_{\mathrm{LL}}(\kappa) \coloneq \mathbb{P}(U_2 < \kappa \mid U_1 < \kappa), 
	& \text{Upper Right:} % &\quad [1-\kappa, 1) \times [1-\kappa, 1), 
	\; \lambda_{\mathrm{UR}}(\kappa) \coloneq \mathbb{P}(U_2 > 1-\kappa \mid U_1 > 1-\kappa), \\
	&   \text{Lower Right:}% &\quad [1-\kappa, 1) \times (0, \kappa], 
	\; \lambda_{\mathrm{LR}}(\kappa) \coloneq \mathbb{P}(U_2 > 1-\kappa \mid U_1 < \kappa),  
	&  \text{Upper Left:} % &\quad (0, \kappa] \times [1-\kappa, 1), 
	\; \lambda_{\mathrm{UL}}(\kappa) \coloneq \mathbb{P}(U_2 < \kappa \mid U_1 > 1-\kappa).
\end{align*}  
These probabilities are computed from the bivariate copula $C(U_1,U_2)$ of the joint 
distribution of $(Y_1,Y_2)$ as in Appendix~\ref{app:quant_dep}. Following~\cite{deng2024large} we measure
the level of asymmetry in the dependence along the major and minor
diagonals of the unit square support of $C$ as  
\begin{equation}\label{eq:asymdep_quant}
\Lambda_\tinytxt{Major}(\kappa) \coloneq \lambda_\tinytxt{UR}(\kappa) - \lambda_\tinytxt{LL}(\kappa),
\quad \text{and} \quad
\Lambda_\tinytxt{Minor}(\kappa) \coloneq \lambda_\tinytxt{UL}(\kappa) - \lambda_\tinytxt{LR}(\kappa)\,.
\end{equation}
In most empirical applications, including when capturing dependence between equity returns, 
$\Lambda_\tinytxt{Major}(\kappa)$ is the more important of these two measures.
%Finally, by integrating over the quantile values we compute the overall measure below. 
%
%\begin{definition}[Total Asymmetric Dependence]\label{def:total_asym}
%	For a small number $\epsilon>0$, the total asymmetric dependence for pair \((U_1, U_2)\) is
%	\begin{equation*}
%		\text{TAD}_\tinytxt{Major} = \int_{\epsilon}^{0.5} \Lambda_\tinytxt{Major}(\kappa) \dd \kappa, 
%		\quad \text{and} \quad
%		\text{TAD}_\tinytxt{Minor} = \int_{\epsilon}^{0.5} \Lambda_\tinytxt{Minor}(\kappa) \dd \kappa.
%	\end{equation*}  
%\end{definition}
%We set $\epsilon=0.01$ because computing 
%$\Lambda_\tinytxt{Major}(\kappa)$ or $\Lambda_\tinytxt{Minor}(\kappa)$ is numerically 
%unstable for lower values of $\kappa<0.01$. 

\subsection{Bayesian Inference for the Copula Parameters}\label{sec:bayescop}
We now outline Bayesian inference for the TrUST copula parameters  $\thetavec=\{\Psi,A,\nu\}$. For simplicity, the marginals of data $Y_{ij}\sim F_{Y_j}$ are assumed known, although joint estimation of these with $\thetavec$ is possible by extending the MCMC scheme discussed here
to also draw $F_{Y_1},\ldots,F_{Y_d}$ from their conditional posteriors.

Let $U_j:=F_{\UST,q}(Z_j;1, \Delta_j, \Sigma, \nu)$ for $j=1,\ldots,d$, where $\mZ\sim \TRUST_q(\Omega,A,\nu)$, and $f_{\UST,q}=\frac{d}{dz_j} F_{\UST,q}$ is the marginal density of $Z_j$ given at~\eqref{eq:trust_unimargin_pdf}. Then the joint density of $(\mU,\mL,W)$ can be evaluated using a change of variables from $\mZ$ to $\mU=(U_1,\ldots,U_d)^\top$ to obtain 
\[
f_{U,L,W}(\uvec,\lvec,w;\thetavec)=f_{Z,L,W}(\zvec,\lvec,w)/\prod_{j=1}^{d} f_{\UST, q}\lrp{z_j; 1, \Delta_j, \Sigma, \nu}\,,  
\] 
where $f_{Z,L,W}$ is defined previously at~\eqref{eq:jnt}, and the denominator arises from the Jacobian of the transformation. This density can be used to define an extended likelihood and 
augmented posterior for $\thetavec$ and the latents $\lvec,\wvec$. The extended likelihood
is more tractable than the likelihood based directly on the copula density at~\eqref{eq:trust_copula_pdf}.

For the observed copula data \(\uvec_{obs}= \{\uvec_i\}_{i=1}^n\), where \(\uvec_i=(u_{i1},\dots,u_{id})^\top\) and \(u_{ij}=F_{Y_j}(y_{ij})\), the extended likelihood is
 simply $\mathcal{L}\lrp{\thetavec,\lvec,\wvec;\uvec_{obs}}
= \prod_{i=1}^n f_{U,L,W}(\uvec_i,\lvec_i,w_i;\thetavec)$. 
When combined with the same prior for $\thetavec$ in Section~\ref{sec:augpost}, the resulting Bayesian
augmented posterior is
\begin{equation} \label{eq:coppost}
	p(\thetavec,\lvec,\wvec|\yvec_{obs}) =
	p(\thetavec,\lvec,\wvec|\uvec_{obs}) \propto \mathcal{L}\lrp{\thetavec,\lvec,\wvec;\uvec_{obs}} p_0(\thetavec)\mathds{1}(\Delta\in {\cal P}(\Omega))\,,
\end{equation}
where the identification constraint for $\Delta$ is imposed through the prior.

The MCMC scheme at Algorithm~1 with some minor adjustments can be used to evaluate this augmented posterior. Generating
 $\lvec$ and $\wvec$ in Steps~1 and~2 is unchanged, except that $\zvec=\{\zvec_i\}_{i=1}^n$, with $\zvec_i=(z_{i1},\ldots,z_{id})^\top$, is computed
from the copula data using the $d$ quantile functions
\[
z_{ij}=F_{\UST,q}^{-1}(u_{ij};1, \Delta_j, \Sigma, \nu)\,,\;\;j=1,\ldots,d\,;\quad i=1,\ldots,n\,.
\]
Evaluating $\zvec$ is the most demanding computation of the sampler. 
The quantile functions above are unavailable in closed form, so that we use the fast and accurate
numerical method in~\cite{Yoshiba_2018} and~\cite{Smith_Maneesoonthorn_2018}. 
In Step~3 of Algorithm~1 each element of $\thetavec$ is generated one element at a time using 
an adaptive random walk Metropolis-Hastings scheme. We use the same approach here, 
except that the target conditional posterior $p(\thetavec|\lvec,\wvec,\uvec_{obs})$ is now given by~\eqref{eq:coppost} (up to proportionality). 
%The same approach to ensure draws of $A,\Psi$ satisfy the conditional $\Delta \in {\cal P}(\Omega)$ is also used.

\subsection{Simulation Example Continued} \label{sec:copsim}
The flexibility of the TrUST copula 
and the efficacy of the Bayesian inference method
are first illustrated using the data simulated from the two DGPs in Section~\ref{sec:sim_dist}.
The copula data $\uvec_{obs}$ was computed using 
the expressions for the univariate marginals $u_{ij}=F_{\UST,q}(z_{ij};1, \Delta_j, \Sigma, \nu)$. The two DGPs have different levels of rank correlation and asymmetric dependence over variable pair. For the data from each DGP, we estimate two TrUST copulas: Copula 1 with \(q=1\) (i.e. AC skew-t copula) and Copula 2 with \(q=2\).

\begin{table}[bh]
	\centering
	\caption{Dependence Metrics for Data Generated using DGP2 (where \( q = 2 \))}
	\resizebox{1.0\textwidth}{!}{
\begin{tabular}{ccrccrccrcc}
    \toprule
    \toprule
    \multicolumn{2}{c}{Pair $\lrp{U_i, U_j}$} &     & \multicolumn{2}{c}{True Values} &     & \multicolumn{2}{c}{Copula 1 Estimates} &     & \multicolumn{2}{c}{Copula 2 Estimates} \\
    \cmidrule{1-2}\cmidrule{4-5}\cmidrule{7-8}\cmidrule{10-11}    &     &     & \multicolumn{8}{l}{Panel A: Rank Correlations} \\
    \cmidrule{4-11}$i$ & $j$ &     & Kendall & Spearman &     & Kendall & Spearman &     & Kendall & Spearman \\
    \multirow{2}[0]{*}{2} & \multirow{2}[0]{*}{1} &     & 0.142 & 0.214 &     & 0.311 & 0.447 &     & 0.158 & 0.238 \\
        &     &     &     &     &     & (0.033) & (0.044) &     & (0.014) & (0.020) \\
    \multirow{2}[0]{*}{3} & \multirow{2}[0]{*}{1} &     & 0.116 & 0.171 &     & 0.240 & 0.340 &     & 0.141 & 0.206 \\
        &     &     &     &     &     & (0.028) & (0.038) &     & (0.018) & (0.025) \\
    \multirow{2}[0]{*}{3} & \multirow{2}[0]{*}{2} &     & 0.409 & 0.578 &     & 0.462 & 0.644 &     & 0.385 & 0.550 \\
        &     &     &     &     &     & (0.026) & (0.031) &     & (0.014) & (0.017) \\
    \midrule
    \midrule
        &     &     & \multicolumn{8}{l}{Panel B: Asymmetric Dependencies} \\
    \cmidrule{4-11}$i$ & $j$ &     & $\Lambda_\tinytxt{Major}(0.05)$ & $\Lambda_\tinytxt{Minor}(0.05)$ &     & $\Lambda_\tinytxt{Major}(0.05)$ & $\Lambda_\tinytxt{Minor}(0.05)$ &     & $\Lambda_\tinytxt{Major}(0.05)$ & $\Lambda_\tinytxt{Minor}(0.05)$ \\
    \multirow{2}[0]{*}{2} & \multirow{2}[0]{*}{1} &     & 0.344 & 0.051 &     & -0.031 & 0.022 &     & 0.353 & 0.038 \\
        &     &     &     &     &     & (0.025) & (0.009) &     & (0.025) & (0.012) \\
    \multirow{2}[0]{*}{3} & \multirow{2}[0]{*}{1} &     & 0.363 & 0.103 &     & -0.107 & 0.056 &     & 0.384 & 0.078 \\
        &     &     &     &     &     & (0.027) & (0.012) &     & (0.025) & (0.016) \\
    \multirow{2}[0]{*}{3} & \multirow{2}[0]{*}{2} &     & 0.457 & 0.001 &     & 0.140 & 0.000 &     & 0.474 & 0.001 \\
        &     &     &     &     &     & (0.050) & (0.001) &     & (0.024) & (0.004) \\
    \bottomrule
    \bottomrule
\end{tabular}%
}
	\label{tab:sim_case_2}
	\caption*{Note: Rank correlations and asymmetric dependence (at quantile $\kappa=0.05$) are reported for the three variable pairs. The true values from DGP2 are reported. The posterior mean estimates for the two fitted copulas are given, with the posterior
		standard deviations in parentheses below.}
\end{table}%

For DGP1, both Copula~1 and Copula~2 provide very similar and accurate estimates of the dependence metrics (see Table~\ref{tab:sim_case_1} in the Online Appendix). Therefore, we focus on DGP2 where greater heterogeneity in dependence over variable pairs is apparent.
Table~\ref{tab:sim_case_2} reports the true rank correlations and asymmetric dependence metrics for DGP2 across the three variable pairs. The posterior mean and posterior standard deviations for these
dependence metrics from the fitted copulas are also given. 
Copula~2 (the TrUST copula with $q=2$) captures these well, whereas Copula~1 (the skew-t copula with $q=1$) is unable to capture the variability in these metrics across variable pairs.
Overall, the results suggest that a TrUST copula with $q\geq 2$ is preferable whenever 
dependence is heterogeneous over variable pairs, as the next section shows is the case for 
intraday equity returns.

  \section{Asymmetric Dependence in Intraday Equity Returns}\label{sec:06}
Tail dependence between the returns on equity pairs is often asymmetric~\citep{patton2004out,harvey2010} and recent studies 
suggest it also
exhibits high levels of heterogeneity over asset pairs, including during the recent COVID-19 pandemic~\citep{le2021covid,ando2022quantile}.    
\cite{deng2024large} found evidence of this in 
15-minute equity returns using an AC~skew-t copula model, and we extend their study here. 
We show that the TrUST copula with $q>1$ is more effective than the skew-t copula 
at capturing the dependence between the market volatility VIX index published by the Chicago Board Options Exchange (CBOE) and returns of leading US stocks.
%In particular, we highlight the increased heterogeneity in asymmetric dependence and the improvement of forecasting performance from our proposed TrUST copula model.

%In particular, we show that the TrUST copula with $q>1$ is more effective at capturing the asymmtric dependence between the market volatility VIX index published by the Chicago Board of Excange (CBOE) and the intraday 15-minute returns of leading US stocks. We consider two examples in our analysis. The 3-dimensional example analyzes the relationship between the VIX and two key banking stocks, Bank of America (BAC) and JP Morgan (JPM), and investigates the ability of the TrUST distribution to capture the complex dependence structure during the COVID turbulation period. In a larger example, we illustrate the benefits of employing the TrUST copula to capture the dependence between the VIX index and the leading stock from five key sectors of the US economy: Apple (AAPL) from Electronics, JPMorgan (JPM) from Banking, Visa (V) from Business Services, Exxon Mobil (XOM) from Energy, and United Parcel Service (UPS) from Freight/Transport. In particular, we highlight the increased degree of flexibility to capture heterogeneous asymmetric dependence and the improvement of forecasting performance from our proposed model.

\subsection{Data and Marginal Models}
Observations on 15-minute equity returns and the VIX index were obtained from the LSEG DataScope Select Database. The intraday GARCH(1,1) model of~\cite{Engle_Sokalska_2012} with student-t innovations is used as a marginal for the equity returns. For the VIX index,
the marginal is a first-order autoregression with a nonparametric disturbance given by a kernel density estimator.
The copula data are computed from these marginal models. 
The same filtering approach was also employed by \cite{deng2024large}, to whom we refer for further details.

%Both series are recorded at a 15-minute intraday frequency: for the VIX, we use the last index reported in each interval, and for equities, we use the last traded price, denoted as \(P_{i,t}\), where \(i\) indexes the interval on day \(t\). The return for the \(i\)th interval is calculated as \(r_{i,t} = \log(P_{i,t}) - \log(P_{i-1,t})\). For the first interval of each day, the return is computed as \(r_{1,t} = \log(P_{1,t}) - \log(P_{26,t-1})\), given there are 26 intervals per day. The returns are organized into 40-day windows, resulting in \(n = 1040\) observations. The equity return margin is filtered using an intraday GARCH(1,1) model from \cite{Engle_Sokalska_2012} with Student‑t innovations, while the VIX marginal model is specified as a first-order autoregression with a nonparametric disturbance via a kernel density estimator.

\subsection{Banking Stocks Example (3-dim)}
The first study is of dependence during the COVID-19 crash between two key banking stocks, Bank of America (BAC) and JP Morgan (JPM),
and the VIX. TrUST copulas with \(q=1\) and \(q=2\) of dimension $d=3$ are
fit to the $n=1040$ intraday observations from February 11 to April 15, 2020. 
The posterior estimates of the copula parameters $\thetavec$, pairwise rank correlations and asymmetric
dependencies are given in Table~\ref{tab:app_3d} in the Online Appendix. They 
show that increasing the number of latent variables in the TrUST copula from $q=1$ to $q=2$ has 
a strong impact on the estimated dependence structure. 
In particular, the DIC when $q=2$ is $-1598$, while for the skew-t copula with $q=1$ it is $-1216$, suggesting an improved fit even though the example is low dimensional.

\begin{figure}[htbp]
	\centering
	\includegraphics[width=0.8\textwidth]{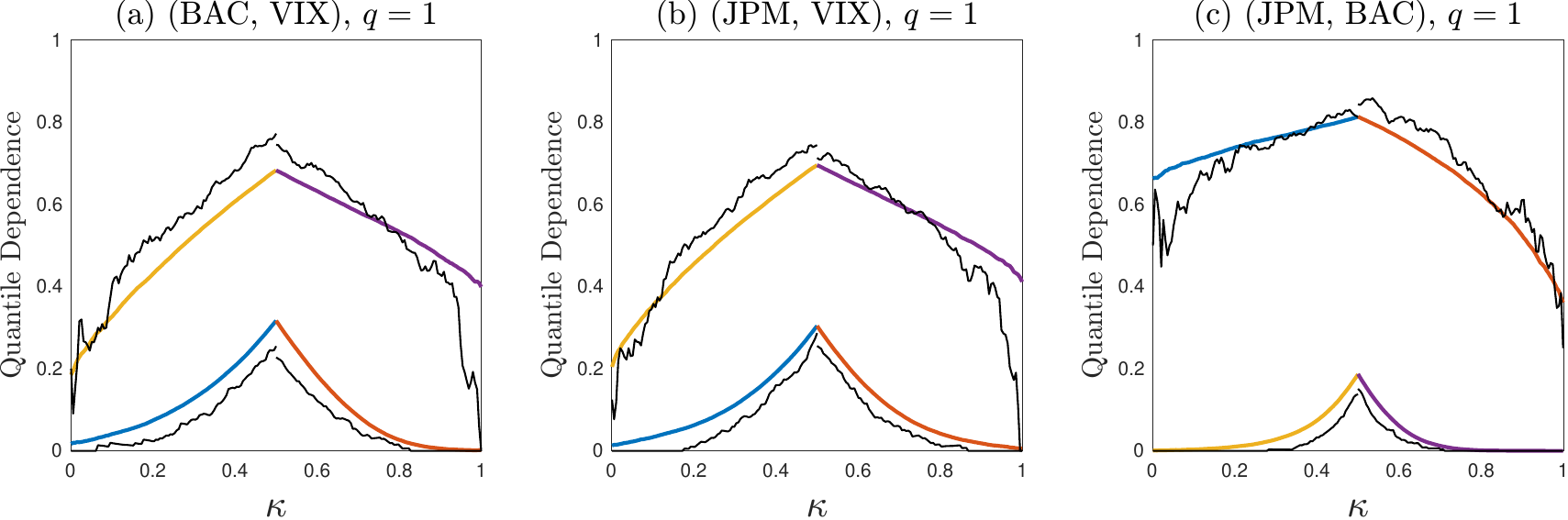}
	\includegraphics[width=0.8\textwidth]{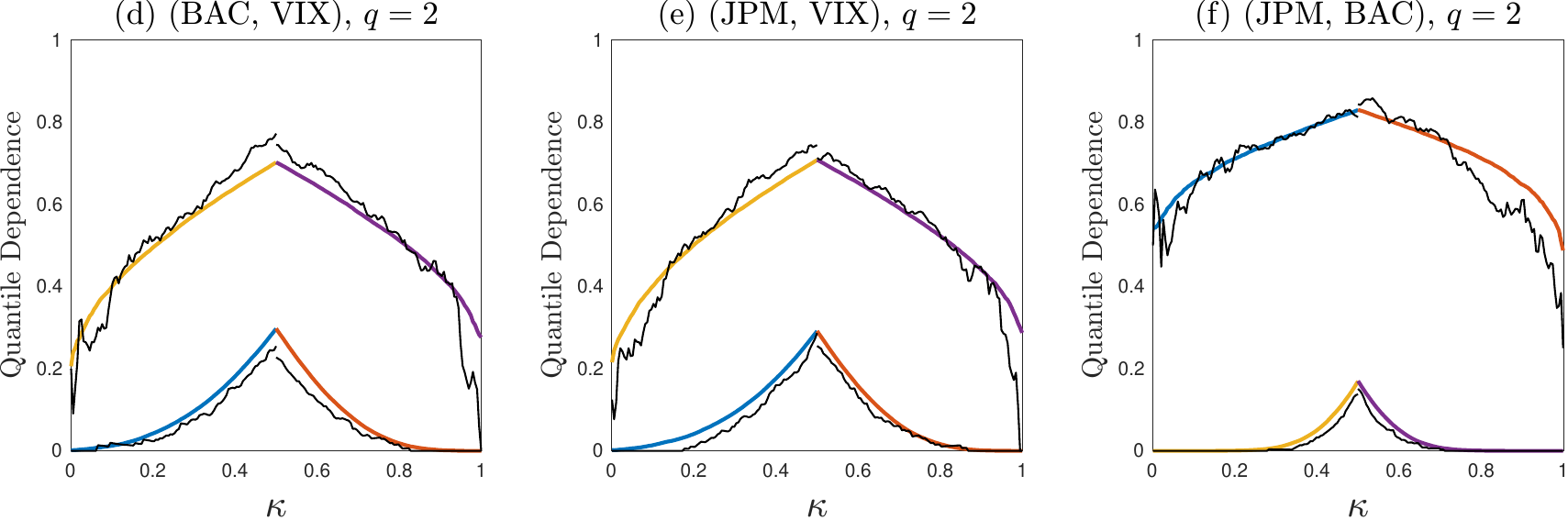}
	\caption{
		Quantile dependence plots for \(\lambda_{\mathrm{LL}}\) (blue), \(\lambda_{\mathrm{UR}}\) (red), \(\lambda_{\mathrm{LR}}\) (yellow), and \(\lambda_{\mathrm{UL}}\) (purple) are shown with empirical dependencies (black) for (BAC, VIX), (JPM, VIX), and (JPM, BAC) from left to right. The top and bottom rows depict TrUST copula estimates for \(q=1\) and \(q=2\) respectively.
	}
	\label{fig:app_3d}
\end{figure}

To visualize the asymmetric
dependence between each pair of variables we use a ``quantile dependence plot'' as in~\cite{patton2012review}. This is where \(\lambda_{\mathrm{LL}}(\kappa)\) (blue line) and 
\(\lambda_{\mathrm{UL}}(\kappa)\) (purple line) are plotted against quantile $\kappa$, and  
\(\lambda_{\mathrm{UR}}(\kappa)\) (red line), \(\lambda_{\mathrm{LR}}(\kappa)\) (yellow line) are plotted against $1-\kappa$. Asymmetric dependence in either the minor or major diagonal results in visual asymmetry around $\kappa=0.5$ in the plot. Moreover, greater consistency with the noisy empirical quantiles (black line) suggest an improved fit. Figure~\ref{fig:app_3d} gives the 
quantile dependence plots for the pairs (BAC, VIX), (JPM, VIX), and (JPM, BAC) for the fitted TrUST copulas with $q=1$ and $q=2$. 
The quantile estimates with $q=2$ presented in panels (d)-(f) more closely align with the empirical quantiles for all three pairs, reinforcing the improvement also indicated by the DIC values. 

\subsection{Cross Industry Example (6-dim)}
Cross industry dependence in U.S. equity returns is estimated using $6$-dimensional TrUST copulas. The filtered copula data was computed for the VIX
and leading stocks from five key sectors of the economy: Apple (AAPL) from Electronics, JPMorgan (JPM) from Banking, Visa (V) from Business Services, Exxon Mobil (XOM) from Energy, and United Parcel Service (UPS) from Freight/Transport. We consider two periods each with $n=1040$ observations: a pre-COVID period from 18 December 2018 to 15 February 2019, and the COVID period from 11 February 2020 to 15 April 2020. Market volatility was high during both periods, allowing investigation of how extreme market conditions influence dependence between the variables.

\begin{table}[tbhp]
	\centering
	\caption{Estimates of the TrUST Copulas for the Cross-Industry Returns in the pre-COVID Period}
	\resizebox{1.0\textwidth}{!}{
		% Table generated by Excel2LaTeX from sheet 'Sheet9'
\begin{tabular}{cccccccccccccrc}
      \toprule
      \toprule
      Variable &     & \multicolumn{6}{c}{$\Omega$}   &     & $\deltavec_1$ & $\deltavec_2$ &     & $\nu$ &     & DIC \\
      \cmidrule{1-1}\cmidrule{3-8}\cmidrule{10-11}\cmidrule{13-13}\cmidrule{15-15}\rowcolor[rgb]{ .906,  .902,  .902} \multicolumn{15}{c}{Panel A: TruST Copula $q=1$} \\
      VIX &     & 1.000 &     &     &     &     &     &     & -0.199 & --- &     & \multirow{6}[0]{*}{14.251} &     & \multirow{6}[0]{*}{	-2054} \\
      AAPL &     & -0.537 & 1.000 &     &     &     &     &     & 0.661 & --- &     &     &     &  \\
      JPM &     & -0.543 & 0.674 & 1.000 &     &     &     &     & 0.719 & --- &     &     &     &  \\
      V   &     & -0.521 & 0.749 & 0.769 & 1.000 &     &     &     & 0.830 & --- &     &     &     &  \\
      XOM &     & -0.576 & 0.594 & 0.674 & 0.657 & 1.000 &     &     & 0.536 & --- &     &     &     &  \\
      UPS &     & -0.459 & 0.658 & 0.715 & 0.745 & 0.613 & 1.000 &     & 0.820 & --- &     &     &     &  \\
      \rowcolor[rgb]{ .906,  .902,  .902} \multicolumn{15}{c}{Panel B: TruST Copula $q=2$} \\
      VIX &     & 1.000 &     &     &     &     &     &     & -0.319 & 0.307 &     & \multirow{6}[1]{*}{27.266} &     & \multirow{6}[1]{*}{\textbf{-2263}} \\
      AAPL &     & -0.485 & 1.000 &     &     &     &     &     & 0.942 & -0.848 &     &     &     &  \\
      JPM &     & -0.481 & 0.933 & 1.000 &     &     &     &     & 0.952 & -0.902 &     &     &     &  \\
      V   &     & -0.435 & 0.952 & 0.963 & 1.000 &     &     &     & 0.987 & -0.918 &     &     &     &  \\
      XOM &     & -0.597 & 0.795 & 0.803 & 0.788 & 1.000 &     &     & 0.745 & -0.628 &     &     &     &  \\
      UPS &     & -0.287 & 0.025 & -0.033 & -0.055 & 0.221 & 1.000 &     & -0.110 & 0.394 &     &     &     &  \\
      \bottomrule
      \bottomrule
      \end{tabular}%
      
	}
	\label{tab:app_6d_estimates_1}
	\caption*{Note: The posterior mean estimates are reported for the TrUST copula with $q=1$ (Panel~A) and $q=2$ (Panel~B). Cells with ``---'' in $\deltavec_2$ indicate that the parameter does not exist when $q=1$. The final column also reports the DIC values.}
\end{table}

The TrUST copula model with \(q=1\) and \(q=2\) is estimated for both periods.
Table~\ref{tab:app_6d_estimates_1} reports the
posterior means of the copula parameters for the pre-COVID period.
The DIC for both copulas is also reported, indicating that the fit with $q=2$ provides a substantial
improvement over the AC skew-t copula with $q=1$. This is also the case for the COVID period data, for which the estimates are given in Table~\ref{tab:app_6d_estimates_2} in the Online Appendix.
% \begin{landscape}
% 	\begin{figure}[htbp]
% 		\centering
% 		% Left composite figure
% 		\begin{minipage}[t]{0.7\textheight}
% 			\centering
% 			\includegraphics[width=\linewidth]{figs/Heatmap_Lambda_Win25_q_1}\\[1ex]
% 			\includegraphics[width=\linewidth]{figs/Heatmap_Lambda_Win25_q_2}
% 			\caption{Heatmap of asymmetric dependence $\Lambda_\tinytxt{Major}^{(i,j)}(\kappa)$ and $\Lambda_\tinytxt{Minor}^{(i,j)}(\kappa)$, computed at $\kappa = 0.05$, across VIX and 5 equities using the TrUST Copula for Pre-Covid period. The top row corresponds to $q=1$ on top and bottom row to $q=2$.}
% 			\label{fig:heatmap_6d_win_1}
% 		\end{minipage}
% 		\hfill
% 		% Right composite figure
% 		\begin{minipage}[t]{0.7\textheight}
% 			\centering
% 			\includegraphics[width=\linewidth]{figs/Heatmap_Lambda_Win39_q_1}\\[1ex]
% 			\includegraphics[width=\linewidth]{figs/Heatmap_Lambda_Win39_q_2}
% 			\caption{Heatmap of asymmetric dependence $\Lambda_\tinytxt{Major}^{(i,j)}(\kappa)$ and $\Lambda_\tinytxt{Minor}^{(i,j)}(\kappa)$, computed at $\kappa = 0.05$, across VIX and 5 equities using the TrUST Copula for During-Covid period. The top row corresponds to $q=1$ on top and bottom row to $q=2$.}
% 			\label{fig:heatmap_6d_win_2}
% 		\end{minipage}
% 	\end{figure}
% \end{landscape}

\begin{sidewaysfigure}[htbp]
%		\begin{landscape}
		\centering
		% Left composite figure
		\begin{minipage}[t]{1\linewidth}
			\centering
			\includegraphics[width=0.49\linewidth]{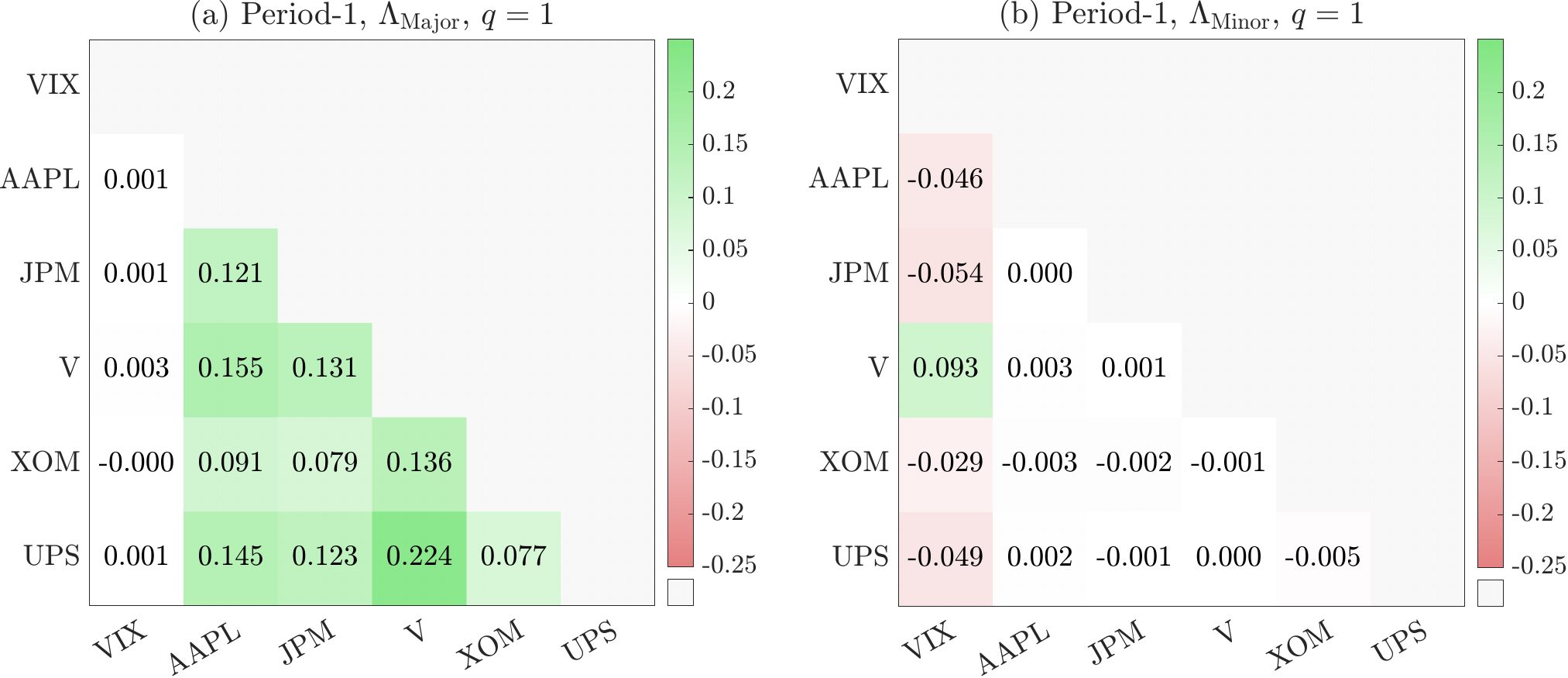} \hfill
			\includegraphics[width=0.49\linewidth]{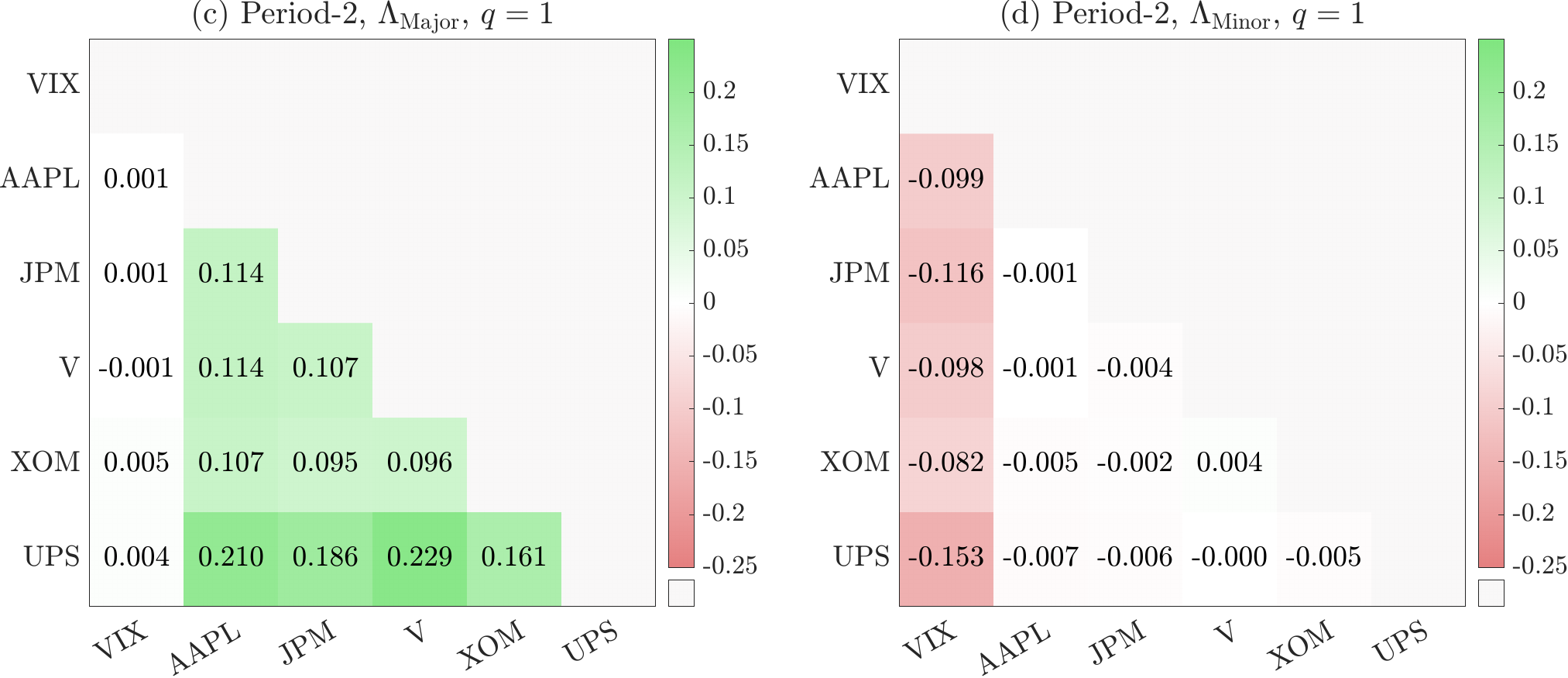} \\[3ex]
			\includegraphics[width=0.49\linewidth]{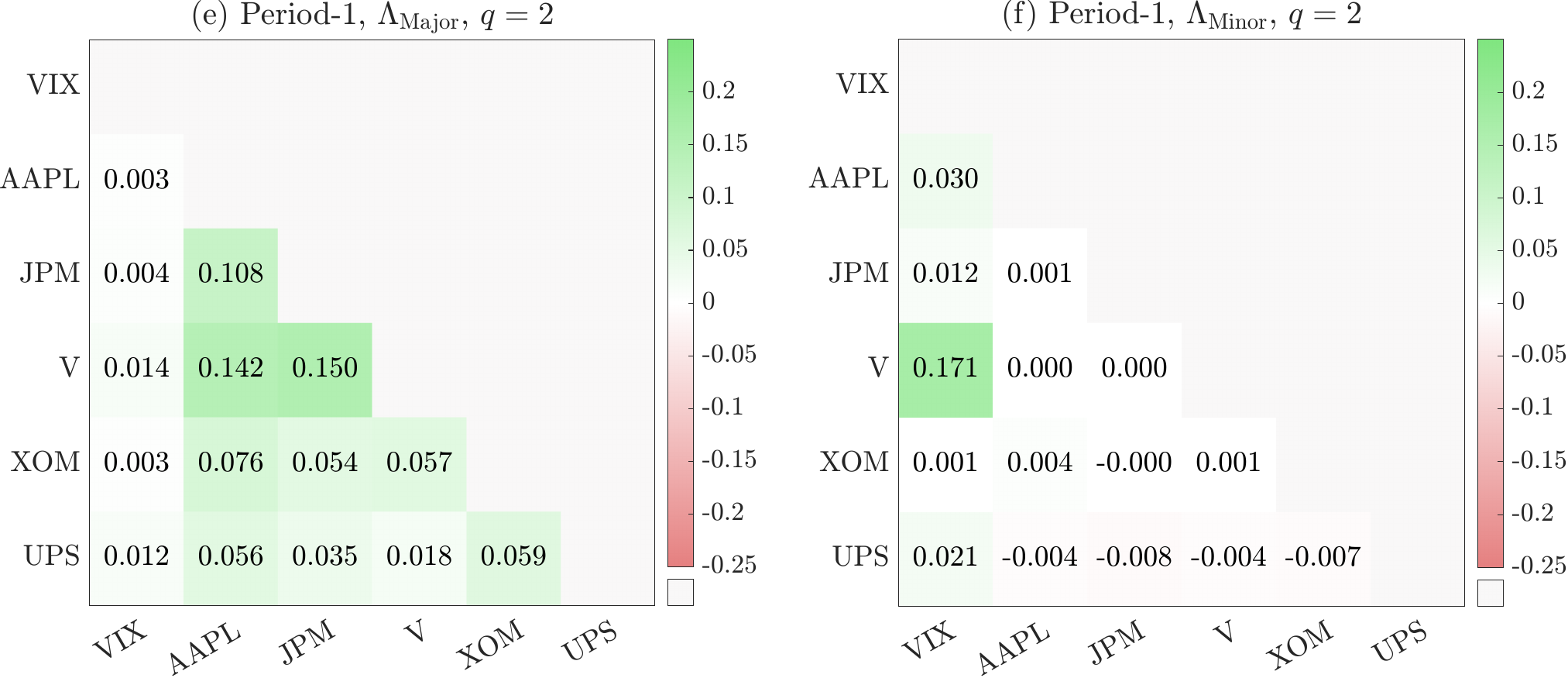} \hfill
			\includegraphics[width=0.49\linewidth]{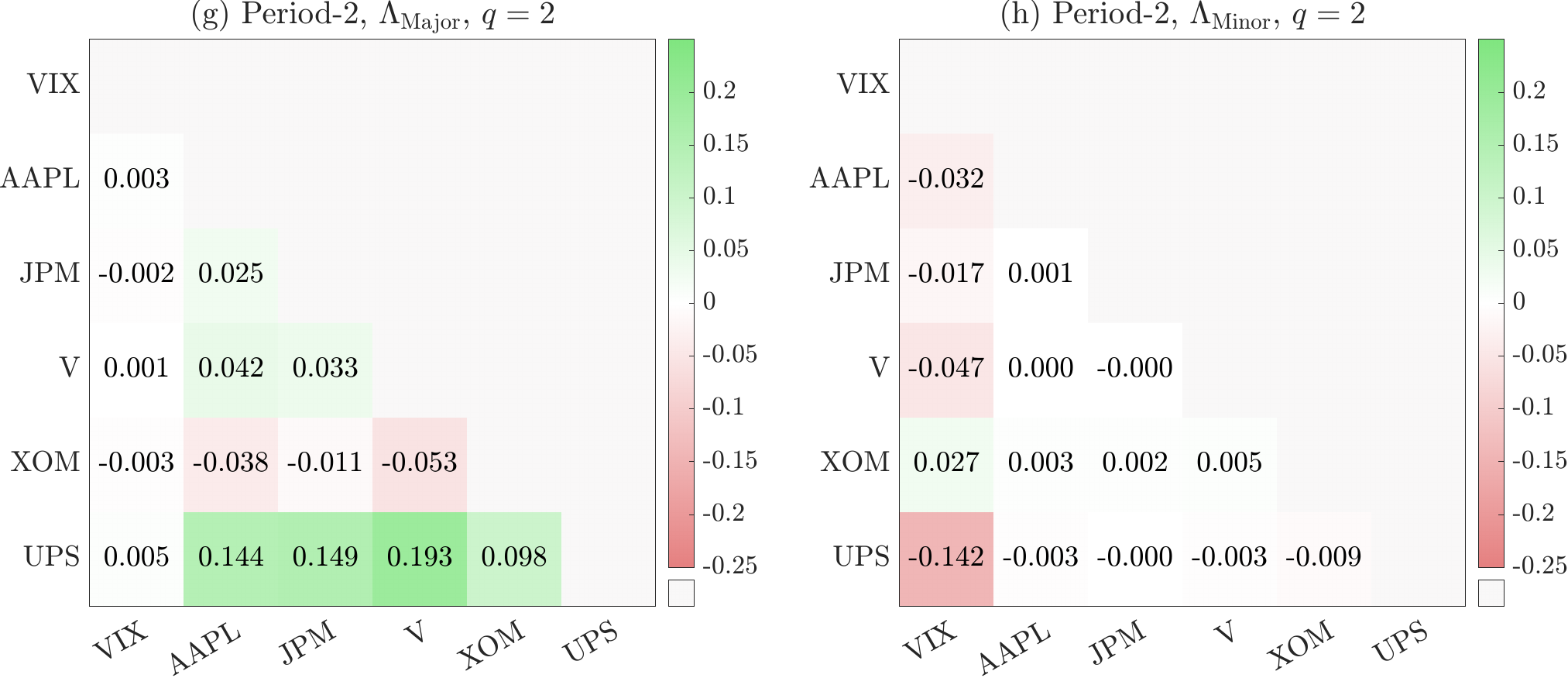} 
		\end{minipage}
		\caption{Pairwise asymmetric dependencies $\Lambda_\tinytxt{Major}(0.05)$ and $\Lambda_\tinytxt{Minor}(0.05)$ of the estimated TrUST copula models for 6-dimensional cross-industry example. Panels~(a, b, e,f) give results for the pre-COVID period data, and panels~(c, d, g, h) for the COVID period. The panels in the top row correspond to $q=1$ and panels in the bottom row when $q=2$.}
		\label{fig:heatmap_6d_win}
%\end{landscape}	
\end{sidewaysfigure}

Estimates of the asymmetric dependence measures $\Lambda_\tinytxt{Major}(0.05)$ and $\Lambda_\tinytxt{Minor}(0.05)$ for all variable pairs are presented as heatmaps in Figure~\ref{fig:heatmap_6d_win}. The level of asymmetry varies across
variable pairs in both periods and for both fitted copulas, particularly for $\Lambda_\tinytxt{Major}(0.05)$ which is the more important measure. 
However, there are substantive differences in direction and degree between the two copulas for some variable pairs due to the improved
fit of the TrUST copula with $q=2$. Equivalent results are given for quantile $\kappa=0.01$ 
in the Online Appendix.

\begin{figure}[tbh]
	\centering
	\includegraphics[width=0.8\textwidth]{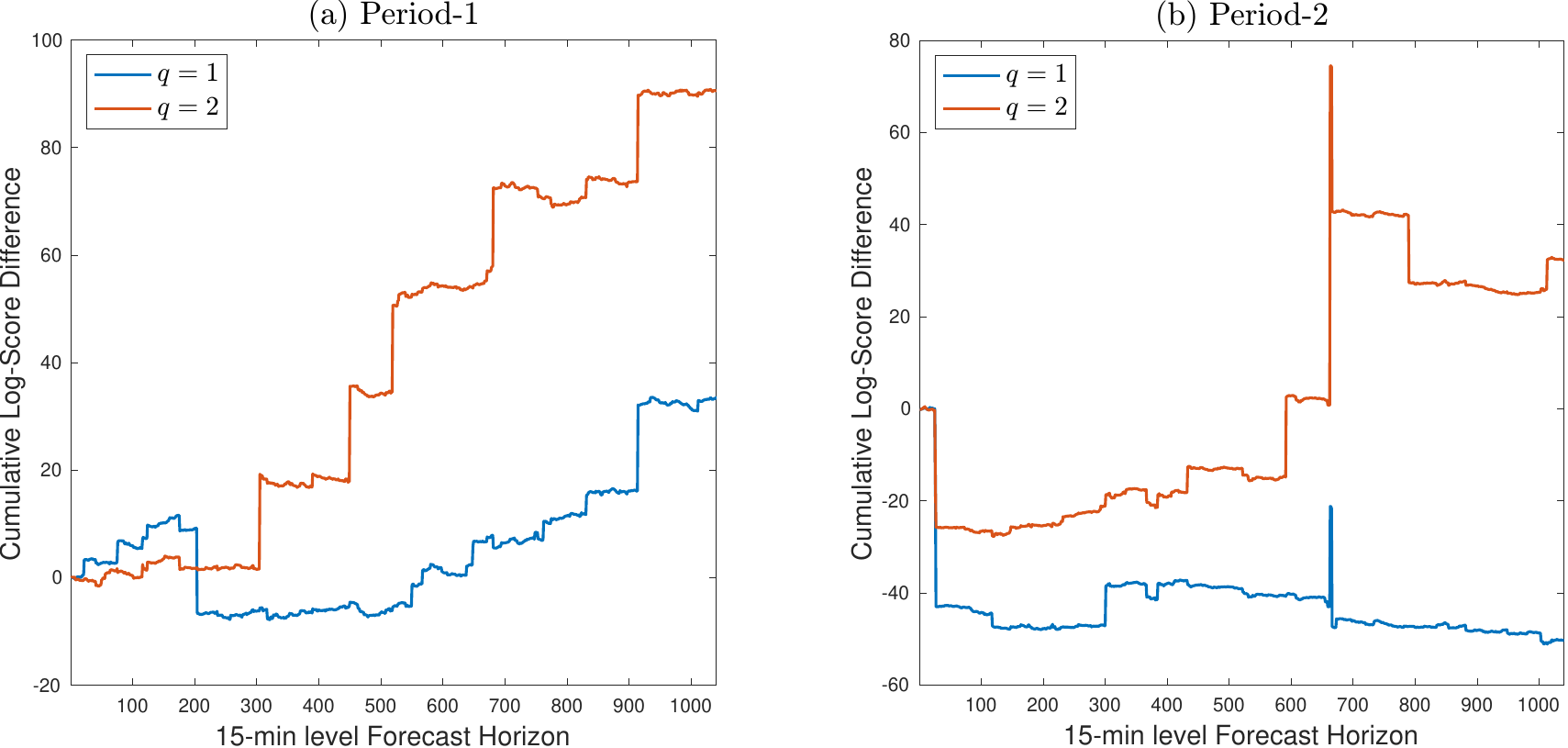}
	\caption{
		Cumulative difference in log-scores between (1) TrUST copula ($q=1$), (2) TrUST copula ($q=2$) and a benchmark Student-t copula model. All copula models have the same marginals. 
	}
	\label{fig:logscore_6d}
\end{figure}

To measure whether these differences are meaningful, we
consider their impact on density forecasts. To do so, the out-of-sample one-step-ahead predictive distributions were computed from both copula models over 40 trading day evaluation periods (equivalent to 1040 intraday observations) following the two estimation periods. Figure~\ref{fig:logscore_6d} presents the cumulative log-score (LS) differences between the TrUST copula models and a Student-t copula model with the same marginals.
Panel~(a) gives the results for the earlier pre-COVID estimation period, and panel~(b) for the later COVID estimation period. Positive values of the cumulative LS difference indicate improved performance, and the TrUST copula with $q=2$ outperforms both the AC skew-t copula and the benchmark Student-t copula in both evaluation periods. 

  \section{Discussion}\label{sec:07}
The UST distribution can capture richer asymmetry than the popular AC skew-t distribution which it nests. However, this potential has not been exploited in data analysis to date because inference is difficult. 
The current paper addresses this by specifying a novel tractable subclass for which likelihood-based 
inference can be computed
using data augmentation methods. The flexibility of this TrUST distribution is illustrated empirically using both simulated and regional Australian electricity price data.

The implicit copulas of skew-elliptical distributions~\citep{demarta_mcneil_2005,smith_gan_kohn_2010,lucas2017,Yoshiba_2018,oh_patton_2023,deng2024large}
are popular because they are both scalable and invariant to permutation of the random vector.
A major contribution of the current paper is to show how the implicit copula of the 
TrUST distribution is also tractable, and how a Bayesian data augmentation method can also be used to compute the posterior of its parameters. The TrUST copula with $q\geq 2$ allows for greater 
heterogeneity in asymmetric
dependence across variable pairs than the AC skew-t copula. An important application 
is the modeling of dependence between equity returns, where an improvement is demonstrated using a TrUST copula model for the intraday returns of large U.S. equities.

We finish by discussing four possible future extensions of our work. First, in this paper we compute exact posterior inference for both the TrUST distribution and its copula using MCMC methods. However, approximate Bayesian inference methods---notably variational inference---are 
faster and have the potential to allow estimation for higher dimensions $d$. Second, 
the prior for $\Omega$ is uniform on its $d(d-1)/2$ hyper-spherical angles. For high values of $d$ these angles can be regularized, such as through Bayesian global-local priors. Alternatively, a factor decomposition can be used for $\Omega$, as in~\cite{Murray_Dunson_Carin_Lucas_2013} or~\cite{deng2024large}, so that parameterization of 
$\Omega$ only increases linearly with $d$. Interestingly, this is similar to the approach used in USE distributions to parametric skewness using $\Delta$. Here, $\Delta$ is a $(d\times q)$ matrix where $q$ is fixed, so that the number of skew parameters only increases linearly with $d$.
Third, adopting a dynamic specification of the TrUST copula, as in the models of \cite{creal2015high,oh2017} and~\cite{oh_patton_2023}, would introduce time variation
in the dependence structure. Finally, parametric skewed distributions are increasingly popular
as fixed form posterior approximations~\citep{smith2020high,pozza2026skew}. \cite{tan2025variational} show that a sparse USN with $q=d$ can improve approximation accuracy in some applications, 
and the TrUST and TrUSN distributions with $q \ll d$ have strong potential as effective alternative approximations for large $d$.
They are particularly attractive for variational inference because the generative representation in Section~\ref{sec:gr} can be used to implement
efficiently stochastic gradient 
optimization using the re-parameterization trick of~\cite{Kingma_2014}.

% In this article, we build a tractable unified skew-elliptical distribution structure allowing for the greater flexibility in modeling skewness with the isolated unconstrained skewness vectors. 

% This methods in this paper suggests
  %\clearpage
% \newpage
\noindent
\appendix  

\section{Quantile Dependence}\label{app:quant_dep}
The pairwise quantile dependencies of the TrUST copula are computed from the bivariate marginal copula function of \( (U_1, U_2) \), which is 
\[ 
C(u_1, u_2) = F_{\UST, q}\lrp{(z_1, z_2)^\top; \Omega, \Delta, \Sigma, \nu} = \frac{T_{2+q}\lrp{(z_1, z_2, \zerovec_q^\top)^\top; R^*, \nu } }{ T_q\lrp{\zerovec^\top; \Sigma, \nu } }
\]
with  
\[
R^* = \left(\begin{array}{cc} 
	\Omega & -\Delta^\top  \\
	-\Delta & \Sigma \\
\end{array}\right), 
\]
which can be derived from Equations~\eqref{eq:sue_rep} and~\eqref{eq:cond_trunc} that 
\[
\dsP\lrp{\mZ \leq \zvec} = \dsP\lrp{\mX \leq \zvec | \mL > \zerovec}  = \frac{\dsP\lrp{\mX \leq \zvec, -\mL \leq \zerovec} }{\dsP\lrp{ -\mL \leq \zerovec} }. 
\]
Each pseudo observation 
\(
z_{ij}=F_{\UST,q}^{-1}(u_{ij};1, \Delta_j, \Sigma, \nu) 
\) 
is obtained via numerical optimization. 

  \newpage
  \singlespacing 
  \bibliography{references}
  % \clearpage
\newpage
\noindent
\begin{center}
	{\bf \Large{Online Appendix for ``A Tractable Unified Skew-t Distribution and Its Copula for Heterogeneous Asymmetries''}}
\end{center}
\spacing{1.5}
\vspace{10pt}
\setcounter{page}{1}
\setcounter{figure}{0}
\setcounter{table}{0}
\setcounter{section}{0}
\setcounter{equation}{0}
\setcounter{algorithm}{1}
\renewcommand{\thetable}{A\arabic{table}}
\renewcommand{\thefigure}{A\arabic{figure}}
\renewcommand{\thealgorithm}{\Alph{section}\arabic{algorithm}}
\renewcommand{\thesection}{Part~\Alph{section}}
\renewcommand{\theequation}{A\arabic{equation}}
\makeatletter
  \renewcommand\@seccntformat[1]{\csname the#1\endcsname\quad}
\makeatother

\noindent

This Online Appendix has the following parts:
\begin{itemize}
	\item[] {\bf Part~A}: Extended Expressions
	\begin{itemize}
		\item[] A.1: Extended TrUST Distribution
		\item[] A.2: Conditional Distribution
	\end{itemize}
    \item[] {\bf Part~B}: Alternative Generative Representations
    \begin{itemize}
    	\item[] B.1: Generative Representation 1
    	\item[] B.2: Generative Representation 2
    \end{itemize}
	\item[] {\bf Part~C}: Supplementary Figures
	\item[] {\bf Part~D}: Supplementary Tables
	\item[] {\bf Part~E}: Proofs
\end{itemize}

\newpage

\section{Extended Expression}\label{app:extended}
\subsection{Extended TrUST Distribution}\label{sec:etrust}
The term ``extended'' follows from \cite{Azzalini_Capitanio_2003} that has hidden truncation on $ L + \tau > 0$. 
\begin{corollary}[Extended TrUST Distribution]\label{theo:etrust}
For joint random vector \( \lrp{\mX^\top, \mL^\top}^\top \sim \T\lrp{\zerovec, R, \nu}\). We denote the Extended-TrUST (ETrUST) distribution for random vector \(\mZ \overset{\dd}{=} \mX | \mL + \tauvec > \zerovec\) with $\tauvec = \lrp{\tau_1,\ldots,\tau_q}^\top $. Then $\mZ \sim \ETRUST_q \lrp{\zerovec, \Omega, A, \nu, \tauvec}$ has density function 
\begin{equation}\label{eq:etrust}
	f_{\ETRUST, q}\lrp{\zvec; \Omega, A, \nu, \tauvec} = t \lrp{\zvec;  \Omega, \nu} \frac{T_q \lrp{ \sqrt{\frac{\nu+d}{\nu+Q \lrp{\zvec}}}  \lrp{\bm{\tautilde} + A^\top \zvec}; I_q, \nu +d } }{ T_q \lrp{\tauvec; \Sigma, \nu} },
\end{equation}
where $\tautilde_k = \tau_k \lrp{1-\deltavec_k^\top \Omega^{-1} \deltavec_k}^{-1/2}$, for $k=1,\ldots,q$, and $\bm{\tautilde}=(\tautilde_1,\tautilde_2,\ldots,\tautilde_q)^\top$.
\end{corollary}

\subsection{Conditional Distribution}\label{sec:condtrust}
\begin{corollary}[Conditional TrUST Distribution]\label{theo:condtrust}
	Let \(\mZ \sim TrUST_{q}(\Omega, A, \nu)\).  Partition
	\[
	\mZ \;=\; (\mZ_1,\mZ_2)^\top 
	\;\overset{d}{=}\; (\mX_1,\mX_2 \mid \mL>\zerovec),
	\]
	where \(\mZ_1\in\mathbb{R}^{d_1}\) and \(\mZ_2\in\mathbb{R}^{d_2}\), and 
	\[
	\Omega = 
	\begin{pmatrix}
	\Omega_1 & \Omega_{12} \\ 
	\Omega_{21} & \Omega_2
	\end{pmatrix}, 
	A = 
	\begin{pmatrix}
		A_1 \\ A_2
	\end{pmatrix}
	\]
	with \( A_1 = \lrp{\alphavec_{k(1)}}_{d_1 \times q},\;  A_2 = \lrp{\alphavec_{k(2)}}_{d_2 \times q} \).
	
	Define the conditional random vector
	\[
	\mZ_{1\mid 2}
	\;:=\;
	(\mZ_1 \mid \mZ_2 = \zvec_2)\,\overset{d}{=}\,
	(\mX_1 \mid \mX_2 = \xvec_2,\;\mL>\zerovec)\,.
	\]
	Then \(\mZ_{1\mid 2}\) admits an "extended" UST distribution, in which the latent location is shifted by conditioning \(\mZ_2\).  Its density is given by:
	\[
	    p_{\mZ_{1|2}}(\zvec_1) = t_{d_1}\lrp{\zvec_1 - \muvec_*; \frac{\nu+Q(\zvec_2)}{\nu+d_2} \Omega_*, \nu+d_2} 
		\frac{  
		T_q\lrp{\sqrt{\frac{\nu+d_2+d_1}{\nu + Q(\zvec)} } (A_1^\top \zvec_1  + A_2^\top \zvec_2); I_q, \nu+d_2+d_1} }{ T_q\lrp{\Delta_2 \Omega_{2}^{-1} \zvec_2; \frac{\nu+Q(\zvec_2)}{\nu+d_2} (\Sigma - \Delta_2 \Omega_{2}^{-1} \Delta_2^\top), \nu+d_2 } },
	\]
	where \( \muvec_* = \Omega_{12}\Omega_2^{-1} \zvec_2\), \( \Omega_{*} = \Omega_1 - \Omega_{12} \Omega_2^{-1} \Omega_{21}\), \(Q(\zvec_2) = \zvec_2^\top \Omega_{2}^{-1} \zvec_2, A=(A_1^\top,A_2^\top)^\top) \).
% \[
% \mZ_{1|2} \sim \ETRUST_q(\muvec_{1|2}, \Omega_{1|2}, A, \nu, \tauvec),
% \]
% where the parameters are defined as:
% \[
% \muvec_{1|2} = \Omega_{12} \Omega_2^{-1} \zvec_2, \quad  
% \Omega_{1|2} = \Omega_1 - \Omega_{12} \Omega_2^{-1} \Omega_{21}, \quad  
% \tauvec = \Delta_2 \Omega_2^{-1} \zvec_2, \quad  
% \Sigma_{|2} = \Sigma - \Delta_2 \Omega_2^{-1} \Delta_2^\top.
% \]
\end{corollary}

\section{Alternative Generative Representations}\label{sm:grs}

This section summarizes three equivalent generative representations for the TrUST distribution. The first representation works directly with the joint Student-t hidden truncation mechanism. The second representation corresponds to the scale mixture of normals construction used in Section~\ref{sec:gr}, where the common Student-t scaling is introduced before hidden truncation. The third representation reverses this ordering by applying hidden truncation in the Gaussian layer first, followed by the common Student-t scaling. Throughout this section, \(W\sim\mathrm{Gamma}(\nu/2,\nu/2)\) denotes a scalar mixing variable under the shape-rate parametrization.

\subsection{Generative Representation 1}\label{sm:gr1}

The first augmentation leverages the hidden truncation mechanism, as described in Eq.~\eqref{eq:sue_rep}, within the framework of the joint Student-t distribution without explicitly introducing the scale mixture of normal. Let
\[
\lrp{\mX^\top,\mL^\top}^\top \sim \T_{d+q}\lrp{\zerovec,R,\nu}, \qquad
R = \lrp{
	\begin{array}{cc}
		\Omega & \Delta^\top \\
		\Delta & \Sigma
	\end{array}}.
\]
The TrUST random vector is obtained by \(\mZ \overset{\dd}{=} \mX \mid \mL>\zerovec\). From the conditional distribution of a partitioned multivariate Student-t distribution,
\begin{equation*}
	\lrp{\mZ \mid \mL=\lvec}
	\sim
	\T_d\lrp{
		\Delta^\top \Sigma^{-1}\lvec,\,
		\frac{\nu+Q_\Sigma\lrp{\lvec}}{\nu+q}
		\lrp{\Omega-\Delta^\top\Sigma^{-1}\Delta},\,
		\nu+q
	},
\end{equation*}
where \(Q_\Sigma\lrp{\lvec}=\lvec^\top\Sigma^{-1}\lvec\). Hence, the augmented density of \(\lrc{\mZ,\mL}\) can be expressed as
\begin{equation}
	p\lrp{\zvec,\lvec} = t_d\lrp{ \zvec;\,
		\Delta^\top \Sigma^{-1}\lvec,\,
		\frac{\nu+Q_\Sigma\lrp{\lvec}}{\nu+q}
		\lrp{\Omega-\Delta^\top\Sigma^{-1}\Delta},\,
		\nu+q
	}
	t_{\lvec>\zerovec}\lrp{\lvec;\zerovec,\Sigma,\nu}.
\end{equation}
The latent variable \(\mL\) can still be sampled through
\begin{equation*}
	\lrp{\mL \mid \mZ,\thetavec} \sim
	\T_q^+\lrp{ \mvec,\,
	\frac{\nu+Q_\Omega\lrp{\mZ}}{\nu+d}H,\, \nu+d},
\end{equation*}
where 
\(Q_\Omega\lrp{\mZ}=\mZ^\top\Omega^{-1}\mZ\), \(H=\Sigma-\Delta\Omega^{-1}\Delta^\top\), 
and
\(\mvec = \Delta\Omega^{-1}\mZ \).
% Therefore, under the TrUST distribution construction, the latent variable \(\mL\) can still be sampled through
% \[
% \lrp{\mL \mid \mZ,\thetavec} 
% \sim T_q^+\lrp{ \sqrt{\frac{\nu+d}{\nu+Q_\Omega(\mZ)}} A^\top \mZ, I_q, \nu+d }.
% \]

\subsection{Generative Representation 2}\label{sm:gr2}

The second representation corresponds to the generative representation in Section~\ref{sec:gr}, where the common Student-t scaling is introduced before the hidden truncation step. Let \(W\sim\mathrm{Gamma}\lrp{\nu/2,\nu/2}\) and
\[
	\lrp{\mX^\top,\mL^\top}^\top \mid W
	\sim
	\N_{d+q}\lrp{\zerovec,W^{-1}R}.
\]
Under this representation,
\begin{equation*}
	\lrp{\mZ \mid \mL,W}
	\sim
	\N_d\lrp{
		\Delta^\top\Sigma^{-1}\mL,\,
		W^{-1}\lrp{\Omega-\Delta^\top\Sigma^{-1}\Delta}
	}.
\end{equation*}
Hence, conditional on \(W\),
\begin{equation*}
	\lrp{\mL \mid \mZ,W,\thetavec} \sim
	\N_q^+\lrp{\mvec,W^{-1}H}.
\end{equation*}
Under the TrUST conditional independence restriction \(H=\diag\lrp{h_1,\ldots,h_q}\), this further implies
\begin{equation*}
	L_k \mid \mZ,W,\thetavec
	\sim
	\N^+\lrp{m_k,W^{-1}h_k}
\end{equation*}
where \(m_k=\deltavec_k^\top\Omega^{-1}\mZ\) and \(h_k=1-\deltavec_k^\top\Omega^{-1}\deltavec_k\), for \(k=1,\ldots,q\).
The posterior of \(W\) also has a closed form under this scaled latent representation:
\begin{equation*}
	W \mid \mZ,\mL,\thetavec \sim
	\mathrm{Gamma}\lrp{ \frac{\nu+d+q}{2},\,
	\frac{1}{2} \lrc{ \nu + Q_\Omega\lrp{\mZ} + \sum_{k=1}^q \frac{\lrp{L_k-m_k}^2}{h_k}
		}
	}.
\end{equation*}

% The posterior of \(W\) also has a closed form under this scaled latent representation. Specifically,
% \begin{equation*}
% 	W \mid \mZ,\mL,\thetavec \sim
% 	\mathrm{Gamma}\lrp{ \frac{\nu+d+q}{2},\,
% 		\frac{1}{2} \lrc{\nu + Q_\Omega\lrp{\mZ} + \lrp{\mL-\Delta\Omega^{-1} \mZ}^\top H^{-1} \lrp{\mL-\Delta\Omega^{-1}\mZ}
% 		}
% 	}.
% \end{equation*}

\subsection{Generative Representation 3}\label{sm:gr3}

The third representation follows the alternative ordering where hidden truncation is applied first in the Gaussian layer, followed by the common Student-t scaling. Let
\[
	\lrp{\mtildeX^\top,\mtildeL^\top}^\top
	\sim
	\N_{d+q}\lrp{\zerovec,R},
\]
where \(W\) is independent of \(\lrp{\mtildeX^\top,\mtildeL^\top}^\top\). Then, given \(W > 0\) and the hidden truncation event \(\mtildeL>\zerovec\), we can express the TrUST random vector as
\[
	\mZ \overset{\dd}{=} 
	\lrp{W^{-1/2}\mtildeX} \mid \lrp{W^{-1/2}\mtildeL>\zerovec} = 
	\lrp{W^{-1/2}\mtildeX} \mid \mtildeL>\zerovec .
\]
Thus, applying the common positive scaling before or after the hidden truncation event leads to the same marginal TrUST distribution.

Conditional on \(W\), the joint distribution of \(W^{-1/2}\mtildeX\) and \(\mtildeL\) is
\begin{equation*}
	\left.
	\lrp{
	\begin{array}{c}
		W^{-1/2}\mtildeX \\
		\mtildeL
	\end{array}}
	\,\right|\, W
	\sim
	\N_{d+q}
	\lrp{
		\zerovec,\,
		\lrp{
		\begin{array}{cc}
			W^{-1}\Omega & W^{-1/2}\Delta^\top \\
			W^{-1/2}\Delta & \Sigma
		\end{array}}
	}.
\end{equation*}
Therefore,
\begin{equation*}
	\lrp{\mZ \mid \mtildeL,W}
	\sim
	\N_d\lrp{
		W^{-1/2}\Delta^\top\Sigma^{-1}\mtildeL,\,
		W^{-1}\lrp{\Omega-\Delta^\top\Sigma^{-1}\Delta}
	}.
\end{equation*}

The full conditional distribution of the Gaussian latent variable \(\mtildeL\) is
\begin{equation*}
	\lrp{\mtildeL \mid \mZ,W,\thetavec}
	\sim
	\N_q^+\lrp{W^{1/2}\mvec,H}.
\end{equation*}
Under the TrUST conditional independence restriction, this becomes
\begin{equation*}
	\widetilde L_k \mid \lrp{ \mZ,W,\thetavec }
	\sim
	\N^+\lrp{W^{1/2}m_k,h_k},
	\qquad k=1,\ldots,q.
\end{equation*}

Unlike the second representation, the full conditional distribution of \(W\) under this augmentation is not available in a Gamma closed form. To see this, let
\(
C=\Omega-\Delta^\top\Sigma^{-1}\Delta
\), 
\(\bvec=\Delta^\top\Sigma^{-1}\mtildeL .
\)
Then, the full conditional density of \(W\) is
\begin{align*}
	p\lrp{W \mid \mZ,\mtildeL,\thetavec}
	&\propto W^{\lrp{\nu+d}/2-1}
	\exp\lrc{
		-\frac{1}{2}
		\lrp{ \nu W + W
			\lrp{\mZ - W^{-1/2}\bvec}^\top
			C^{-1}
			\lrp{\mZ-W^{-1/2}\bvec}
		}
	} \\
	&\propto W^{\lrp{\nu+d}/2-1}
	\exp\lrc{
		-\frac{1}{2}
		\lrp{\nu+\mZ^\top C^{-1}\mZ}W + W^{1/2}\mZ^\top C^{-1}\bvec
	}.
\end{align*}
The appearance of the \(W^{1/2}\) term prevents this full conditional from reducing to a Gamma distribution. Therefore, \(W\) can be updated using a Metropolis--Hastings step. 
% For example, if \(\tildeW=\log W\) is updated by a random-walk proposal, the corresponding log-likelihood is
% \begin{equation*}
% 	\ell\lrp{\tildeW} =
% 	\frac{\nu+d}{2}\tildeW -
% 	\frac{1}{2} \lrp{\nu+\zvec^\top C^{-1}\zvec} \exp\lrp{\tildeW} + 
% 	\lrp{\zvec^\top C^{-1}\bvec}
% 	\exp\lrp{\tildeW/2}.
% \end{equation*}
 
Both GR1 and GR3 provide alternative augmentations to the scale mixture representation in GR2. The first representation avoids explicitly introducing the scale mixture variable, whereas the third representation applies hidden truncation in the Gaussian layer before scaling. These representations are distributionally equivalent for generating the TrUST random vector, but they lead to different augmented posterior structures and therefore different computational implications.

\section{Supplementary Figures}
% \subsection{Monte Carlo Proof of Theorem \ref{theo:kendall} and \ref{theo:spearman}}
% \begin{figure}[H] 
% 	\centering
% 	\includegraphics[width=0.85\textwidth]{figs/RankCorr_FSN_9_plot}	
% 	\caption{
% 		Contour density plots of the bivariate FST distribution with share the same degrees of freedom $\nu = 5$ and the first skewness vector $\alphavec_1 = (5, 5)^\top$. Each row corresponds to a different correlation $\omega = \lrc{-0.5, 0, 0.5 }$ from top to bottom. The columns have different second skewness vectors $\alphavec_2 = \lrc{ (5, 5)^\top, (0, 5)^\top, (-5, 5)^\top }$ from left to right. 
% 		}
% 	\label{fig:rankcorr} 
% \end{figure}
  
\begin{figure}[H]
	\centering
	\includegraphics[width=1\textwidth]{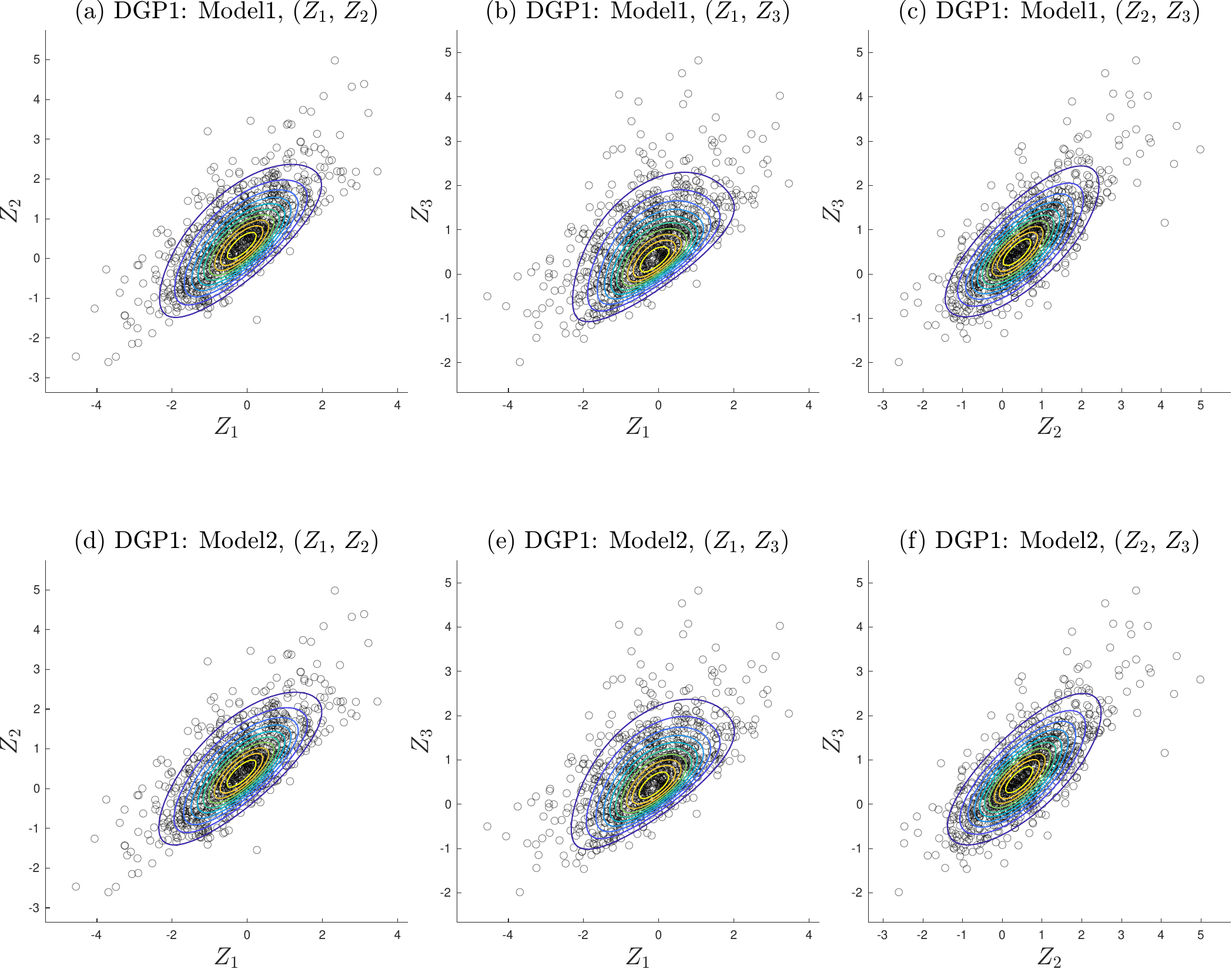}
	\caption{Estimated contours of DGP1 as described in Sec.~\ref{sec:sim_dist}. The top row displays results from Model1, whereas the bottom row shows those from Model2.}
	\label{fig:sim_dist_case_1}
\end{figure}

\begin{figure}[H]
	\centering
	\includegraphics[width=1\textwidth]{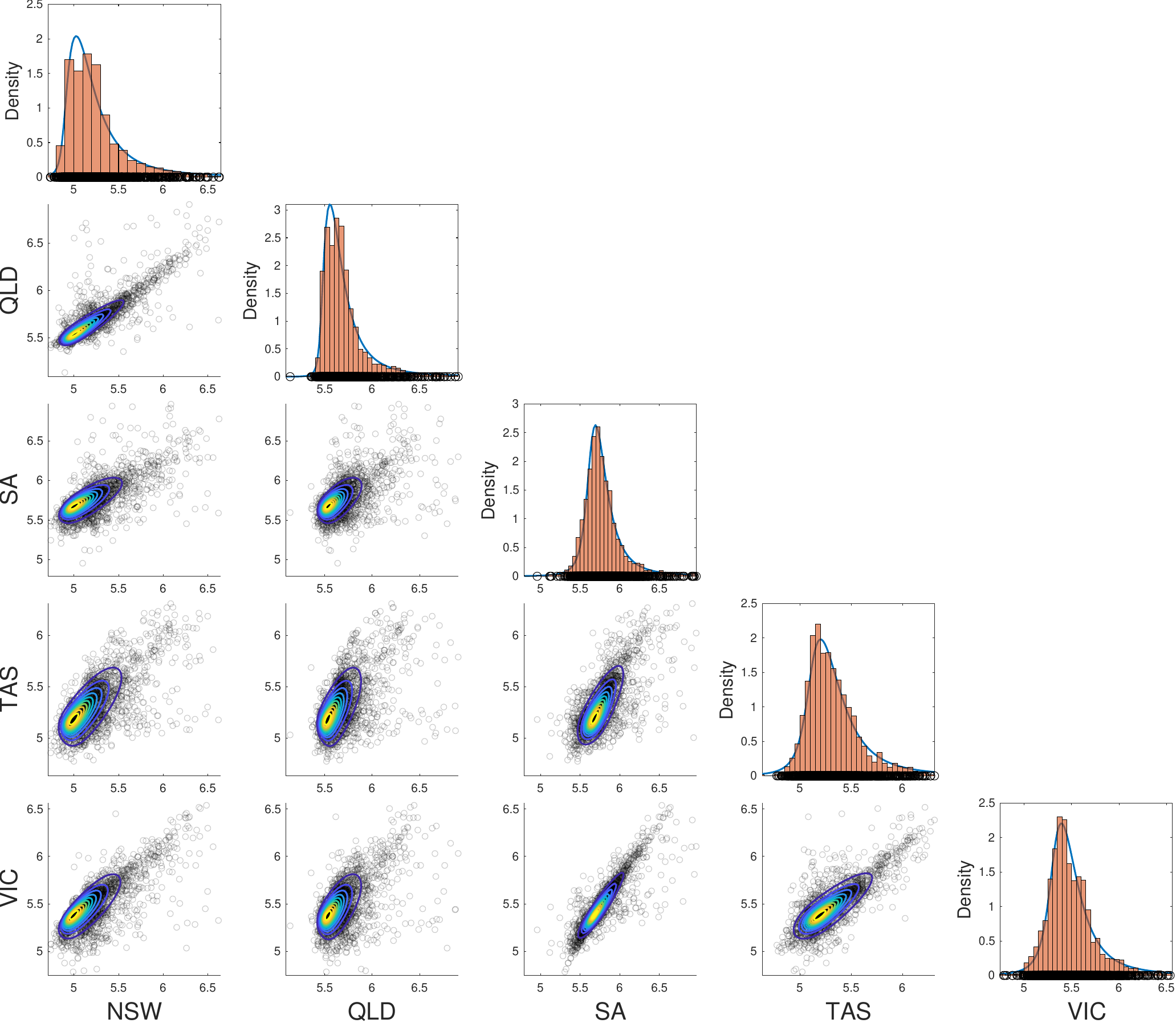}
	\caption{Estimated contours for the distribution application outlined in Sec.~\ref{sec:dist_app}. Diagonal panels show the marginal fits, while the lower triangular panels display the bivariate contours.}
	\label{fig:nem_pairs} 
\end{figure}

\begin{figure}[H]
	\centering
	\includegraphics[width=0.7\textwidth]{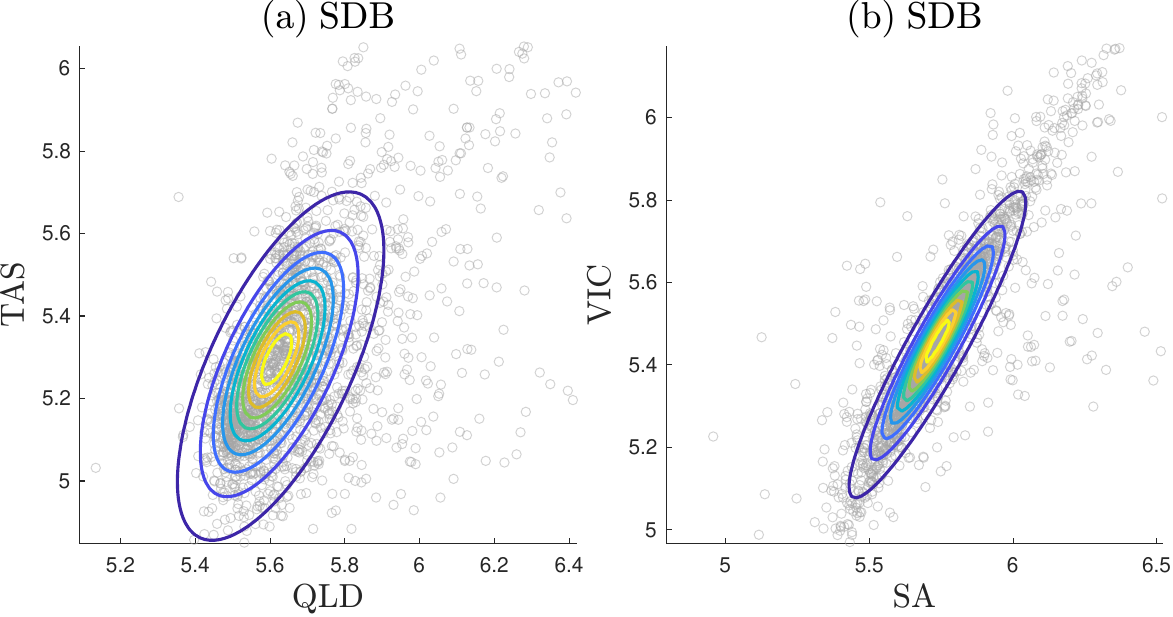}
	\caption{ Estimated contours of the SDB skew-t distribution for the distribution application outlined in Sec.~\ref{sec:dist_app}.}	
	\label{fig:sdb_nem_pairs}
\end{figure}

\begin{figure}[H]
	\centering
	\includegraphics[width=1\textwidth]{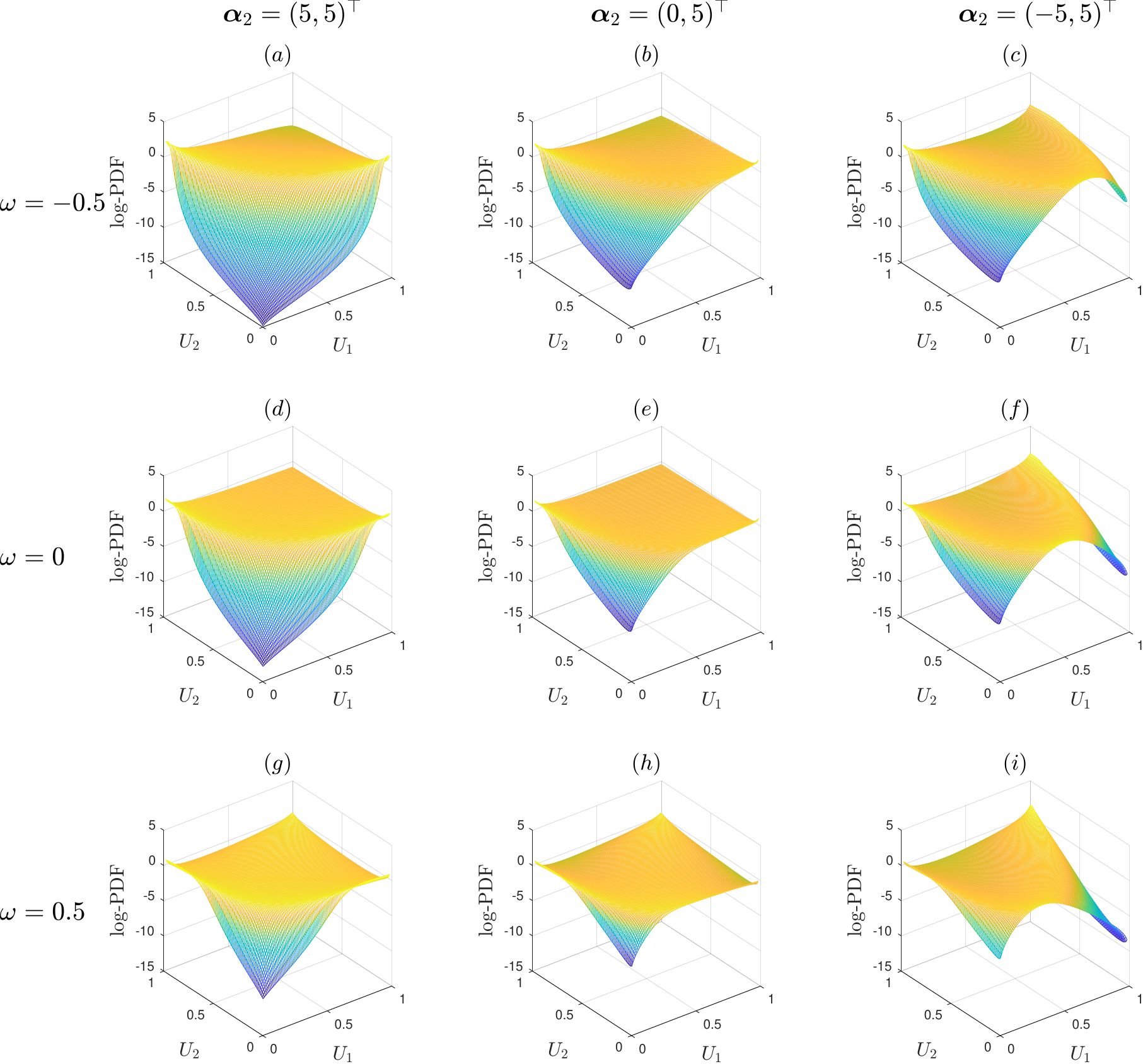}
	\caption{Surface plots of the bivariate TrUST copula log-density (\(q = 2\)) with parameters specified in Fig.~\ref{fig:fst_pdfs}. With \(\nu = 5\) and \(\alphavec_1 = (5, 5)^\top\) held constant, rows represent correlation values \(\omega \in \{-0.5, 0, 0.5\}\) (from top to bottom), while columns correspond to second skewness vectors \(\alphavec_2 \in \{(5, 5)^\top, (0, 5)^\top, (-5, 5)^\top\}\) (from left to right).}
	\label{fig:trust_copula_surface}
\end{figure}

\begin{figure}[H]
	\centering
	\includegraphics[width=1\textwidth]{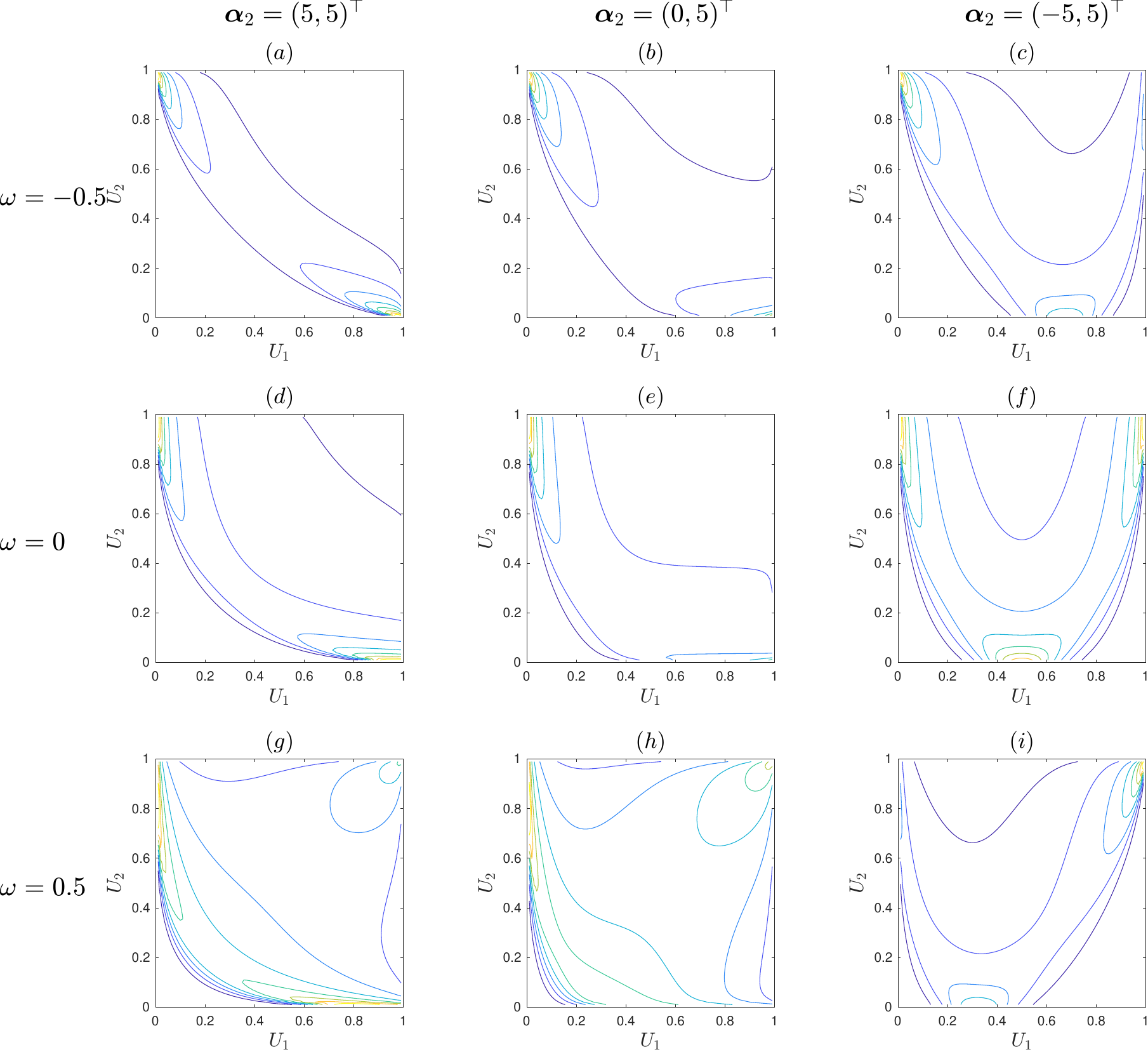}
	\caption{  
		Contour plots of the bivariate TrUST \((q = 2)\) copula density with \(\nu = 5\) and \(\alphavec_1 = (5, 5)^\top\) fixed. Rows vary by correlation \(\omega \in \{-0.5, 0, 0.5\}\) (top to bottom), and columns by parameter vector \(\alphavec_2 \in \{(5, 5)^\top, (0, 5)^\top, (-5, 5)^\top\}\) (left to right).  These are the implicit copulas of the bivariate distributions depicted in Figure~\ref{fig:fst_pdfs}.}	\label{fig:fstc_pdfs}
\end{figure}

\begin{landscape}
	\begin{figure}[htbp]
		\centering
		% Left composite figure
		\begin{minipage}[t]{1\linewidth}
			\centering
			\includegraphics[width=0.49\linewidth]{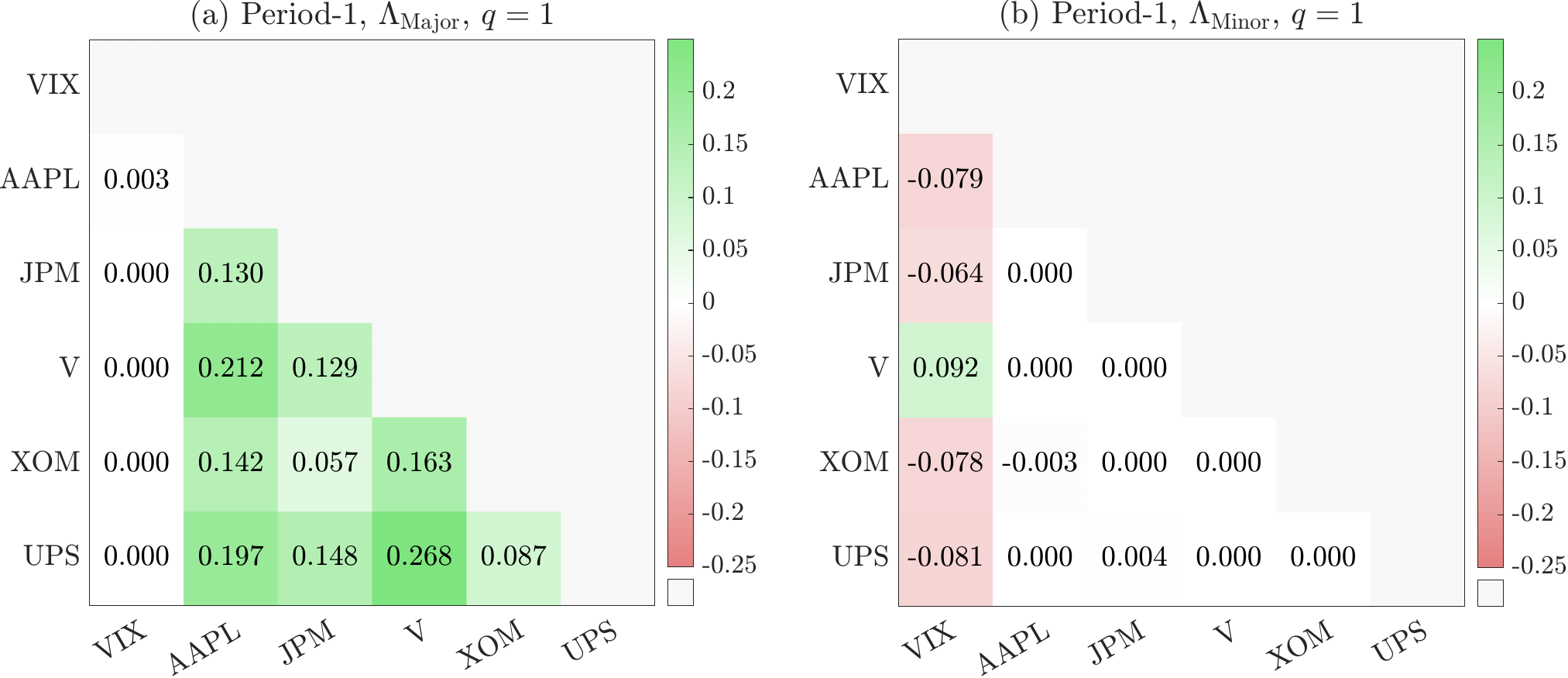} \hfill
			\includegraphics[width=0.49\linewidth]{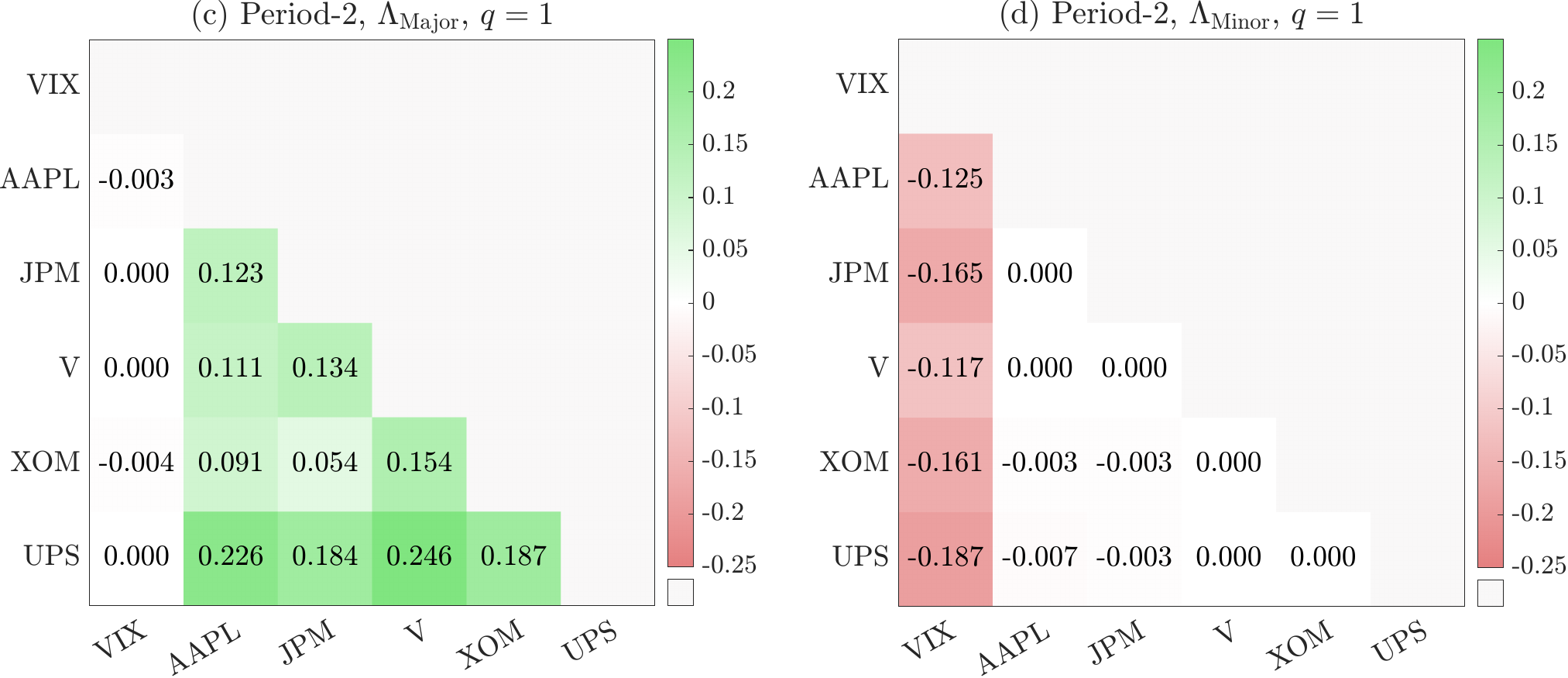} \\[3ex]
			\includegraphics[width=0.49\linewidth]{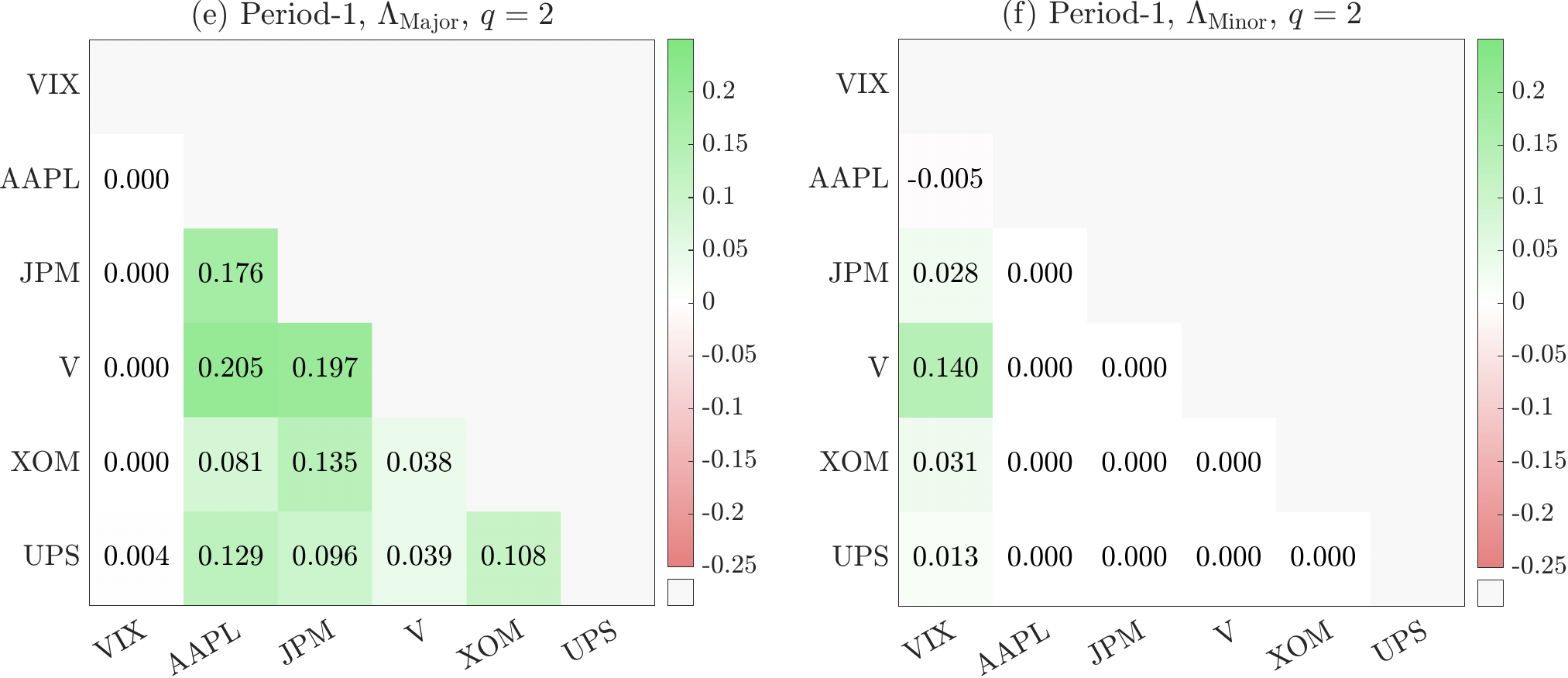} \hfill
			\includegraphics[width=0.49\linewidth]{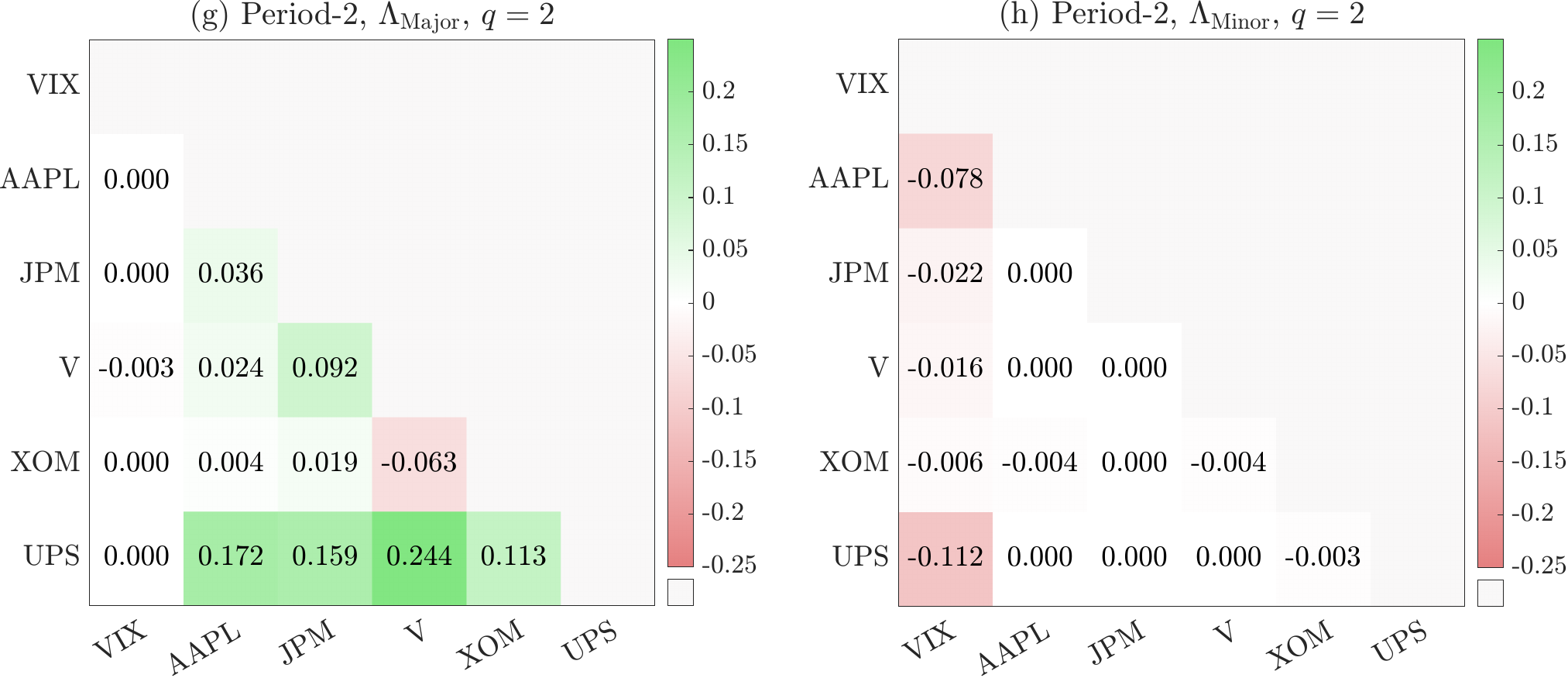} 
		\end{minipage}
		\caption{Pairwise asymmetric dependencies $\Lambda_\tinytxt{Major}(0.01)$ and $\Lambda_\tinytxt{Minor}(0.01)$ of the estimated TrUST copula models for 6-dimensional cross-industry example. Panels~(a,b,e,f) give results for the pre-COVID period data, and panels~(c,d,g,h) for the COVID period. The panels in the top row correspond to $q=1$ and panels in the bottom row when $q=2$. The results are presented in the same format as
		those at quantile $\kappa=0.05$ given in the manuscript.}
		\label{fig:heatmap_6d_win001}
	\end{figure}
\end{landscape}

\section{Supplementary Tables}\label{app:tables} 
% \subsection{Estimates of Distribution Example}
\begin{table}[H]
    \centering
    \caption{Bayesian MCMC Estimates of Different Distribution Models}
    \resizebox{1\textwidth}{!}{

\begin{tabular}{ccccccccccccccc}
\midrule
\midrule
Dist. Name &     & Gaussian &     & Skew-Normal &     & \multicolumn{2}{c}{TrUSN} &     & Student-t &     & Skew-t &     & \multicolumn{2}{c}{TrUST} \\
$q$ &     & 0   &     & 1   &     & 2   & 3   &     & 0   &     & 1   &     & 2   & 3 \\
\cmidrule{1-1}\cmidrule{3-3}\cmidrule{5-5}\cmidrule{7-8}\cmidrule{10-10}\cmidrule{12-12}\cmidrule{14-15}\rowcolor[rgb]{ .906,  .902,  .902} $p$ &     & 20  &     & 25  &     & 30  & 35  &     & 21  &     & 26  &     & 31  & 36 \\
DIC &     & -9324 &     & -10657 &     & -11163 & -11484 &     & -18510 &     & -19540 &     & \textbf{-19967} & -19697 \\
\rowcolor[rgb]{ .906,  .902,  .902} $\muvec$ &     & 5.240 &     & 4.890 &     & 4.992 & 4.846 &     & 5.133 &     & 4.931 &     & 4.886 & 4.891 \\
\rowcolor[rgb]{ .906,  .902,  .902}     &     & 5.700 &     & 5.476 &     & 5.461 & 5.379 &     & 5.622 &     & 5.486 &     & 5.456 & 5.460 \\
\rowcolor[rgb]{ .906,  .902,  .902}     &     & 5.781 &     & 5.641 &     & 5.544 & 5.461 &     & 5.736 &     & 5.661 &     & 5.634 & 5.631 \\
\rowcolor[rgb]{ .906,  .902,  .902}     &     & 5.324 &     & 5.180 &     & 5.233 & 5.025 &     & 5.284 &     & 5.178 &     & 5.128 & 5.142 \\
\rowcolor[rgb]{ .906,  .902,  .902}     &     & 5.493 &     & 5.316 &     & 5.304 & 5.219 &     & 5.451 &     & 5.363 &     & 5.328 & 5.336 \\
$\svec$ &     & 0.324 &     & 0.476 &     & 0.485 & 0.930 &     & 0.189 &     & 0.267 &     & 0.264 & 0.637 \\
    &     & 0.243 &     & 0.330 &     & 0.884 & 1.194 &     & 0.125 &     & 0.179 &     & 0.173 & 0.435 \\
    &     & 0.260 &     & 0.296 &     & 0.911 & 0.890 &     & 0.135 &     & 0.150 &     & 0.169 & 0.275 \\
    &     & 0.268 &     & 0.304 &     & 0.303 & 0.838 &     & 0.182 &     & 0.204 &     & 0.238 & 0.594 \\
    &     & 0.270 &     & 0.323 &     & 0.636 & 0.599 &     & 0.162 &     & 0.179 &     & 0.207 & 0.408 \\
\rowcolor[rgb]{ .906,  .902,  .902} $\vecl\lrp{\Omegabar}$ &     & 0.773 &     & 0.883 &     & 0.837 & 0.862 &     & 0.911 &     & 0.958 &     & 0.954 & 0.992 \\
\rowcolor[rgb]{ .906,  .902,  .902}     &     & 0.605 &     & 0.710 &     & -0.497 & -0.531 &     & 0.746 &     & 0.800 &     & 0.808 & 0.904 \\
\rowcolor[rgb]{ .906,  .902,  .902}     &     & 0.637 &     & 0.729 &     & 0.204 & 0.584 &     & 0.726 &     & 0.793 &     & 0.804 & 0.970 \\
\rowcolor[rgb]{ .906,  .902,  .902}     &     & 0.695 &     & 0.797 &     & -0.366 & 0.375 &     & 0.773 &     & 0.816 &     & 0.819 & 0.947 \\
\rowcolor[rgb]{ .906,  .902,  .902}     &     & 0.450 &     & 0.613 &     & -0.844 & -0.841 &     & 0.619 &     & 0.723 &     & 0.727 & 0.877 \\
\rowcolor[rgb]{ .906,  .902,  .902}     &     & 0.511 &     & 0.652 &     & -0.180 & 0.145 &     & 0.642 &     & 0.742 &     & 0.746 & 0.960 \\
\rowcolor[rgb]{ .906,  .902,  .902}     &     & 0.488 &     & 0.671 &     & -0.763 & -0.104 &     & 0.624 &     & 0.725 &     & 0.725 & 0.923 \\
\rowcolor[rgb]{ .906,  .902,  .902}     &     & 0.611 &     & 0.700 &     & 0.524 & 0.244 &     & 0.753 &     & 0.810 &     & 0.852 & 0.925 \\
\rowcolor[rgb]{ .906,  .902,  .902}     &     & 0.817 &     & 0.862 &     & 0.966 & 0.525 &     & 0.930 &     & 0.946 &     & 0.956 & 0.978 \\
\rowcolor[rgb]{ .906,  .902,  .902}     &     & 0.733 &     & 0.800 &     & 0.635 & 0.912 &     & 0.816 &     & 0.860 &     & 0.894 & 0.967 \\
$\deltavec_1$ &     &     &     & 0.983 &     & 0.761 & 0.694 &     &     &     & 0.945 &     & 0.960 & 0.992 \\
    &     &     &     & (0.979, 0.987) &     & ( 0.569,  0.845) & ( 0.607,  0.744) &     &     &     & (0.927, 0.960) &     & (0.926, 0.973) & ( 0.988,  0.995) \\
    &     &     &     & 0.906 &     & 0.987 & 0.959 &     &     &     & 0.954 &     & 0.946 & 0.989 \\
    &     &     &     & (0.883, 0.925) &     & ( 0.981,  0.991) & ( 0.939,  0.973) &     &     &     & (0.933, 0.966) &     & (0.893, 0.967) & ( 0.985,  0.994) \\
    &     &     &     & 0.634 &     & -0.908 & -0.926 &     &     &     & 0.619 &     & 0.818 & 0.907 \\
    &     &     &     & (0.594, 0.670) &     & (-0.926, -0.888) & (-0.940, -0.915) &     &     &     & (0.558, 0.696) &     & (0.778, 0.840) & ( 0.884,  0.941) \\
    &     &     &     & 0.633 &     & -0.308 & -0.132 &     &     &     & 0.634 &     & 0.876 & 0.983 \\
    &     &     &     & (0.592, 0.671) &     & (-0.606,  0.231) & (-0.347,  0.118) &     &     &     & (0.565, 0.710) &     & (0.834, 0.905) & ( 0.982,  0.992) \\
    &     &     &     & 0.731 &     & -0.844 & -0.368 &     &     &     & 0.610 &     & 0.836 & 0.952 \\
    &     &     &     & (0.701, 0.757) &     & (-0.879, -0.782) & (-0.548, -0.131) &     &     &     & (0.547, 0.686) &     & (0.795, 0.862) & ( 0.943,  0.967) \\
\rowcolor[rgb]{ .906,  .902,  .902} $\deltavec_2$ &     &     &     &     &     & -0.623 & -0.851 &     &     &     &     &     & 0.719 & 0.898 \\
\rowcolor[rgb]{ .906,  .902,  .902}     &     &     &     &     &     & (-0.729, -0.391) & (-0.904, -0.768) &     &     &     &     &     & (0.669, 0.846) & ( 0.838,  0.930) \\
\rowcolor[rgb]{ .906,  .902,  .902}     &     &     &     &     &     & -0.912 & -0.950 &     &     &     &     &     & 0.763 & 0.917 \\
\rowcolor[rgb]{ .906,  .902,  .902}     &     &     &     &     &     & (-0.933, -0.890) & (-0.961, -0.933) &     &     &     &     &     & (0.727, 0.872) & ( 0.863,  0.945) \\
\rowcolor[rgb]{ .906,  .902,  .902}     &     &     &     &     &     & 0.977 & 0.874 &     &     &     &     &     & 0.267 & 0.654 \\
\rowcolor[rgb]{ .906,  .902,  .902}     &     &     &     &     &     & ( 0.967,  0.983) & ( 0.818,  0.908) &     &     &     &     &     & (0.157, 0.475) & ( 0.541,  0.756) \\
\rowcolor[rgb]{ .906,  .902,  .902}     &     &     &     &     &     & 0.392 & -0.226 &     &     &     &     &     & 0.275 & 0.813 \\
\rowcolor[rgb]{ .906,  .902,  .902}     &     &     &     &     &     & (-0.151,  0.678) & (-0.478, -0.020) &     &     &     &     &     & (0.154, 0.446) & ( 0.744,  0.878) \\
\rowcolor[rgb]{ .906,  .902,  .902}     &     &     &     &     &     & 0.923 & 0.061 &     &     &     &     &     & 0.229 & 0.721 \\
\rowcolor[rgb]{ .906,  .902,  .902}     &     &     &     &     &     & ( 0.878,  0.945) & (-0.210,  0.302) &     &     &     &     &     & (0.118, 0.462) & ( 0.625,  0.794) \\
$\deltavec_3$ &     &     &     &     &     &     & 0.686 &     &     &     &     &     &     & -0.889 \\
    &     &     &     &     &     &     & ( 0.510,  0.825) &     &     &     &     &     &     & (-0.926, -0.872) \\
    &     &     &     &     &     &     & 0.278 &     &     &     &     &     &     & -0.897 \\
    &     &     &     &     &     &     & ( 0.045,  0.520) &     &     &     &     &     &     & (-0.939, -0.880) \\
    &     &     &     &     &     &     & 0.090 &     &     &     &     &     &     & -0.739 \\
    &     &     &     &     &     &     & (-0.147,  0.317) &     &     &     &     &     &     & (-0.823, -0.686) \\
    &     &     &     &     &     &     & 0.981 &     &     &     &     &     &     & -0.898 \\
    &     &     &     &     &     &     & ( 0.959,  0.991) &     &     &     &     &     &     & (-0.936, -0.884) \\
    &     &     &     &     &     &     & 0.852 &     &     &     &     &     &     & -0.827 \\
    &     &     &     &     &     &     & ( 0.766,  0.930) &     &     &     &     &     &     & (-0.868, -0.792) \\
\rowcolor[rgb]{ .906,  .902,  .902} $\nu$ &     &     &     &     &     &     &     &     & 2.189 &     & 2.024 &     & 2.029 & 2.031 \\
\bottomrule
\bottomrule
\end{tabular}%
}
    \label{tab:app_NEM}
    \caption*{Note: MCMC estimates for the distributions of the log-price are provided. The latent dimension \(q\), the number of free parameters \(p\), and the DIC values are given along with estimates for the location parameter \(\muvec\), scale parameter \(\svec\), the lower triangular elements of the correlation parameter \(\Omega\), the joint skewness vectors \(\deltavec_1\), \(\deltavec_2\), \(\deltavec_3\), and the degrees of freedom \(\nu\).}
\end{table}%

\begin{table}[H]
	\centering
	\caption{Dependence Metrics for Data Generated using DGP1 (where $q = 1$) }
	\resizebox{1.0\textwidth}{!}{
\begin{tabular}{ccrccrccrcc}
\toprule
\toprule
\multicolumn{2}{c}{Pair $\lrp{U_i, U_j}$} &     & \multicolumn{2}{c}{True Values} &     & \multicolumn{2}{c}{Copula~1 Estimates} &     & \multicolumn{2}{c}{Copula~2 Estimates} \\
\cmidrule{1-2}\cmidrule{4-5}\cmidrule{7-8}\cmidrule{10-11}    &     &     & \multicolumn{8}{l}{Panel A: Rank Correlations} \\
\cmidrule{4-11}$i$ & $j$ &     & Kendall & Spearman &     & Kendall & Spearman &     & Kendall & Spearman \\
\multirow{2}[0]{*}{2} & \multirow{2}[0]{*}{1} &     & 0.505 & 0.686 &     & 0.504 & 0.685 &     & 0.499 & 0.678 \\
    &     &     &     &     &     & (0.013) & (0.015) &     & (0.015) & (0.018) \\
\multirow{2}[0]{*}{3} & \multirow{2}[0]{*}{1} &     & 0.405 & 0.564 &     & 0.385 & 0.538 &     & 0.379 & 0.529 \\
    &     &     &     &     &     & (0.016) & (0.020) &     & (0.016) & (0.020) \\
\multirow{2}[0]{*}{3} & \multirow{2}[0]{*}{2} &     & 0.533 & 0.722 &     & 0.526 & 0.713 &     & 0.525 & 0.712 \\
    &     &     &     &     &     & (0.013) & (0.014) &     & (0.013) & (0.014) \\
\midrule
\midrule
    &     &     & \multicolumn{8}{l}{Panel B: Asymmetric Dependencies} \\
\cmidrule{4-11}$i$ & $j$ &     & $\Lambda_\tinytxt{Major}(0.05)$ & $\Lambda_\tinytxt{Minor}(0.05)$ &     & $\Lambda_\tinytxt{Major}(0.05)$ & $\Lambda_\tinytxt{Minor}(0.05)$ &     & $\Lambda_\tinytxt{Major}(0.05)$ & $\Lambda_\tinytxt{Minor}(0.05)$ \\
\multirow{2}[0]{*}{2} & \multirow{2}[0]{*}{1} &     & -0.036 & 0.005 &     & -0.045 & 0.007 &     & -0.030 & 0.007 \\
    &     &     &     &     &     & (0.028) & (0.004) &     & (0.027) & (0.004) \\
\multirow{2}[0]{*}{3} & \multirow{2}[0]{*}{1} &     & -0.091 & 0.019 &     & -0.092 & 0.023 &     & -0.085 & 0.029 \\
    &     &     &     &     &     & (0.024) & (0.008) &     & (0.027) & (0.009) \\
\multirow{2}[0]{*}{3} & \multirow{2}[0]{*}{2} &     & 0.098 & 0.000 &     & 0.112 & 0.001 &     & 0.107 & 0.000 \\
    &     &     &     &     &     & (0.030) & (0.002) &     & (0.029) & (0.002) \\
\bottomrule
\bottomrule
\end{tabular}%
}
	\label{tab:sim_case_1}
	% \caption*{Note: In }
\end{table}%

\begin{table}[H]
    \centering
    \caption{Estimates of TrUST Copula for VIX and two equities}
    \resizebox{1.0\textwidth}{!}{
\begin{tabular}{ccccccccccccc}
      \toprule
      \toprule
      \multicolumn{13}{c}{Panel A: Estimates with $q = 1$ and DIC = -1216} \\
      \midrule
      \multicolumn{2}{c}{Pair $\lrp{U_i, U_j}$} &     & \multicolumn{2}{c}{Rank Correlation} &     & \multicolumn{2}{c}{Asymmetric Dependence} &     & \multicolumn{4}{c}{$\thetavec_{(i,j)}$} \\
      \cmidrule{1-2}\cmidrule{4-5}\cmidrule{7-8}\cmidrule{10-13}$i$ & $j$ &     & Kendall & Spearman &     & $\Lambda_\tinytxt{Major}(0.05)$ & $\Lambda_\tinytxt{Minor}(0.05)$ &     & $\omega_{ij}$ & $\delta_i$ & $\delta_j$ & $\nu$ \\
      \multirow{2}[0]{*}{BAC} & \multirow{2}[0]{*}{VIX} &     & -0.368 & -0.517 &     & -0.030 & -0.190 &     & \multirow{2}[0]{*}{-0.650} & \multirow{2}[0]{*}{-0.983} & \multirow{2}[0]{*}{0.572} &  \\
          &     &     & (0.025) & (0.034) &     & (0.019) & (0.071) &     &     &     &     &  \\
      \multirow{2}[0]{*}{JPM} & \multirow{2}[0]{*}{VIX} &     & -0.389 & -0.541 &     & -0.014 & -0.170 &     & \multirow{2}[0]{*}{-0.677} & \multirow{2}[0]{*}{-0.840} & \multirow{2}[0]{*}{0.572} & \multirow{2}[0]{*}{3.753} \\
          &     &     & (0.023) & (0.032) &     & (0.012) & (0.046) &     &     &     &     &  \\
      \multirow{2}[0]{*}{JPM} & \multirow{2}[0]{*}{BAC} &     & 0.617 & 0.801 &     & -0.247 & 0.003 &     & \multirow{2}[0]{*}{0.902} & \multirow{2}[0]{*}{-0.840} & \multirow{2}[0]{*}{-0.983} &  \\
          &     &     & (0.017) & (0.019) &     & (0.060) & (0.005) &     &     &     &     &  \\
      \midrule
      \midrule
      \multicolumn{13}{c}{Panel B: Estimates with $q = 2$ and DIC = -1598} \\
      \midrule
      \multicolumn{2}{c}{Pair $\lrp{U_i, U_j}$} &     & \multicolumn{2}{c}{Rank Correlation} &     & \multicolumn{2}{c}{Asymmetric Dependence} &     & \multicolumn{4}{c}{$\thetavec_{(i,j)}$} \\
      \cmidrule{1-2}\cmidrule{4-5}\cmidrule{7-8}\cmidrule{10-13}$i$ & $j$ &     & Kendall & Spearman &     & $\Lambda_\tinytxt{Major}(0.05)$ & $\Lambda_\tinytxt{Minor}(0.05)$ &     & $\omega_{ij}$ & $\Delta_i^\top$ & $\Delta_j^\top$ & $\nu$ \\
      \multirow{2}[0]{*}{BAC} & \multirow{2}[0]{*}{VIX} &     & -0.402 & -0.568 &     & -0.004 & -0.048 &     & \multirow{2}[0]{*}{-0.636} & \multirow{2}[0]{*}{[-0.204, -0.920]} & \multirow{2}[0]{*}{[-0.021, 0.479]} &  \\
          &     &     & (0.020) & (0.025) &     & (0.004) & (0.060) &     &     &     &     &  \\
      \multirow{2}[0]{*}{JPM} & \multirow{2}[0]{*}{VIX} &     & -0.406 & -0.573 &     & -0.006 & -0.059 &     & \multirow{2}[0]{*}{-0.591} & \multirow{2}[0]{*}{[-0.564, -0.678]} & \multirow{2}[0]{*}{[-0.021, 0.479]} & \multirow{2}[0]{*}{12.043} \\
          &     &     & (0.019) & (0.025) &     & (0.004) & (0.046) &     &     &     &     &  \\
      \multirow{2}[0]{*}{JPM} & \multirow{2}[0]{*}{BAC} &     & 0.652 & 0.841 &     & -0.053 & 0.000 &     & \multirow{2}[0]{*}{0.861} & \multirow{2}[0]{*}{[-0.564, -0.678]} & \multirow{2}[0]{*}{[-0.204, -0.920]} &  \\
          &     &     & (0.009) & (0.008) &     & (0.062) & (0.000) &     &     &     &     &  \\
      \bottomrule
      \bottomrule
\end{tabular}%

        }
    \label{tab:app_3d}
    % \caption*{Note: }
\end{table}

\begin{table}[H]
    \centering
    \caption{Estimates of TrUST Copula of VIX and Five equities in During-COVID Period}
    \resizebox{1\textwidth}{!}{
        % Table generated by Excel2LaTeX from sheet 'Sheet9'
\begin{tabular}{cccccccccccccrc}
    \toprule
    \toprule
    Variable &     & \multicolumn{6}{c}{$\Omega$}   &     & $\deltavec_1$ & $\deltavec_2$ &     & $\nu$ &     & DIC \\
    \cmidrule{1-1}\cmidrule{3-8}\cmidrule{10-11}\cmidrule{13-13}\cmidrule{15-15}\rowcolor[rgb]{ .906,  .902,  .902} \multicolumn{15}{c}{Panel A: TruST Copula $q=1$} \\
    VIX &     & 1.000 &     &     &     &     &     &     & -0.596 & --- &     & \multirow{6}[0]{*}{7.680} &     & \multirow{6}[0]{*}{-2424} \\
    AAPL &     & -0.732 & 1.000 &     &     &     &     &     & 0.676 & --- &     &     &     &  \\
    JPM &     & -0.686 & 0.745 & 1.000 &     &     &     &     & 0.654 & --- &     &     &     &  \\
    V   &     & -0.716 & 0.818 & 0.759 & 1.000 &     &     &     & 0.680 & --- &     &     &     &  \\
    XOM &     & -0.634 & 0.643 & 0.707 & 0.648 & 1.000 &     &     & 0.612 & --- &     &     &     &  \\
    UPS &     & -0.629 & 0.703 & 0.686 & 0.704 & 0.631 & 1.000 &     & 0.996 & --- &     &     &     &  \\
    \rowcolor[rgb]{ .906,  .902,  .902} \multicolumn{15}{c}{Panel B: TruST Copula $q=2$} \\
    VIX &     & 1.000 &     &     &     &     &     &     & -0.726 & 0.672 &     & \multirow{6}[1]{*}{10.871} &     & \multirow{6}[1]{*}{\textbf{-2652}} \\
    AAPL &     & -0.853 & 1.000 &     &     &     &     &     & 0.794 & -0.734 &     &     &     &  \\
    JPM &     & -0.821 & 0.871 & 1.000 &     &     &     &     & 0.759 & -0.701 &     &     &     &  \\
    V   &     & -0.841 & 0.913 & 0.875 & 1.000 &     &     &     & 0.774 & -0.752 &     &     &     &  \\
    XOM &     & -0.756 & 0.801 & 0.803 & 0.810 & 1.000 &     &     & 0.608 & -0.965 &     &     &     &  \\
    UPS &     & -0.753 & 0.808 & 0.789 & 0.790 & 0.637 & 1.000 &     & 0.996 & -0.523 &     &     &     &  \\
    \bottomrule
    \bottomrule
    \end{tabular}%
    }
    \label{tab:app_6d_estimates_2}
    \caption*{Note: Blank cells in $\Omega$ indicate the repeated elements. Cells with ``---'' in $\deltavec_2$ indicate that the parameter does not exist in model $q=1$.}
\end{table}%

\section{Proofs}\label{app:proof}
 
\subsection{Proof of Lemma~\ref{lem1}}
Assumption~\ref{asmp:CI} asserts that the scale matrix $H=\Sigma-\Delta \Omega^{-1}\Delta^\top$ of the elliptical distribution 
$\bm{L}|\bm{X}$ is diagonal. Let $M=\Delta \Omega^{-1}\Delta^\top$ and recall that $\mbox{diag}(\Sigma)=I_q$, where \(\diag(A)\) is the diagonal matrix containing the diagonal elements of \(A\). Then, the off-diagonal elements of $H$ are zero iff 
\[H-\mbox{diag}(H)=(\Sigma-I_q)-(M-\mbox{diag}(M))=\zerovec\,.\]
The solution to this equation is  $\Sigma=I_q+(M-\mbox{diag}(M))$, so that $\Sigma$ is a function of \(\lrc{\Omega, \Delta}\).
%
%Applying Assumption~\ref{asmp:CI}, the off-diagonal elements of correlation parameter \(\Sigma\) satisfying \(\vecl\lrp{\Sigma} \equiv \vecl\lrp{\Delta \Omega^{-1} \Delta^\top}\), the deterministic expression for \(\Sigma\) is given by:
%	\[
%	\Sigma = I_q + \lrp{M - \diag\lrp{M}},
%	\]
%where  \(I_q\) is the identity matrix of size \(q\), \(M = \Delta \Omega^{-1} \Delta^\top\), and \(\diag(M)\) is a diagonal matrix containing the diagonal elements of \(M\). This construction ensures that .

\subsection{Proof of Lemma~\ref{lem2}}
	From the partitioned elliptical distribution in~\eqref{eq:sue_rep}, the conditional distribution 
	\[
	\bm{L}\mid \bm{X}=\bm{z}
	\sim
	\mbox{EL}_q\left(\Delta\Omega^{-1}\bm{z},H,g_{\bm{X}}\right).
	\]
	By Assumption~\ref{asmp:CI}, \(H\) is diagonal with 
	$H=\operatorname{diag}(h_1,\ldots,h_q)$, and we set 
	$D_H=\operatorname{diag}\left(\sqrt{h_1},\ldots,\sqrt{h_q}\right)$.
	Because \(R\) is a correlation matrix, \(\operatorname{diag}(\Sigma)=I_q\). Hence, with
	\(M=\Delta\Omega^{-1}\Delta^\top\),
	\[
	H=\Sigma-M.
	\]
Lemma~\ref{lem1} establishes that $\Sigma=I_q+\{M-\operatorname{diag}(M)\}$, so that
	\[
	H
	=
	\Sigma-M
	=
	I_q-\operatorname{diag}(M).
	\]
	Writing \(\Delta^\top=[\bm{\delta}_1|\cdots|\bm{\delta}_q]\), the \(k\)-th diagonal element of
	\(M\) is $\bm{\delta}_k^\top\Omega^{-1}\bm{\delta}_k$, so that
	\[
	H=
	\operatorname{diag}\left(
	1-\bm{\delta}_1^\top\Omega^{-1}\bm{\delta}_1,\ldots,
	1-\bm{\delta}_q^\top\Omega^{-1}\bm{\delta}_q
	\right).
	\]
	
	Now standardize the conditional distribution of \(\bm{L}\mid \bm{X}=\bm{z}\). Let
	\[
	\widetilde{\bm{L}}=D_H^{-1}\bm{L}.
	\]
	Then
	\[
	\widetilde{\bm{L}}\mid \bm{X}=\bm{z}
	\sim
	EL_q\left(D_H^{-1}\Delta\Omega^{-1}\bm{z},I_q,g_{\bm{X}}\right).
	\]
	Consequently,
	\[
	F_{EL}\left(\Delta\Omega^{-1}\bm{z};H,g_{\bm{X}}\right)
	=
	F_{EL}\left(D_H^{-1}\Delta\Omega^{-1}\bm{z};I_q,g_{\bm{X}}\right).
	\]
	The \(k\)-th element of \(D_H^{-1}\Delta\Omega^{-1}\bm{z}\) is
	\[
	h_k^{-1/2}\bm{\delta}_k^\top\Omega^{-1}\bm{z}.
	\]
	Define
	\[
	\bm{\alpha}_k
	=
	h_k^{-1/2}\Omega^{-1}\bm{\delta}_k
	=
	\left(1-\bm{\delta}_k^\top\Omega^{-1}\bm{\delta}_k\right)^{-1/2}
	\Omega^{-1}\bm{\delta}_k,
	\qquad k=1,\ldots,q.
	\]
	Since \(\Omega\) is symmetric,
	\[
	h_k^{-1/2}\bm{\delta}_k^\top\Omega^{-1}\bm{z}
	=
	\bm{\alpha}_k^\top\bm{z}.
	\]
	Therefore, with \(A=[\bm{\alpha}_1|\cdots|\bm{\alpha}_q]\),
	\[
	D_H^{-1}\Delta\Omega^{-1}\bm{z}
	=
	A^\top\bm{z}.
	\]
	Hence,
	\[
	F_{EL}\left(\Delta\Omega^{-1}\bm{z};H,g_{\bm{X}}\right)
	=
	F_{EL}\left(A^\top\bm{z};I_q,g_{\bm{X}}\right),
	\]
	which completes the proof.

\subsection{Proof of Lemma~\ref{theo:PI}}\label{proof_lem3}
For any permutation \(\pi\in G(q)\), let the permutation matrix \(P_{\pi}\) be defined by
\[
  P_{\pi} = (p_{ij})_{q\times q}, 
  \quad
  p_{ij} = \mathds{1}\lrc{\pi(i) = j}.
\]
Following~Wang et al.~(2023, Proposition~3),
% \citet[Proposition 3]{wang2023non}, 
if 
\(\mY\sim\USE_q(\Omega,\Delta,\Sigma,g)\), then
\[
  \mY\overset{d}{=}\mY_{\pi},
\]
where
\(\mY_{\pi}\sim\USE_q\bigl(\Omega,\,P_{\pi}\Delta,\,P_{\pi}\Sigma\,P_{\pi}^\top,\,g\bigr)\).
Hence, neither \(\Delta\) nor \(\Sigma\) can be uniquely recovered; they are identifiable only up to permutations in \(G(q)\), thereby establishing un-identification under latent variable label switching \(P_{\pi} \mL = \mL_{\pi} = \lrp{L_{\pi(1)}, \ldots, L_{\pi(q)}}^\top \).

By Lemma~\ref{lem2}, permuting the rows of constrained skewness parameter \(\Delta\) via \(P_{\pi}\) yields  
\(\;A P_{\pi}^\top = A_{\pi} = \lrp{\alphavec_{\pi(1)},\ldots,\alphavec_{\pi(q)}}\),  
so that  
\(\mZ_{\pi} \overset{\dd}{=} (\mX\mid \mL_{\pi} >\zerovec)\sim \TRUSE_q\lrp{\Omega, A_{\pi}, g}\)  
has density  
\begin{equation*}
	f_{\TRUSE, q}\lrp{\zvec; \Omega, A_{\pi}, g }  
	= f_{\EL}\lrp{\zvec; \Omega, g} 
	\frac{  F_{\EL} \lrp{A_\pi^\top \zvec;I_q, g_\mX } }{ F_{\EL} \lrp{\zerovec;  P_{\pi} \Sigma P_{\pi}^\top, g } },
\end{equation*}
where $A_\pi$ is $A$ with its columns permuted consistently with the rows of $\Delta$ and the components of $\bm{L}$.
This coincides exactly with the right-hand side of equation~\eqref{eq:fse_pdf} for
\(\mZ\sim\TRUSE_q(\Omega,\,A,\,g)\).  
Hence, for any \(\pi\), \(\mZ_{\pi}\) is observationally equivalent to \(\mZ\), establishing un-identification of parameter $A$ under the latent vector permutation \(\mL_{\pi}\).

\subsection{Proof of Theorem~\ref{thm1}}
%Following the proof in~\ref{proof_lem3}, enforcing a permutation on \(\Sigma\) does not uniquely recover \(A\), since Lemma~\ref{lem2} shows \(A\) is a function of both \(\Delta\) and \(\Omega\).  
%We therefore focus instead on permuting \(H\), which corresponds to permuting the latent vector \(\bigl(\mL\mid\mX\bigr)\).

From Lemma~\ref{lem2} the scale matrix of the elliptical distribution of  \(\bigl(\mL\mid\mX\bigr)\) is
\[
H = \Sigma - \Delta \Omega^{-1} \Delta^\top
= \diag(h_{1},\dots,h_{q}),
\]
where each diagonal value \(h_{k} = 1 - \deltavec_{k}^\top \Omega^{-1} \deltavec_{k}\) coincides with an eigenvalue \(\lambda_{k}\in[0,1]\). Then under Assumption~\ref{asmp:LP}, for any \(\pi\in G(q)\) define the weighted sum
\[
S(\pi)
= \sum_{k=1}^{q} \pi(k)\,h_{\pi(k)}.
\]
By the rearrangement inequality, the maximizer \(\pi^{*}=\arg\max_{\pi\in G(q)}  S(\pi)\) satisfies  
\(
S(\pi^{*}) \ge S(\pi)
\)
with inequality only when  
\[
0 \le h_{\pi^{*}(1)} \le \cdots \le h_{\pi^{*}(q)} \le 1.
\]

%Under Lemma~\ref{lem1}, \(\Sigma\) is determined by \(\Omega\) and \(A\).  Lemma~\ref{lem2} then gives  
%\[
%  H = \Sigma - \Delta \Omega^{-1} \Delta^\top
%    = \diag(h_{1},\dots,h_{q}),
%\]
%each \(h_{k} = 1 - \deltavec_{k}^\top \Omega^{-1} \deltavec_{k}\) coincides with an eigenvalue \(\lambda_{k}\in[0,1]\), and the values \(\{h_{k}\} \) are strictly distinct.  
%Then under Assumption~\ref{asmp:LP}, for any \(\pi\in G(q)\) define the weighted sum
%\[
%  S(\pi)
%  = \sum_{k=1}^{q} \pi(k)\,h_{\pi(k)}.
%\]
%By the rearrangement inequality, the maximizer \(\pi^{*}=\arg\max_{\pi\in G(q)}  S(\pi)\) satisfies  
%\(
%  S(\pi^{*}) \ge S(\pi)
%\)
%with inequality only when  
%\[
%  0 \le h_{\pi^{*}(1)} \le \cdots \le h_{\pi^{*}(q)} \le 1.
%\]
Hence \(\pi^{*}\) uniquely aligns the latent coordinates with the diagonal entries of \(H\) for \(\lrp{\mL_{\pi^*}\mid \mX}\).  Since there is a one‑to‑one correspondence between \(\deltavec_{k}\) and \(\alphavec_{k}\), \(\pi^{*}\) simultaneously identifies \(A\), \(\Delta\), and \(\Sigma\).

\subsection{Proof of Lemma~\ref{lem:marginal}}
For any index set \( J \subset \lrc{1,\ldots,d} \) of size $|J| = d_J<d$, the joint random vector \(\mZ \overset{\dd}{=} \mX | \mL > \zerovec \sim TrUST_q(\Omega, A, \nu)\), following the stochastic representation in Equations~\eqref{eq:sue_rep} and~\eqref{eq:cond_trunc}, the marginal random vector \( \mZ_J = \lrp{Z_{J(1)}, \ldots, Z_{J(d_J)}}^\top \) has the representation: 
\[
\mZ_J \overset{\dd}{=} \mX_J | \mL > \zerovec 
\]
with \( \mX_J = \lrp{X_{J(1)},\ldots, X_{J(d_J)}}^\top\). It admits the joint expression 
\[
\begin{pmatrix}
	\mX_J \\
	\mL
\end{pmatrix}
\sim \T\lrp{\zerovec, R_J, \nu}, \quad
R_J = 
\begin{pmatrix}
	\Omega_J & \Delta_J^\top \\
	\Delta_J & \Sigma
\end{pmatrix},
\]
where let \( \Omega_J = \lrc{\omega_{ij}}_{i,j \in J} \) denote the \(d_J \times d_J\) principal submatrix of \(\Omega = \lrp{\omega_{ij} }_{d \times d}\), and let \( \Delta_J = \lrp{\Delta_{J(1)}, \ldots, \Delta_{J(d_J)}}\) denote the \(q \times d_J\) submatrix of \(\Delta\) formed by its columns indexed by \(J\). 

Using the Bayes' theorem, the random vector \(\mZ_J \) follows density function that 
\[
p_{\mZ_J}(\zvec_J) = t_{d_J}(\zvec_J ; \Omega_J, \nu) \frac{T_q\lrp{ \sqrt{\frac{\nu+d_J}{\nu+Q(\zvec_J)}} \Delta_J \Omega^{-1}_J \zvec_J; \Sigma - \Delta_J \Omega^{-1}_J \Delta^\top_J,\nu+d_J }}{T_q\lrp{\zerovec; \Sigma, \nu}}. 
\]
The conditional distribution of \( \mL | \mX_J \) has scale matrix 
\[
H_J = \Sigma - \Delta_J \Omega_J^{-1} \Delta_J^\top.
\]
%By the semi-graphoid axioms, %see~\cite[Chapter~2]{studeny2006probabilistic},
%see Chapter~2 of Studeny~(2006), 
% the Schur-complement cannot in general cancel all off-diagonal entries of \(\Sigma\) unless Assumption~\ref{asmp:CI} is strengthened to require 
%\[
%\mL \independent \mX_{-J} | \mX_J, 
%\] 
%which asserts the graphical structure of TrUST model.
Application of Bayes theorem then completes the proof that \( \mZ_J \sim \UST_q\lrp{\Omega_J, \Delta_J, \Sigma, \nu} \).

\subsection{Proof of Lemma~\ref{theo:kendall}}
From Equation~2.41 of Joe~(2014), the Kendall correlation for continuous marginals can be represented as
\begin{equation}\label{eq:proof_kendall_start}
	\begin{aligned}
		\rho_K\lrp{Z_1,Z_2}
		&=4\bbP\lrp{Z_1>Z_1',Z_2>Z_2'}-1\\
		&=4\bbP\lrp{\mZ-\mZ'>\zerovec}-1,
	\end{aligned}
\end{equation}
where \(\mZ'=(Z_1',Z_2')^\top\) is an independent copy of
\(\mZ=(Z_1,Z_2)^\top\).
Use the hidden-truncation representation
\(\mZ=\mX\mid\mL>\zerovec\) and
\(\mZ'=\mX'\mid\mL'>\zerovec\), with independent copies throughout. Then
\begin{equation}\label{eq:proof_kendall_ratio}
	\begin{aligned}
		\rho_{K,\UST}\lrp{Z_1,Z_2}
		&=4\bbP\lrp{\mX-\mX'>\zerovec\mid \mL>\zerovec,\mL'>\zerovec}-1\\
		&=4\frac{
			\bbP\lrp{\mX-\mX'>\zerovec,\mL>\zerovec,\mL'>\zerovec}
		}{
			\bbP\lrp{\mL>\zerovec,\mL'>\zerovec}
		}-1.
	\end{aligned}
\end{equation}

Let \(W,W'\overset{\iid}{\sim}\Gammadst(\nu/2,\nu/2)\). Conditional on
\(W=w,W'=w'\), write the Gaussian mixture variables as
\((\widetilde{\mX},\widetilde{\mL})^\top\) and
\((\widetilde{\mX}',\widetilde{\mL}')^\top\), with independent scale matrices
\(w^{-1}R\) and \({w'}^{-1}R\), where
\[
R=
\begin{pmatrix}
	\Omega &\Delta^\top \\ 
	\Delta &\Sigma
\end{pmatrix}.
\]
For \(\widetilde{\mX}^*=\widetilde{\mX}-\widetilde{\mX}'\), define the joint Gaussian vector
\[
\mY_K=
\begin{pmatrix}
	\widetilde{\mX}^*\\
	\widetilde{\mL}\\
	\widetilde{\mL}'
\end{pmatrix}.
\]
The numerator in \eqref{eq:proof_kendall_ratio}, conditional on \(W=w,W'=w'\), is the orthant probability
\(\bbP\lrp{\mY_K>\zerovec}\). Since the two copies are independent,
\[
\mY_K\sim N_{2+2q}\lrp{\zerovec,\,V_K(w,w')},
\]
where
\[
V_K(w,w')=
\begin{pmatrix}
	\Omega/w+\Omega/w' & \Delta^\top/w & -\Delta^\top/w'\\
	\Delta/w & \Sigma/w & \mathbf{0}_{q\times q}\\
	-\Delta/w' & \mathbf{0}_{q\times q} & \Sigma/w'
\end{pmatrix}.
\]
Equivalently, with
\[
D_K(w,w')=\diag\lrp{
	\sqrt{w^{-1}+{w'}^{-1}}\,I_2,\;
	w^{-1/2}I_q,\;
	{w'}^{-1/2}I_q
},
\]
we have \(h_k(w,w')=D_K(w,w')^{-1} V_K(w,w')D_K(w,w')^{-1}\). Because all diagonal entries of \(D_K(w,w')\) are positive, the zero-threshold orthant probability is unchanged by this componentwise standardization. Hence
\[
h_k(w,w')=
\begin{pmatrix}
	\Omega&B_K(w,w')^\top\\
	B_K(w,w')&I_2\otimes\Sigma
\end{pmatrix},
\quad
B_K(w,w')=
\begin{pmatrix}
	\sqrt{\frac{w'}{w+w'}}\Delta\\
	-\sqrt{\frac{w}{w+w'}}\Delta
\end{pmatrix}.
\]
The denominator in \eqref{eq:proof_kendall_ratio} is
\(\lrb{\Phi_q(\zerovec;\Sigma)}^2\), because centered Gaussian orthant probabilities are invariant to positive scalar rescaling. Averaging over \(W,W'\) gives
\[
\rho_{K,\UST}\lrp{Z_1,Z_2}
=4\frac{
	\mathbb{E}_{W,W'}\lrb{
	\Phi_{2+2q}\lrp{\zerovec;h_k(W,W')}
	}
}{\lrb{\Phi_q(\zerovec;\Sigma)}^2}-1.
\]

\subsection{Proof of Lemma~\ref{theo:spearman}}
Following Chapter~2.12.5 of Joe~(2014), Spearman's rank correlation can be written as
\begin{equation}\label{eq:proof_spearman_start}
	\begin{aligned}
		\rho_S\lrp{Z_1,Z_2}
		&=12\bbP\lrp{Z_1>Z_1'',Z_2>Z_2''}-3\\
		&=12\bbP\lrp{\mZ-\mZ''>\zerovec}-3,
	\end{aligned}
\end{equation}
where \(Z_1''\) is an independent copy of the first marginal and \(Z_2''\) is an independent copy of the second marginal, with \(Z_1''\independent Z_2''\).
Using the hidden-truncation representation,
\[
Z_1''=X_1''\mid \mL_{(1)}''>\zerovec,\qquad
Z_2''=X_2''\mid \mL_{(2)}''>\zerovec,
\]
where the two marginal copies are independent. Hence
\begin{equation}\label{eq:proof_spearman_ratio}
	\begin{aligned}
		\rho_{S,\UST}\lrp{Z_1,Z_2}
		&=12\bbP\lrp{\mX-\mX''>\zerovec\mid
		\mL>\zerovec,\mL_{(1)}''>\zerovec,\mL_{(2)}''>\zerovec}-3\\
		&=12\frac{
			\bbP\lrp{\mX-\mX''>\zerovec,\mL>\zerovec,
			\mL_{(1)}''>\zerovec,\mL_{(2)}''>\zerovec}
		}{
			\bbP\lrp{\mL>\zerovec,\mL_{(1)}''>\zerovec,\mL_{(2)}''>\zerovec}
		}-3.
	\end{aligned}
\end{equation}

Let \(W,W_1,W_2\overset{\iid}{\sim}\Gammadst(\nu/2,\nu/2)\) be the independent scale variables for
\((\mX,\mL)\), \((X_1'',\mL_{(1)}'')\), and \((X_2'',\mL_{(2)}'')\), respectively. Conditional on \(W=w,W_1=w_1,W_2=w_2\), set
\[
\widetilde{\mX}^*=(\widetilde X_1-\widetilde X_1'',\widetilde X_2-\widetilde X_2'')^\top
\]
and define the full joint Gaussian vector
\[
\mY_S(w,w_1,w_2)=
\begin{pmatrix}
	\widetilde{\mX}^*\\
	\widetilde{\mL}\\
	\mtildeL_{(1)}''\\
	\mtildeL_{(2)}''
\end{pmatrix}.
\]
The numerator in \eqref{eq:proof_spearman_ratio}, conditional on the scale variables, is
\(\bbP\{\mY_S(w,w_1,w_2)>\zerovec\}\), with
\[
\mY_S(w,w_1,w_2)\sim N_{2+3q}\lrp{\zerovec,\,V_S(w,w_1,w_2)}.
\]
Writing \(\Delta_j\) for column \(j\) of \(\Delta\), the covariance matrix is
\[
V_S(w,w_1,w_2)=
\begin{pmatrix}
	\Omega/w+\diag(1/w_1,1/w_2) & \Delta^\top/w & B_1(w_1)^\top & B_2(w_2)^\top\\
	\Delta/w & \Sigma/w & \mathbf{0}_{q\times q} & \mathbf{0}_{q\times q}\\
	B_1(w_1) & \mathbf{0}_{q\times q} & \Sigma/w_1 & \mathbf{0}_{q\times q}\\
	B_2(w_2) & \mathbf{0}_{q\times q} & \mathbf{0}_{q\times q} & \Sigma/w_2
\end{pmatrix},
\]
where
\[
B_1(w_1)=\left(-\Delta_1/w_1,\zerovec\right),
\qquad
B_2(w_2)=\left(\zerovec,-\Delta_2/w_2\right).
\]
Let
\[
D_S(w,w_1,w_2)=\diag\lrp{
	\sqrt{w^{-1}+w_1^{-1}},\;
	\sqrt{w^{-1}+w_2^{-1}},\;
	w^{-1/2}I_q,\;
	w_1^{-1/2}I_q,\;
	w_2^{-1/2}I_q
}.
\]
Then \(R_S(w,w_1,w_2)=D_S(w,w_1,w_2)^{-1}V_S(w,w_1,w_2)D_S(w,w_1,w_2)^{-1}\). Again, positive componentwise standardization leaves the zero-threshold orthant probability unchanged. Thus
\[
R_S(w,w_1,w_2)=
\begin{pmatrix}
	\Omega^\star_S(w,w_1,w_2)&B_S(w,w_1,w_2)^\top\\
	B_S(w,w_1,w_2)&I_3\otimes\Sigma
\end{pmatrix},
\]
where \(\Omega^\star_S(w,w_1,w_2)\) has unit diagonal and off-diagonal element
\[
\omega\sqrt{\frac{w_1w_2}{(w+w_1)(w+w_2)}},
\]
and
\[
B_S(w,w_1,w_2)=
\begin{pmatrix}
	\sqrt{\frac{w_1}{w+w_1}}\Delta_1&
	\sqrt{\frac{w_2}{w+w_2}}\Delta_2\\
	-\sqrt{\frac{w}{w+w_1}}\Delta_1&\zerovec\\
	\zerovec&-\sqrt{\frac{w}{w+w_2}}\Delta_2
\end{pmatrix}.
\]
The denominator in \eqref{eq:proof_spearman_ratio} is
\(\lrb{\Phi_q(\zerovec;\Sigma)}^3\). Averaging over the independent scale variables gives
\[
\rho_{S,\UST}\lrp{Z_1,Z_2}
=12\frac{
	\mathbb{E}_{W,W_1,W_2}\lrb{
	\Phi_{2+3q}\lrp{\zerovec;R_S(W,W_1,W_2)}
	}
}{\lrb{\Phi_q(\zerovec;\Sigma)}^3}-3.
\]

% \section{Angular Correlation Specification}\label{app:parameter}
% Define polar matrix $\Psi$ as the strict lower triangular matrix:
% \begin{equation}
%     \Psi = 
%     \begin{pmatrix}
%         0 & 0 & 0 & \cdots & 0 & 0 \\
%         \psi_{2,1} & 0 & 0 & \cdots & 0 & 0 \\
%         \psi_{3,1} & \psi_{3,2} & 0 & \cdots & 0 & 0\\
%         \vdots & \vdots & \vdots & \ddots & \vdots & \vdots\\
%         \psi_{d,1}& \psi_{d,2} & \psi_{d,3} & \cdots & \psi_{d,d-1} & 0
%     \end{pmatrix},
% \end{equation}
% with two corresponding matrix 
% \begin{equation}
% 	\mC  = 
%         \begin{pmatrix}
%             1 & 1 & 1 & \cdots & 1 & 1 \\
%             c_{2,1} & 1 & 1 & \cdots & 1 & 1 \\
%             c_{3,1} & c_{3,2} & 1 & \cdots & 1 & 1 \\
%             \vdots & \vdots & \vdots & \ddots & \vdots & \vdots\\
%             c_{d,1}& c_{d,2} & c_{d,3} & \cdots & c_{d,d-1} & 1
%         \end{pmatrix}, 
% 	\quad \text{and} \quad
%     \mS = 
%         \begin{pmatrix}
%             1 & 0 & 0 & 0 & \cdots & 0  \\
%             1 & s_{2,1} & 0 & 0 & \cdots & 0  \\
%             1 & s_{3,1} & s_{3,2} & 0 & \cdots & 0 \\
%             \vdots & \vdots & \vdots & \ddots & \vdots & \vdots\\
%             1 & s_{d,1}& s_{d,2} & s_{d,3} & \cdots & s_{d,d-1} 
%         \end{pmatrix},
%     %   \end{tabular}}%
%     \label{tab:addlabel}% 
% \end{equation}%
% where $s_{i,j} \coloneq \sin(\psi_{i,j}), c_{i,j} \coloneq \cos(\psi_{i,j})$ for all \( i > j\). Then, the Cholesky matrix $B$ of Pearson-correlation parameter $\Omega = B B^\top$ has parameterization

\subsection{Proof of Corollary~\ref{theo:etrust}}
The ``extended-TrUST'' means that given the joint random vector follows student-t distribution 
\( \lrp{\mX^\top, \mL^\top}^\top \sim \T\lrp{\zerovec, R, \nu} \), the observable vector \(\mX\) is selected by the hidden vector with location \(\tauvec = \lrp{\tau_1,\ldots,\tau_q}^\top\) follows the stochastic representation 
\[
\mZ = \mX | \mL + \tauvec > \zerovec.
\]
One can derive the density function using Bayes' theorem that 
\begin{equation}\label{eq:extend_trust}
	\begin{aligned}
		p_{\mZ}(\zvec) 
		&= p_{\mX}\lrp{\zvec} \frac{\dsP\lrp{\mL + \tauvec > \zerovec | \mX = \zvec} }{\dsP\lrp{\mL + \tauvec > \zerovec}} \\
		&= t_d\lrp{\zvec;  \Omega, \nu} 
		\frac{ T_q \lrp{ \sqrt{\frac{\nu+d}{\nu+Q \lrp{\zvec}}} \lrp{\tauvec + \Delta \Omega^{-1} \zvec}; \Sigma - \Delta \Omega^{-1} \Delta^\top, \nu +d } }{ T_q \lrp{\tauvec; \Sigma, \nu} }, \\
		% &= t_d\lrp{\zvec;  \Omega, \nu} \frac{\prod_{k=1}^{q} T_1 \lrp{ \sqrt{\frac{\nu+d}{\nu+Q \lrp{\zvec}}}  \lrp{\tautilde_k + \alphavec_k^\top \zvec}; 1, \nu +d } }{ T_q \lrp{\tauvec; \Sigma, \nu} },
	\end{aligned}
\end{equation}
while location-scale transformation still allows the Lemmas~\ref{lem1} and~\ref{lem2}, we complete the proof that \( \mZ \sim \ETRUST\lrp{\Omega, A, \Sigma, \nu, \tauvec} \) with density function:
\begin{equation*}
	\begin{aligned}
		f_{\ETRUST, q}\lrp{\zvec; \Omega, A, \nu, \tauvec} 
		% &= p_{\mX}\lrp{\zvec} \frac{\dsP\lrp{\mL + \tauvec > \zerovec | \mX = \zvec} }{\dsP\lrp{\mL + \tauvec > \zerovec}} \\
		% &= t_d\lrp{\zvec;  \Omega, \nu} 
		% \frac{ T_q \lrp{ \sqrt{\frac{\nu+d}{\nu+Q \lrp{\zvec}}} \lrp{\tauvec_k + \Delta \Omega^{-1} \zvec}; \Sigma - \Delta \Omega^{-1} \Delta^\top, \nu +d } }{ T_q \lrp{\tauvec; \Sigma, \nu} }, \\
		&= t_d\lrp{\zvec;  \Omega, \nu} \frac{ T_q \lrp{ \sqrt{\frac{\nu+d}{\nu+Q \lrp{\zvec}}}  \lrp{\bm{\tautilde} + A^\top \zvec}; I_q, \nu +d } }{ T_q \lrp{\tauvec; \Sigma, \nu} }
	\end{aligned}
\end{equation*}
with \( \tautilde_k = \tau_k \lrp{1-\deltavec_k^\top \Omega^{-1} \deltavec_k}^{-1/2}\) for 
$k=1,\ldots,q$ and $\bm{\tautilde}=(\tautilde_1,\ldots,\tautilde_q)^\top$. 

% \begin{equation*}
% 	f_{\mZ}\lrp{\zvec} = t_d\lrp{\zvec; \zerovec, \Omega, \nu} \frac{
% 		\prod_{k=1}^{q} T_1\lrp{\tautildevec \alphavec_k^\top \zvec; 1}
% 	}{
% 		T_q\lrp{\tauvec; \Sigma, \nu}
% 	}
% \end{equation*}

\subsection{Proof of Corollary~\ref{theo:condtrust}}
Let the joint distribution \(\mZ = (\mZ_1^\top, \mZ_2^\top)^\top \sim TrUST_{d_1+d_2,q}(\zerovec, \Omega, A, \nu)\), and the scale matrix \(\Omega\) and skewness matrix \(A\) are conformably partitioned as:
\[
\Omega = 
\begin{pmatrix}
\Omega_1 & \Omega_{12} \\ 
\Omega_{21} & \Omega_2
\end{pmatrix}, 
A = 
\begin{pmatrix}
	A_1 \\ A_2
\end{pmatrix}
\]
with \( A_1 = \lrp{\alphavec_{k(1)}}_{d_1 \times q},  A_2 = \lrp{\alphavec_{k(2)}}_{d_2 \times q} \).

Under the stochastic representation in Equation~\ref{eq:cond_trunc}, we may write \( \mZ = \mX | \mL > \zerovec =\lrp{\mX_1^\top, \mX_2^\top}^\top | \mL > \zerovec \). Then the conditional distribution 
\(\mZ_{1|2} \coloneq (\mZ_1 | \mZ_2 = \zvec_2)\).
It follows that the hidden truncation that
\[
\mZ_{1|2} \overset{\dd}{=} \lrp{\mX_* \mid \mL_* > \zerovec }
= \Bigl(\lrp{\mX_1 | \mX_2 = \zvec_2} \mid \lrp{\mL > \zerovec | \mX_2 = \zvec_2}\Bigr).
\]
Using successive conditioning and Bayes’ theorem, the conditional density of \(\mZ_{1\mid2}\) can be written as
\begin{equation*} 
	\begin{aligned}
		p_{\mZ_{1|2}}(\zvec_1) 
		&= p_{\mX_*}(\zvec_1) \frac{ \dsP\lrp{\mL_* > \zerovec | \mX_* = \zvec_1} }{ \dsP\lrp{\mL_* > \zerovec} } \\
		&= p_{\mX_*}(\zvec_1) \frac{ \dsP\lrp{\mL > \zerovec | \mX_{1} = \zvec_1, \mX_2 = \zvec_2} }{ \dsP\lrp{\mL > \zerovec | \mX_2 = \zvec_2} } \\
		&= t_{d_1}\lrp{\zvec_1 - \muvec_*; \frac{\nu+Q(\zvec_2)}{\nu+d_2} \Omega_*, \nu+d_2} 
		\frac{T_q\lrp{\sqrt{\frac{\nu+d_2+d_1}{\nu + Q(\zvec)} } (A_1^\top \zvec_1  + A_2^\top \zvec_2); I_q, \nu+d_2+d_1} }{ T_q\lrp{\Delta_2 \Omega_{2}^{-1} \zvec_2; \frac{\nu+Q(\zvec_2)}{\nu+d_2} (\Sigma - \Delta_2 \Omega_{2}^{-1} \Delta_2^\top), \nu+d_2 } },
	\end{aligned}
\end{equation*}
where \( \muvec_* = \Omega_{12}\Omega_2^{-1} \zvec_2\), \( \Omega_{*} = \Omega_1 - \Omega_{12} \Omega_2^{-1} \Omega_{21}\), \(Q(\zvec_2) = \zvec_2^\top \Omega_{2}^{-1} \zvec_2  \). The numerator follows by applying Lemma~\ref{lem2} to the full conditioning event $\bm{X}=(\mX_1^\top,\mX_2^\top)^\top=\zvec$. 
%Hence, no further parameterization is required, which keeps the model tractable in both identification and computation. This completes the proof.

\noindent {\LARGE \bf References}
\begin{itemize}
	\item Joe, H. (2014). Dependence Modeling with Copulas. Chapman and Hall/CRC.
	\item Studeny, M. (2006). {\em Probabilistic conditional independence structures}. Springer Science \& Business Media. 
	\item Wang, K, Arellano-Valle, R.B, Azzalini, A. and Genton, M.G. (2023). On the non-identifiability of unified skew-normal distributions. {\em Stat}, 12(1): e597.
\end{itemize}
\end{document}